\documentclass[twocolumn, astrosymb]{aastex631} 

\usepackage{graphicx}
\usepackage{amsmath}
\usepackage{amssymb}
\usepackage{makecell}
\usepackage{dirtytalk}
\usepackage{threeparttable}
\usepackage{multirow}
\usepackage{enumitem}
\usepackage{pifont}

\usepackage{soul}

\usepackage{array}

\makeatletter
\def\switch@array{}
\makeatother

\usepackage{bm}
\usepackage[utf8]{inputenc}
\DeclareUnicodeCharacter{2212}{\bm{\text{--}}}

\shorttitle{The Stellar Populations and Rest-Frame Colors of Star-Forming Galaxies at $z \approx 8$}
\shortauthors{Helton et al.}

\graphicspath{{./}{files/}}

\begin{document}

\title{The Stellar Populations and Rest-Frame Colors of Star-Forming Galaxies at \boldmath$z \approx 8$: \\ Exploring the Impact of Filter Choice and Star Formation History Assumption with JADES}


\author[0000-0003-4337-6211]{Jakob M. Helton}
\affiliation{Steward Observatory, University of Arizona, 933 N. Cherry Ave., Tucson, AZ 85721, USA}

\author[0000-0002-8909-8782]{Stacey Alberts}
\affiliation{Steward Observatory, University of Arizona, 933 N. Cherry Ave., Tucson, AZ 85721, USA}
\affiliation{AURA for the European Space Agency (ESA), Space Telescope Science Institute, 3700 San Martin Dr., Baltimore, MD 21218, USA}

\author[0000-0003-2303-6519]{George H. Rieke}
\affiliation{Steward Observatory, University of Arizona, 933 N. Cherry Ave., Tucson, AZ 85721, USA}

\author[0000-0003-4565-8239]{Kevin N. Hainline}
\affiliation{Steward Observatory, University of Arizona, 933 N. Cherry Ave., Tucson, AZ 85721, USA}

\author[0000-0001-7673-2257]{Zhiyuan Ji}
\affiliation{Steward Observatory, University of Arizona, 933 N. Cherry Ave., Tucson, AZ 85721, USA}

\author[0000-0002-7893-6170]{Marcia J. Rieke}
\affiliation{Steward Observatory, University of Arizona, 933 N. Cherry Ave., Tucson, AZ 85721, USA}

\author[0000-0002-9280-7594]{Benjamin D. Johnson}
\affiliation{Center for Astrophysics $|$ Harvard \& Smithsonian, 60 Garden St., Cambridge, MA 02138, USA}

\author[0000-0002-4271-0364]{Brant Robertson}
\affiliation{Department of Astronomy and Astrophysics, University of California, Santa Cruz, 1156 High St., Santa Cruz, CA 95064, USA}

\author[0000-0002-8224-4505]{Sandro Tacchella}
\affiliation{Kavli Institute for Cosmology, University of Cambridge, Madingley Rd., Cambridge CB3 OHA, UK}
\affiliation{Cavendish Laboratory, University of Cambridge, 19 JJ Thomson Ave., Cambridge CB3 0HE, UK}

\author[0000-0003-1432-7744]{Lily Whitler}
\affiliation{Steward Observatory, University of Arizona, 933 N. Cherry Ave., Tucson, AZ 85721, USA}



%

\author[0000-0003-0215-1104]{William M. Baker}
\affiliation{DARK, Niels Bohr Institute, University of Copenhagen, Jagtvej 128, DK-2200, Copenhagen, Denmark}

\author[0000-0003-0883-2226]{Rachana Bhatawdekar}
\affiliation{European Space Agency (ESA), European Space Astronomy Centre (ESAC), Camino Bajo del Castillo s/n, 28692 Villanueva de la Ca\~{n}ada, Madrid, Spain}

\author[0000-0003-4109-304X]{Kristan Boyett}
\affiliation{School of Physics, University of Melbourne, Parkville 3010, VIC, Australia}
\affiliation{ARC Centre of Excellence for All Sky Astrophysics in 3 Dimensions (ASTRO 3D), Australia}

\author[0000-0002-8651-9879]{Andrew J. Bunker}
\affiliation{Department of Physics, University of Oxford, Denys Wilkinson Building, Keble Rd., Oxford OX1 3RH, UK}

\author[0000-0002-1617-8917]{Phillip A. Cargile}
\affiliation{Center for Astrophysics $|$ Harvard \& Smithsonian, 60 Garden St., Cambridge, MA 02138, USA}

\author[0000-0002-6719-380X]{Stefano Carniani}
\affiliation{Scuola Normale Superiore, Piazza dei Cavalieri 7, I-56126 Pisa, Italy}

\author[0000-0003-3458-2275]{Stephane Charlot}
\affiliation{Sorbonne Universit\'{e}, CNRS, UMR 7095, Institut d'Astrophysique de Paris, 98 bis bd Arago, 75014 Paris, France}

\author[0000-0002-7636-0534]{Jacopo Chevallard}
\affiliation{Department of Physics, University of Oxford, Denys Wilkinson Building, Keble Rd., Oxford OX1 3RH, UK}

\author[0000-0002-9551-0534]{Emma Curtis-Lake}
\affiliation{Centre for Astrophysics Research, Department of Physics, Astronomy and Mathematics, University of Hertfordshire, Hatfield AL10 9AB, UK}

\author[0000-0003-1344-9475]{Eiichi Egami}
\affiliation{Steward Observatory, University of Arizona, 933 N. Cherry Ave., Tucson, AZ 85721, USA}

\author[0000-0002-2929-3121]{Daniel J. Eisenstein}
\affiliation{Center for Astrophysics $|$ Harvard \& Smithsonian, 60 Garden St., Cambridge, MA 02138, USA}

\author[0000-0002-8543-761X]{Ryan Hausen}
\affiliation{Department of Physics and Astronomy, The Johns Hopkins University, 3400 N. Charles St., Baltimore, MD 21218, USA}

\author[0000-0002-6221-1829]{Jianwei Lyu}
\affiliation{Steward Observatory, University of Arizona, 933 N. Cherry Ave., Tucson, AZ 85721, USA}

\author[0000-0002-4985-3819]{Roberto Maiolino}
\affiliation{Kavli Institute for Cosmology, University of Cambridge, Madingley Rd., Cambridge CB3 OHA, UK}
\affiliation{Cavendish Laboratory, University of Cambridge, 19 JJ Thomson Ave., Cambridge CB3 0HE, UK}
\affiliation{Department of Physics and Astronomy, University College London, Gower St., London WC1E 6BT, UK}

\author[0000-0002-7524-374X]{Erica Nelson}
\affiliation{Department for Astrophysical and Planetary Science, University of Colorado, Boulder, CO 80309, USA}

\author[0000-0003-4528-5639]{Pablo G. P\'{e}rez-Gonz\'{a}lez}
\affiliation{Centro de Astrobiolog\'{i}a (CAB), CSIC–INTA, Ctra. de Ajalvir km 4, Torrej\'{o}n de Ardoz, E-28850, Madrid, Spain}

\author[0000-0002-5104-8245]{Pierluigi Rinaldi}
\affiliation{Steward Observatory, University of Arizona, 933 N. Cherry Ave., Tucson, AZ 85721, USA}

\author[0000-0002-9720-3255]{Meredith Stone}
\affiliation{Steward Observatory, University of Arizona, 933 N. Cherry Ave., Tucson, AZ 85721, USA}

\author[0000-0002-4622-6617]{Fengwu Sun}
\affiliation{Center for Astrophysics $|$ Harvard \& Smithsonian, 60 Garden St., Cambridge, MA 02138, USA}

\author[0000-0003-2919-7495]{Christina C. Williams}
\affiliation{NSF’s National Optical-Infrared Astronomy Research Laboratory, 950 N. Cherry Ave., Tucson, AZ 85719, USA}

\author[0000-0001-9262-9997]{Christopher N. A. Willmer}
\affiliation{Steward Observatory, University of Arizona, 933 N. Cherry Ave., Tucson, AZ 85721, USA}

\author[0000-0002-4201-7367]{Chris Willott}
\affiliation{NRC Herzberg, 5071 W. Saanich Rd., Victoria, BC V9E 2E7, Canada}

\author[0000-0002-7595-121X]{Joris Witstok}
\affiliation{Cosmic Dawn Center (DAWN), Copenhagen, Denmark}
\affiliation{Niels Bohr Institute, University of Copenhagen, Jagtvej 128, DK-2200, Copenhagen, Denmark}

\correspondingauthor{Jakob M. Helton}
\email{jakobhelton@arizona.edu}


\begin{abstract}

Our understanding of the physical properties of star-forming galaxies during the Epoch of Reionization (EoR; at $z > 6$) suffers from degeneracies among the apparent properties of the stars, nebular gas, and dust. These degeneracies are most prominent with photometry, which has insufficient (1) spectral resolution and (2) rest-frame spectral coverage. We explore ways to break these degeneracies with a sample of $N = 22$ high-redshift star-forming galaxies at $7 < z_{\mathrm{phot}} \leq 9$, using some of the deepest existing imaging from JWST/NIRCam and JWST/MIRI with JADES. Key to this study is the imaging from JWST/MIRI at $7.7\ \mu\mathrm{m}$, which provides coverage of the rest-frame $I$-band at the observed redshifts. We infer stellar population properties and rest-frame colors using a variety of filter sets and star formation history assumptions to explore the impact of these choices. Evaluating these quantities both with and without the $7.7\ \mu\mathrm{m}$ data point shows that dense spectral coverage with JWST/NIRCam (eight or more filters, including at least one medium-band at $\lambda_{\mathrm{obs}} \approx 4-5\ \mu\mathrm{m}$) can compensate for lacking the rest-frame $I$-band coverage for the vast majority ($\approx 80\%$) of our sample. Furthermore, these galaxy properties are most consistently determined by assuming the delayed-tau star formation history, which provides the smallest offsets and scatters around these offsets when including JWST/MIRI. Within extragalactic surveys like JADES and CEERS, our findings suggest that robust characterization of the stellar population properties and rest-frame colors for typical high-redshift star-forming galaxies (i.e., galaxies with $M_{\mathrm{UV}} \approx -20$ and $\beta_{\mathrm{UV}} \approx -2$) is possible with JWST/NIRCam alone at $z \approx 8$.
 
\end{abstract}




\section{Introduction}
\label{SectionOne}

Prior to the launch of JWST in December 2021, the photometric redshift frontier was at $z \approx 9-11$ \citep[e.g.,][]{Oesch:2014, Bouwens:2021, Finkelstein:2022, Bagley:2024}. This frontier was primarily driven by observations with the Hubble Space Telescope (HST), particularly with the Wide Field Camera 3 (WFC3) which delivers imaging at observed wavelengths of $\lambda_{\mathrm{obs}} = 0.2-1.7\ \mu\mathrm{m}$. Although these observations provided useful constraints on the properties of the rest-frame ultraviolet (UV) continua, it was the addition of the rest-frame optical coverage afforded by the Spitzer Space Telescope Infrared Array Camera (IRAC) that allowed for the first robust constraints on the stellar population properties of star-forming galaxies at $z \approx 6-8$ \citep[e.g.,][]{Stark:2013, Duncan:2014, Grazian:2015}. Using the full suite of HST/WFC3 and Spitzer/IRAC filters at these redshifts, it was possible to disentangle the relative contributions to the rest-optical emission from the stars, nebular gas, and dust. These studies were primarily limited by the spectral and spatial resolution of Spitzer, since bright nebular emission lines can make extreme contributions to rest-optical broadband filters, potentially making the interpretation of the photometry ambiguous \citep[e.g.,][]{Schaerer:2009, Stark:2013, Smit:2014, Grazian:2015}.  Expanding on this previous work, identifying and understanding the physical properties of galaxies deep into the Epoch of Reionization (EoR; at $z > 6$) was one of the primary science drivers for JWST.

Beyond the difficulties in disentangling the relative contributions to the rest-optical emission, another complication in determining the stellar population properties is that we know very little about typical star-forming galaxies in the EoR, which forces us to make simplifying assumptions about the complex physical processes governing these early systems. For example, stellar mass measurements are crucial for understanding these early galaxies since they encode valuable information about the build-up of their first stars. However, assuming an initial mass function (IMF) that slowly evolves with redshift produces stellar mass estimates two to three times smaller than those inferred from assuming a local IMF \citep{Woodrum:2024}, and up to ten times smaller when using other physically motivated assumptions about the IMF \citep{Wang:2024}. Additionally, assuming a star formation history (SFH) that disfavors burstiness produces stellar mass estimates two to five times larger than those inferred from assuming bursty models \citep[e.g.,][]{Tacchella:2023, Endsley:2024}, which is especially important, since such bursty SFHs are seemingly ubiquitous at $z = 7-9$ \citep[e.g.,][]{Endsley:2023, Endsley:2024, Boyett:2024, Helton:2024b, Simmonds:2024}. Bursty SFHs are possibly the result of the available large-scale gas supplies at high redshifts and the outflows that are subsequently driven by strong stellar feedback. Although these simplifying assumptions about high-redshift galaxies can have profound implications on their stellar population properties, they often do not produce an observable difference in the resulting photometric or spectroscopic values, as a result of recently formed stars outshining the older stellar population \citep[e.g.,][]{Narayanan:2024}.

Early work with JWST by \citet{Papovich:2023} estimated the stellar population properties for a sample of high-redshift star-forming galaxies using a combination of existing data from HST and Spitzer alongside new imaging from the mid-infrared instrument (MIRI). These authors found that including JWST/MIRI photometry reduced stellar mass and star formation rate estimates by a factor of two to three at $z = 7-9$. These reductions were the result of a better understanding of the contributions from the nebular emission lines to the rest-optical broadband filters from Spitzer. The sensitivity of JWST/MIRI is limited by the increase in zodiacal emission at $\lambda_{\mathrm{obs}} > 5 \mu\mathrm{m}$. Therefore, if JWST/MIRI observations were required for accurate stellar mass and star formation rate estimates, it would significantly increase the observing time required to derive physical properties for low-mass high-redshift galaxies. However, this early work by \citet{Papovich:2023} lacked observations from JWST/NIRCam, which might be able to compensate for the lack of JWST/MIRI data.

Similar work by \citet{Wang_2025_Massive} found nearly identical conclusions to the early work by \citet{Papovich:2023} for another sample of high-redshift star-forming galaxies at $z = 7-9$ by combining JWST/NIRCam and JWST/MIRI observations \citep[see also][]{Williams:2024}. However, \citet{Wang_2025_Massive} focused on massive galaxies ($M_{\ast} > 10^{10}\ M_{\odot}$), making it difficult to interpret their results in the context of more typical star-forming galaxies during the EoR. In contrast, \citet{Alberts:2024a} combined JWST/NIRCam and JWST/MIRI observations to study lower-mass quiescent and post-starburst galaxies at later times than the EoR. They found that densely sampled JWST/NIRCam data (i.e., eight or more filters, including at least one medium-band filter) can compensate for the lack of JWST/MIRI imaging when determining the stellar population properties of galaxies at $z = 3-6$. This is because densely sampled JWST/NIRCam and/or JWST/MIRI imaging provides valuable constraints on the continuum at rest-frame optical wavelengths.  Similar work by \citet{Iani:2024}, who also combined JWST/NIRCam and JWST/MIRI observations,  found nearly identical conclusions to those of \citet{Alberts:2024a}, but for a sample of Lyman-$\alpha$ emitters at $z = 3-6$. 

In summary, existing studies have explored the impact of JWST/MIRI in understanding the physical properties of galaxies during the EoR. However, these studies have focused on subsets of high-redshift galaxies, which has made it difficult to interpret these results in the general context of typical star-forming galaxies during the EoR. In this work, we address this issue more generally. We combine some of the deepest existing JWST/NIRCam and JWST/MIRI imaging in eight wide-band and seven medium-band filters spanning observed wavelengths of $\lambda_{\mathrm{obs}} = 0.8-8.9\ \mu\mathrm{m}$, with typical exposure times of $40$ hours per filter. These data were acquired as part of the JWST Advanced Deep Extragalactic Survey \citep[JADES;][]{Eisenstein:2026}, which assembles roughly $770$ hours of observing time from the JWST/NIRCam and JWST/NIRSpec Instrument Development Teams, and represents the largest program from JWST Cycle 1. Key to this work are the ultra-deep observations in the MIRI/F770W filter (with $5\sigma$ limiting magnitudes of $m \approx 28.1\ \mathrm{AB\ mag}$, assuming circular apertures with diameters of $0.7^{\prime\prime}$), which provide coverage of the rest-frame $I$-band at $z \approx 8$. Using these data, we explore ways to break the degeneracies among the inferred properties of the stars, nebular gas, and dust for a sample of $N = 22$ high-redshift star-forming galaxies at $z = 7-9$ that have significant detections of MIRI/F770W. To accomplish this, we infer stellar population properties and rest-frame colors using four different filter sets and four different SFH assumptions in order to evaluate these quantities both with and without the $7.7\ \mu\mathrm{m}$ data point. We demonstrate that dense spectral coverage with JWST/NIRCam (i.e., eight or more filters, including at least one medium-band filter at $\lambda_{\mathrm{obs}} \approx 4-5\ \mu\mathrm{m}$) can compensate for lacking the rest-frame $I$-band (observed-frame MIRI/F770W) coverage for the vast majority of our sample, which means that robust characterization of the stellar population properties and rest-frame colors is possible with JWST/NIRCam alone. But in the absence of dense spectral coverage with JWST/NIRCam, we find that observations with JWST/MIRI are needed to robustly determine the properties of the stellar populations and rest-frame colors.

This paper proceeds as follows. In Section~\ref{SectionTwo}, we describe the data and observations that are used in our analysis. In Section~\ref{SectionThree}, we present our sample selection, including the derivation of photometric redshifts (Section~\ref{SectionThreeOne}) and the available spectroscopic redshifts (Section~\ref{SectionThreeTwo}). In Section~\ref{SectionFour}, we present our inferred physical properties, including the properties of the rest-UV continua (Section~\ref{SectionFourOne}) and of the stellar populations (Sections~\ref{SectionFourTwo} and \ref{SectionFourThree}) using a variety of filter sets and SFHs. In Section~\ref{SectionFive}, we provide comparisons of the inferred physical properties, including the properties of the stellar populations (Section~\ref{SectionFiveOne}) and the rest-frame colors (Section~\ref{SectionFiveTwo}). In Section~\ref{SectionSix}, we summarize our results and their broader implications for future extragalactic science with JWST. All magnitudes are in the AB system \citep{Oke:1983}. Uncertainties are quoted as $68\%$ confidence intervals, unless otherwise stated. Throughout this work, we report wavelengths in vacuum and adopt the standard flat $\Lambda$CDM cosmology from Planck18 with $H_{0} = 67.4\ \mathrm{km/s/Mpc}$ and $\Omega_{m} = 0.315$ \citep[][]{Planck:2020}.


\section{Data \& Observations}
\label{SectionTwo}

The primary dataset used in this work consists of near-infrared imaging from JWST/NIRCam and mid-infrared imaging from JWST/MIRI in the Great Observatories Origins Deep Survey South \citep[GOODS-S;][]{Giavalisco:2004} field, near the Hubble Ultra Deep Field \citep[HUDF;][]{Beckwith:2006} and the JADES Origins Field \citep[JOF;][]{Eisenstein:2025}. The secondary dataset consists of optical imaging from the Advanced Camera for Surveys (ACS) on HST with five photometric filters (F435W, F606W, F775W, F814W, and F850LP), which were produced as part of the Hubble Legacy Fields (HLF) project v2.0 and include observations covering a $25^{\prime} \times 25^{\prime}$ area over the GOODS-S field \citep[][]{Illingworth:2016, Whitaker:2019}. All of the image mosaics considered here are registered to the \textsc{Gaia} DR3 frame \citep[][]{GaiaDR3} and resampled onto the same World Coordinate System (WCS) with a $30\,\mathrm{mas/pixel}$ grid.

The near-infrared imaging from JWST/NIRCam includes (1) a shallow mosaic with three photometric filters (F182M, F210M, and F444W) across an area of roughly 62 square arcminutes (PID: 1895; PI: P. Oesch), (2) a medium mosaic with eight filters (F090W, F115W, F150W, F200W, F277W, F356W, F410M, and F444W) across an area of roughly 40 square arcminutes (PID: 1180; PI: D. Eisenstein), (3) another medium mosaic with ten filters (F070W, F090W, F115W, F150W, F200W, F277W, F335M, F356W, F410M, and F444W) across an area of roughly 10 square arcminutes (PID: 1286; PI: N. L\"{u}tzgendorf), and (4) a deep mosaic with fourteen filters (F090W, F115W, F150W, F162M, F182M, F200W, F210M, F250M, F277W, F300M, F335M, F356W, F410M, and F444W) across an area of roughly 10 square arcminutes (PID: 1210, 3215; PI: N. L\"{u}tzgendorf, D. Eisenstein). The shallow mosaic was observed as part of the First Reionization Epoch Spectroscopic COmplete Survey \citep[FRESCO;][]{Oesch:2023} while the remaining mosaics were all observed as part of the JWST Advanced Deep Extragalactic Survey \citep[][]{Eisenstein:2026}. A detailed description of the reduction, mosaicking, source detection, and photometric measurements for the JWST/NIRCam data is provided in \citet{Rieke:2023} as part of the first JADES data release in GOODS-S. 

The mid-infrared imaging from JWST/MIRI includes a deep mosaic with one photometric filter (F770W) taken in parallel with JWST/NIRCam across an area of roughly 9 square arcminutes (PID: 1180; PI: D. Eisenstein), observed as part of JADES. A detailed description of the reduction, mosaicking, source detection, and photometric measurements for the JWST/MIRI survey data is provided in \citet{Alberts:2024b} while a more detailed description for this specific dataset will be presented in a forthcoming paper from the JADES collaboration (S. Alberts et al. in preparation). Assuming circular apertures with diameters of $0.7^{\prime\prime}$ (i.e., CIRC5), the $5\sigma$ detection limit is $m \approx 28.1\ \mathrm{AB\ mag}$ after applying aperture corrections for point-source morphologies. The definitions for the complete set of circular apertures are provided in \citet{Rieke:2023} as part of the first JADES data release in GOODS-S.

Following the methodology of \citet{Alberts:2024a} and \citet{Williams:2024}, the photometric measurements of the JWST/NIRCam data include forced circular aperture photometry convolved to the point spread function (PSF) of the F444W filter assuming circular apertures with diameters of $0.5^{\prime\prime}$ (i.e., CIRC3). We have validated by eye that this aperture is suitable for each of the high-redshift star-forming galaxies considered in this work. However, to correct for potential missing light, we rescale the JWST/NIRCam photometry by the flux ratio of CIRC5 (i.e., circular apertures with diameters of $0.7^{\prime\prime}$, which roughly corresponds to the 65\% encircled energy of F770W) to CIRC3 in the F444W filter. Aperture corrections are subsequently applied assuming point-source morphologies and using model encircled energy curves from the \texttt{WebbPSF} \citep{Perrin:2014} package as described in \citet{Ji:2024}. Photometric uncertainties are estimated by combining in quadrature the Poisson uncertainty for each detected source with the flux variance as calculated by placing random apertures across regions of the image mosaics \citep[e.g.,][]{Labbe:2005a, Quadri:2007, Whitaker:2011}.

Since the full width at half maximum (FWHM) for the PSF of JWST/MIRI (FWHM of $0.269^{\prime\prime}$ for F770W) is so much larger than that of JWST/NIRCam (FWHM of $0.145^{\prime\prime}$ for F444W), we choose to measure the F444W$-$F770W color of sources in the JWST/NIRCam catalog after rebinning and convolving the F444W mosaic to the F770W pixel size and PSF. This color is measured assuming the CIRC5 circular apertures, but prior to applying aperture corrections. The F770W flux is therefore found by subtracting this F444W$-$F770W color from the total F444W flux found using the CIRC5 circular apertures. With this methodology, we are able to take advantage of the higher spatial resolution offered by JWST/NIRCam when compared to JWST/MIRI while still deriving accurate integrated colors that are robust to any potential color gradients.


\section{Sample Selection}
\label{SectionThree}

Using the data and observations from Section~\ref{SectionTwo}, we assembled a flux-limited sample of high-redshift star-forming galaxies. The photometric redshift measurements described in Section~\ref{SectionThreeOne} were used to select the final sample of galaxies with MIRI/F770W detections in the desired redshift range. We additionally selected a secondary sample of galaxies without MIRI/F770W detections, to explore any physical differences between these two populations of galaxies. Spectroscopic redshifts from the literature are available for a subset of the final sample and are described in Section~\ref{SectionThreeTwo}.

\subsection{Photometric Redshifts}
\label{SectionThreeOne}

Using forced circular aperture photometry assuming diameters of $0.2^{\prime\prime}$ (i.e., CIRC1) without convolution to the PSF of the NIRCam/F444W filter, we measure photometric redshifts with the template-fitting code \texttt{EAZY} \citep{Brammer:2008}. Adopting the unconvolved photometry from the smaller circular apertures reduces the background noise associated with the use of the convolved photometry from the larger circular apertures in Section~\ref{SectionTwo}. Because of the difficulty in properly matching photometric measurements from JWST/NIRCam and JWST/MIRI (see discussion in Section~\ref{SectionTwo}), we do not use the available MIRI/F770W data when measuring photometric redshifts with \texttt{EAZY}. This code uses a chi-square ($\chi^{2}$) minimization technique to model the spectral energy distributions (SEDs) of galaxies using linear combinations of various galaxy templates, fitting across the redshift range of $z = 0.2-22$ with a redshift step size of $\Delta z = 0.01$. For more information on the photometric redshift estimation, we refer the reader to \citet{Rieke:2023} and \citet{Hainline:2024}.

The primary measurements used here are the \texttt{EAZY} ``$z_{\mathrm{a}}$'' redshifts and the ``$P(z > 7)$'' probabilities. The former corresponds to the fit where the $\chi^{2}$ is minimized. The latter corresponds to the summed probability that the galaxy is at $z > 7$, assuming the uniform redshift prior $P(z) = \mathrm{exp}[-\chi^{2}(z)/2]$, normalized such that $\int P(z)\,dz = 1$. We also make use of the first \texttt{EAZY} confidence interval ($\Delta z_{1}$), defined to be the difference between the 16th and 84th percentiles of photometric redshift posterior distribution, roughly equal to twice the standard deviation ($\approx \pm 1\sigma$).

\begin{figure*}
    \centering
    \includegraphics[width=0.7\linewidth]{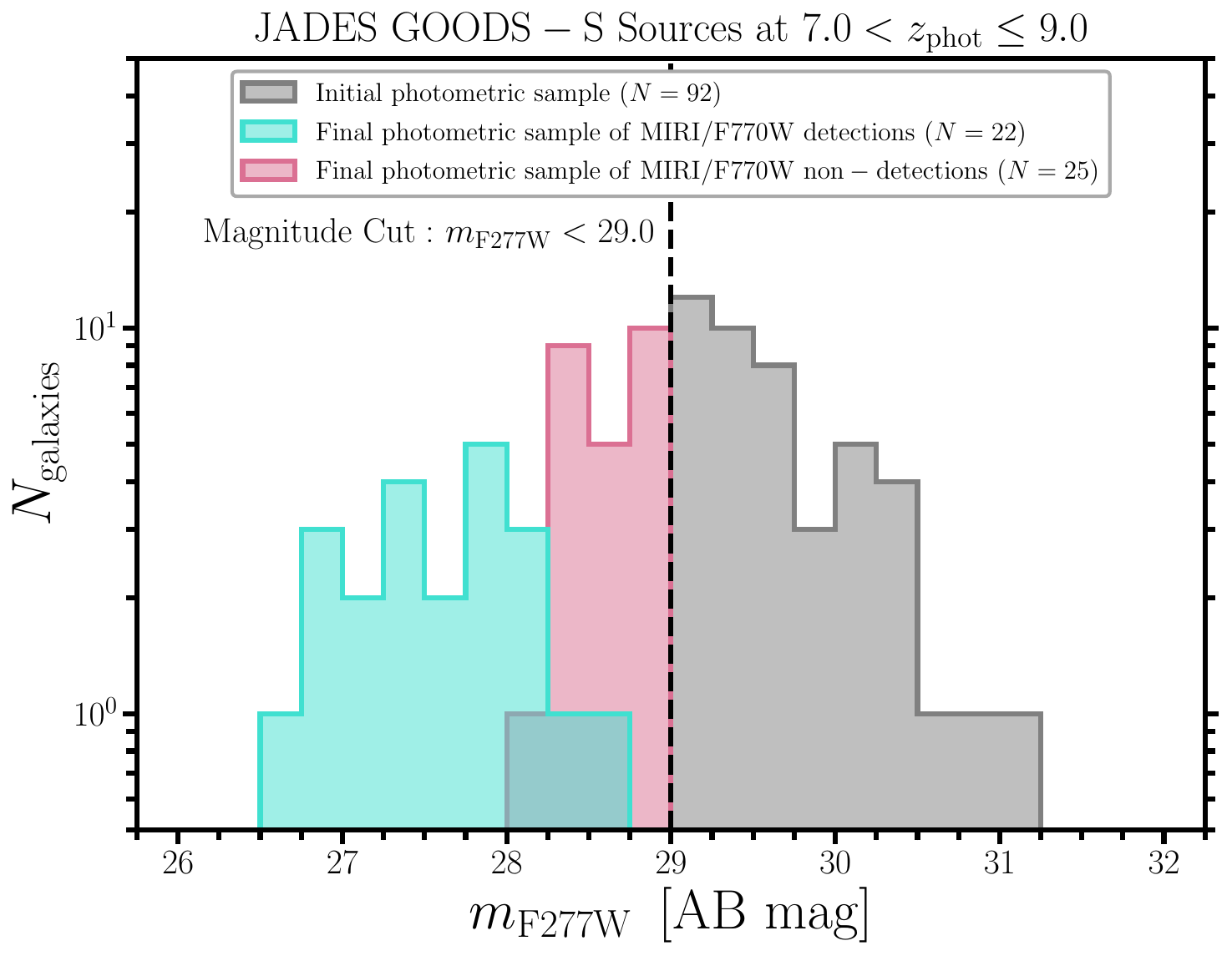}
    \caption{Histograms showing the distribution of apparent magnitudes in F277W for galaxies at $7 < z_{\mathrm{phot}} \leq 9$ within the ultra-deep JADES MIRI/F770W parallel region in GOODS-S, assuming the fiducial photometry described in Section~\ref{SectionTwo}. The gray shaded regions represent galaxies from the initial photometric sample while, the blue (pink) shaded regions represent the subset of galaxies from the initial photometric sample with MIRI/F770W detections (non-detections). The black vertical dashed line shows the adopted apparent magnitude cut for the final photometric samples, which is similar to a rest-UV absolute magnitude cut at these redshifts (see Section~\ref{SectionFourOne} and Figure \ref{fig:RestUV_Properties} for more discussion). \label{fig:m_F277W_Distribution}}
\end{figure*}

This paper's analysis is based on comparing the flux-limited sample of high-redshift star-forming galaxies that have robust detections of MIRI/F770W with the analogous sample of galaxies that do not have significant detections in that filter. To accomplish this, we adopt the following selection criteria:
\begin{enumerate}
    \topsep 0pt
    \parsep 0pt
    \itemsep 0pt
    \item The \texttt{EAZY} photometric redshift corresponding to the fit where the $\chi^{2}$ is minimized, $z_{\mathrm{a}}$, must be greater than seven but less than or equal to nine (i.e., $7 < z_{\mathrm{a}} \leq 9$).
    \item The \texttt{EAZY} summed probability of being at $z > 7$, $P(z > 7)$, must be greater than or equal to 99\% (i.e., $P(z > 7) = \int_{7}^{22} P(z)\,dz \geq 0.99$).
    \item The apparent magnitude in F277W, assuming the fiducial photometry described in Section~\ref{SectionTwo}, $m_{\mathrm{F277W}}$, must be brighter than the shallowest region of JADES (i.e., $m_{\mathrm{F277W}} < 29.0\ \mathrm{AB\ mag}$).
\end{enumerate}

\begin{figure}
    \centering
    \includegraphics[width=1.0\linewidth]{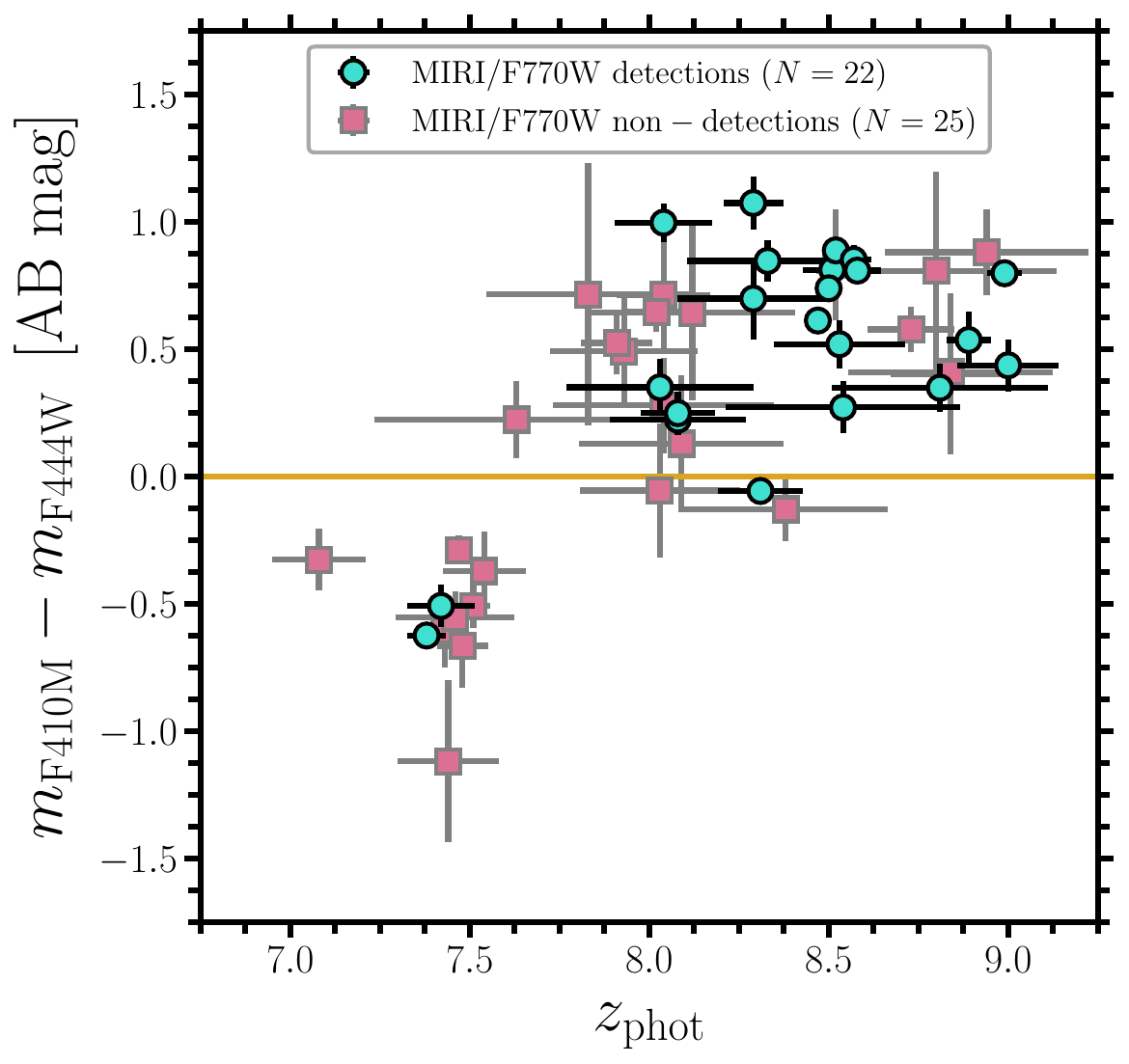}
    \caption{Observed color in NIRCam/F410M with respect to NIRCam/F444W as a function of photometric redshift for the final photometric sample described in Section~\ref{SectionThreeOne}. The blue (pink) points represent galaxies with MIRI/F770W detections (non-detections). The rest-optical nebular emission lines $\mathrm{H}\beta$ and $\mathrm{[OIII]}\lambda\lambda 4960{,}5008$ produce excess flux in F410M relative to F444W at $z \lesssim 7.6$ (F444W relative to F410M at $z \gtrsim 7.6$). The vast majority of objects in the final photometric sample are high-redshift star-forming galaxies since they show evidence for bright rest-optical emission lines in the NIRCam/F444W filter. \label{fig:Color_F410M_F444W}}
\end{figure}

The lower bound on redshift of $z = 7$ was chosen since galaxies at these redshifts will be partial or complete dropouts in ACS/F814W and NIRCam/F090W due to absorption by the intervening intergalactic medium (IGM). Dropouts in these two filters are similar to the $z$-band dropouts previously discovered by HST using ACS/F850LP \citep[e.g.,][]{Bouwens:2008, Bouwens:2010, Bunker:2010, McLure:2010, Oesch:2010}. Galaxies at these redshifts will also have bright rest-optical emission lines in NIRCam/F444W (see Figure~\ref{fig:Color_F410M_F444W}). The photometric redshift distributions of star-forming galaxies at $z > 7$ are well constrained by the measured dropouts and the emission-line excess. The upper bound on redshift of $z = 9$ was chosen since this was the highest redshift with a detection in MIRI/F770W that was visible in the final MIRI/F770W image mosaic \citep[not counting the detection of JADES-GS-z14-0 at $z > 14$ which was previously reported by][]{Helton:2025}. Imposing such an upper limit on the redshift is important for deriving an analogous sample of galaxies that do not have significant detections in MIRI/F770W. The summed probability requirement was chosen to prevent contamination from low-redshift interlopers. The cut on apparent magnitude was chosen to produce a flux-limited sample that is representative for the entirety of JADES. NIRCam/F277W was chosen for the apparent magnitude cut since it is the deepest filter in the JOF \citep[][]{Robertson:2024} and, therefore, the deepest filter available in the MIRI/F770W parallel region. For galaxies at the redshift considered here, NIRCam/F277W provides spectral coverage at roughly rest-frame $3000\ \mathrm{\AA}$. Objects that are deblended in JWST/NIRCam but blended in JWST/MIRI are counted as multiple sources in the photometric catalog, but we will consider these as singular objects throughout.

\begin{figure}
    \centering
    \includegraphics[width=0.9625\linewidth]{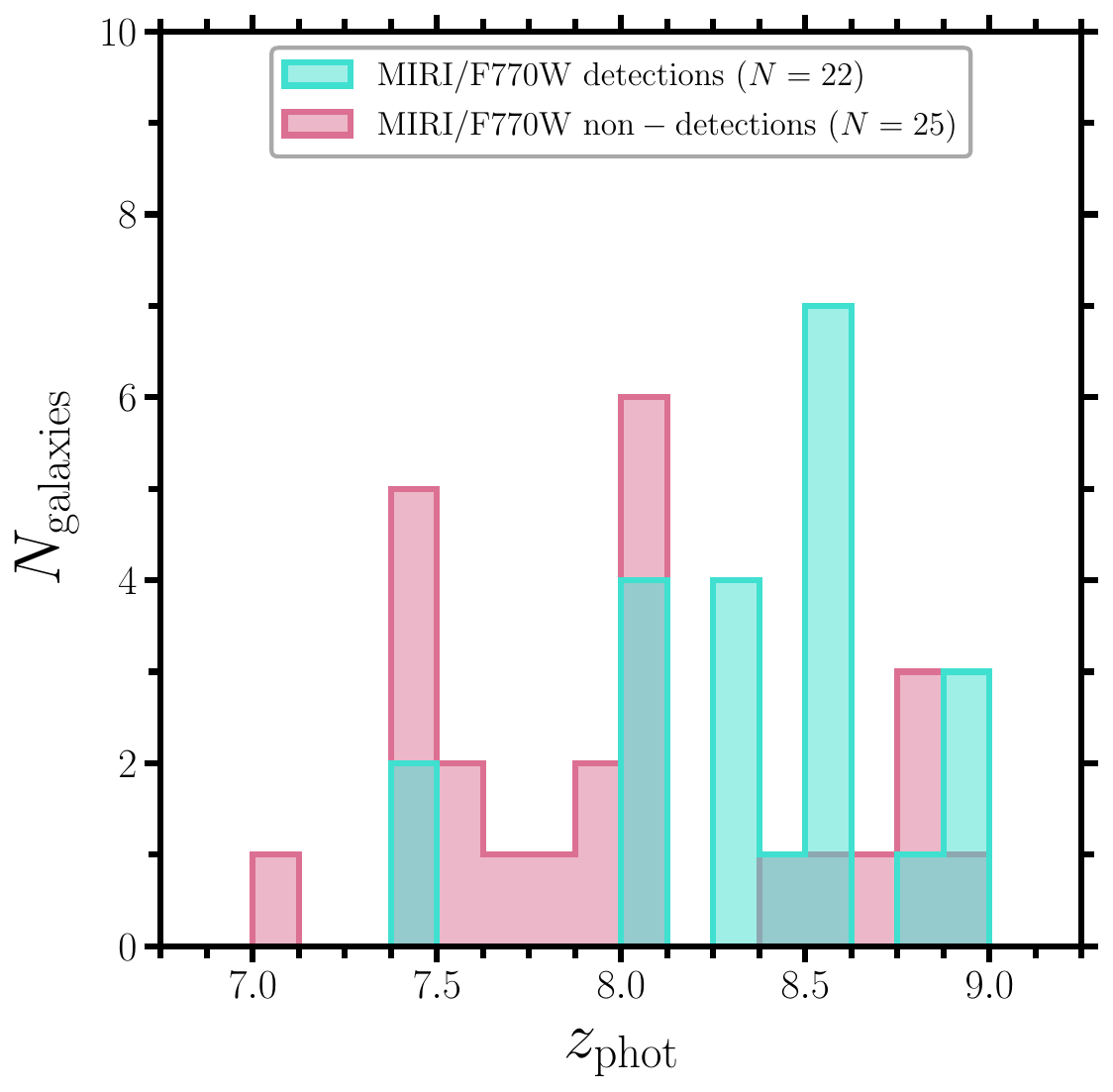}
    \caption{Histograms showing the distribution of photometric redshifts for the final photometric sample described in Section~\ref{SectionThreeOne}. The blue (pink) shaded regions represent galaxies with MIRI/F770W detections (non-detections). It is interesting to note that at $z \lesssim 8$ ($z \gtrsim 8$), the vast majority of galaxies in the final photometric sample have MIRI/F770W non-detections (detections). However, there do not appear to be any prominent emission features entering or leaving the MIRI/F770W filter at $z \approx 8$. These two distributions of photometric redshifts are statistically different from one another at the $\approx 2.5-2.6\sigma$ level. \label{fig:RedshiftDistribution}}
\end{figure}

The aforementioned selection criteria yielded $N = 47$ star-forming galaxies at $7 < z_{\mathrm{phot}} \leq 9$, which we separated into two subsamples. The first of these contains $N = 22$ galaxies and includes sources with significant detections (i.e., signal-to-noise ratios of $\mathrm{S/N} > 1.5$) in MIRI/F770W, which we refer to as our ``final photometric sample of JWST/MIRI detections.'' The second subsample contains $N = 25$ galaxies and includes galaxies without significant detections (i.e., $\mathrm{S/N} \leq 1.5$) of MIRI/F770W, which we refer to as our ``final photometric sample of JWST/MIRI non-detections.'' This separation was based on visual inspecting each of the galaxies from the initial sample using all available JWST/NIRCam and JWST/MIRI image mosaics. The adopted threshold for the signal-to-noise ratio (SNR) was chosen as the minimum value that produced visible detections in the final MIRI/F770W image mosaic. It is somewhat unexpected that such low SNRs correspond with these detections, but we attribute this effect to the use of circular apertures that correspond to the $65\%$ encircled energy of MIRI/F770W. Such large apertures do not maximize the measured SNR; rather, they avoid any bias from potential color gradients at the expense of SNR. For reference, there are $N = 14$ galaxies with SNRs of $\mathrm{S/N} > 3.0$ in MIRI/F770W for the final photometric sample of JWST/MIRI detections, which is roughly two-thirds of the sample. For galaxies with SNRs near the adopted threshold, more than half clearly have visible flux and, therefore, are robust detections, while the remaining few have marginal detections.

Based on the aforementioned visual inspections, we found that the vast majority ($\approx 60\%$) of galaxies from the final photometric sample of JWST/MIRI detections appear to be point sources. However, many of these galaxies ($\approx 40\%$) are morphologically extended and typically appear as multiple star-forming clumps in the short-wavelength JWST/NIRCam images. The most extreme of these spatially extended galaxies have already been identified and discussed in the literature by \citet[][]{Hainline:2024}, who provided color thumbnails for four of our extended galaxies: JADES$-$GS$-$ID$-$$25526$, JADES$-$GS$-$ID$-$$165595$, JADES$-$GS$-$ID$-$$179485$, and JADES$-$GS$-$ID$-$$380203$.

Figure~\ref{fig:m_F277W_Distribution} shows the distribution of apparent magnitudes in NIRCam/F277W for galaxies at $7 < z_{\mathrm{phot}} \leq 9$ within the ultra-deep JADES MIRI/F770W parallel region in GOODS-S, assuming the fiducial photometry described in Section~\ref{SectionTwo}. The gray histograms represent galaxies from the initial photometric sample while the blue (pink) histograms represent galaxies from the final photometric sample of MIRI/F770W detections (non-detections). Galaxies within the flux-limited photometric sample of detections (non-detections) typically have apparent magnitudes $m_{\mathrm{F277W}} \approx 26.5-28.5\ \mathrm{AB\ mag}$ ($\approx 28.0-29.0\ \mathrm{AB\ mag}$). For comparison, we provide a black vertical dashed line to illustrate the apparent magnitude cut used to select the two final photometric samples ($m_{\mathrm{F277W}} < 29.0\ \mathrm{AB\ mag}$), which is similar to a rest-UV absolute magnitude cut at these redshifts (see Section~\ref{SectionFourOne} and Figure \ref{fig:RestUV_Properties} for more about this). This is the expected behavior for a flux-limited galaxy sample.

\begin{table*}
    \caption{A summary of physical quantities for the final photometric sample of flux-limited high-redshift star-forming galaxies with MIRI/F770W detections ($\mathrm{S/N} > 1.5$), as described in Section~\ref{SectionThreeOne}.}
    \label{tab:FinalSampleTable}
    \begin{threeparttable}
        \makebox[\textwidth]{
        \begin{tabular}{c|cc|ccc|ccc|c}
            \hline
            \hline
		$\mathrm{ID}$\tnote{a} & $\mathrm{R.A.}$\tnote{b} & $\mathrm{Decl.}$\tnote{c} & $z_{\mathrm{spec}}$\tnote{d} & $z_{\mathrm{phot}}$\tnote{e} & $P\left( z > 7 \right)$\tnote{f} & $m_{\mathrm{F277W}}$\tnote{g} & $M_{\mathrm{UV}}$\tnote{h} & $\beta_{\mathrm{UV}}$\tnote{i} & $\mathrm{Ref.}$\tnote{j} \\
		\hline
            $14647$ & $53.08646$ & $-27.88925$ & n/a & $7.38 \pm 0.05$ & $1.000$ & $27.96 \pm 0.07$ & $-19.21 \pm 0.06$ & $-2.42 \pm 0.24$ & n/a \\
            $21468$ & $53.10108$ & $-27.88310$ & $8.808$ & $8.89 \pm 0.06$ & $1.000$ & $28.02 \pm 0.09$ & $-19.56 \pm 0.07$ & $-2.40 \pm 0.17$ & n/a \\
            $25526$ & $53.09942$ & $-27.88038$ & $7.957$ & $8.04 \pm 0.14$ & $1.000$ & $27.43 \pm 0.05$ & $-19.95 \pm 0.05$ & $-2.28 \pm 0.10$ & n/a \\
            $27503$ & $53.07581$ & $-27.87938$ & $8.196$ & $8.51 \pm 0.08$ & $1.000$ & $26.96 \pm 0.03$ & $-20.29 \pm 0.05$ & $-2.01 \pm 0.10$ & $2$ \\
            $30333$ & $53.05373$ & $-27.87789$ & $7.891$ & $8.52 \pm 0.03$ & $1.000$ & $27.60 \pm 0.03$ & $-19.67 \pm 0.05$ & $-2.16 \pm 0.10$ & n/a \\
            $37458$ & $53.08932$ & $-27.87269$ & $8.225$ & $8.57 \pm 0.05$ & $1.000$ & $27.85 \pm 0.04$ & $-19.65 \pm 0.05$ & $-2.41 \pm 0.10$ & $4$ \\
            $57378$ & $53.08650$ & $-27.85920$ & $7.950$ & $8.50 \pm 0.03$ & $1.000$ & $27.33 \pm 0.03$ & $-19.73 \pm 0.05$ & $-1.85 \pm 0.10$ & $1$ \\
            $66293$ & $53.04601$ & $-27.85399$ & $8.065$ & $8.33 \pm 0.22$ & $1.000$ & $27.81 \pm 0.09$ & $-19.24 \pm 0.14$ & $-1.81 \pm 0.23$ & n/a \\
            $164055$ & $53.08168$ & $-27.88858$ & $7.397$ & $7.42 \pm 0.09$ & $1.000$ & $28.31 \pm 0.09$ & $-19.00 \pm 0.07$ & $-1.90 \pm 0.21$ & n/a \\
            $165595$ & $53.05830$ & $-27.88486$ & $8.585$ & $8.99 \pm 0.05$ & $1.000$ & $27.64 \pm 0.03$ & $-19.69 \pm 0.04$ & $-2.06 \pm 0.10$ & n/a \\
            $173624$ & $53.07688$ & $-27.86967$ & $8.270$ & $8.58 \pm 0.07$ & $1.000$ & $27.38 \pm 0.03$ & $-19.77 \pm 0.05$ & $-1.92 \pm 0.10$ & n/a \\
            $173679$ & $53.07277$ & $-27.86929$ & n/a & $8.08 \pm 0.19$ & $1.000$ & $26.56 \pm 0.01$ & $-20.85 \pm 0.05$ & $-2.26 \pm 0.10$ & $1$ \\
            $174121$ & $53.05567$ & $-27.86882$ & $7.623$ & $8.31 \pm 0.12$ & $1.000$ & $27.14 \pm 0.05$ & $-19.86 \pm 0.06$ & $-1.92 \pm 0.10$ & n/a \\
            $174693$ & $53.06058$ & $-27.86795$ & $7.882$ & $8.81 \pm 0.30$ & $1.000$ & $28.13 \pm 0.09$ & $-19.35 \pm 0.09$ & $-2.37 \pm 0.16$ & n/a \\
            $175729$ & $53.06058$ & $-27.86603$ & $7.883$ & $8.53 \pm 0.18$ & $1.000$ & $27.49 \pm 0.06$ & $-19.86 \pm 0.06$ & $-2.23 \pm 0.11$ & n/a \\
            $175837$ & $53.06021$ & $-27.86572$ & $7.884$ & $9.00 \pm 0.14$ & $1.000$ & $27.77 \pm 0.09$ & $-19.15 \pm 0.11$ & $-1.71 \pm 0.19$ & n/a \\
            $177322$ & $53.06036$ & $-27.86355$ & $7.885$ & $8.29 \pm 0.21$ & $1.000$ & $27.98 \pm 0.12$ & $-19.39 \pm 0.10$ & $-2.24 \pm 0.21$ & $3$ \\
            $179485$ & $53.08738$ & $-27.86031$ & $7.955$ & $8.47 \pm 0.03$ & $1.000$ & $26.91 \pm 0.02$ & $-20.18 \pm 0.05$ & $-1.89 \pm 0.10$ & $2$ \\
            $180446$ & $53.08626$ & $-27.85932$ & $7.956$ & $8.54 \pm 0.33$ & $1.000$ & $28.62 \pm 0.10$ & $-18.47 \pm 0.10$ & $-1.93 \pm 0.18$ & $1$ \\
            $300287$ & $53.08812$ & $-27.90817$ & n/a & $8.08 \pm 0.10$ & $0.990$ & $26.94 \pm 0.05$ & $-20.05 \pm 0.06$ & $-1.71 \pm 0.11$ & n/a \\
            $300391$ & $53.08513$ & $-27.90636$ & n/a & $8.29 \pm 0.08$ & $1.000$ & $27.01 \pm 0.05$ & $-20.41 \pm 0.05$ & $-2.25 \pm 0.10$ & n/a \\
            $380203$ & $53.08172$ & $-27.89881$ & n/a & $8.03 \pm 0.26$ & $1.000$ & $28.02 \pm 0.06$ & $-19.15 \pm 0.08$ & $-2.02 \pm 0.14$ & $5$ \\
            \hline
	\end{tabular}
        }
	\begin{tablenotes}
	    \footnotesize
            \item \textbf{Notes.}
            \item[a] Identification number, from the JADES internal catalog and public data releases.
            \item[b] Right ascension, in degrees from the epoch J2000.
            \item[c] Declination, in degrees from the epoch J2000.
            \item[d] Spectroscopic redshift, when available.
            \item[e] Best-fit photometric redshift from \texttt{EAZY}, where the $\chi^{2}$ is minimized, and the associated $1\sigma$ uncertainty.
            \item[f] Summed probability of being at $z > 7$ from \texttt{EAZY}, assuming the uniform redshift prior $P(z) = \mathrm{exp}[-\chi^{2}(z)/2]$.
            \item[g] Apparent magnitude in F277W, assuming the fiducial photometry described in Section~\ref{SectionTwo}, and the associated $1\sigma$ uncertainty.
            \item[h] Rest-UV absolute magnitude, as measured in Section~\ref{SectionFourOne}, and the associated $1\sigma$ uncertainty.
            \item[i] Rest-UV continuum slope, as measured in Section~\ref{SectionFourOne}, and the associated $1\sigma$ uncertainty.
            \item[j] Original reference, when available, as determined by \texttt{astroquery}.
            \item \textbf{References.}
            \item[1] \citet{Grazian:2012}.
            \item[2] \citet{Oesch:2012}.
            \item[3] \citet{Yan:2012}.
            \item[4] \citet{McLure:2013}.
            \item[5] \citet{Bouwens:2015}.
        \end{tablenotes}
    \end{threeparttable}
\end{table*}

Figure~\ref{fig:RedshiftDistribution} shows the distribution of photometric redshifts for the final photometric sample described in Section~\ref{SectionThreeOne}. The blue histograms represent galaxies with MIRI/F770W detections while the pink histograms represent galaxies with MIRI/F770W non-detections. It seems that the vast majority of galaxies in the final photometric sample at $z \lesssim 8$ have MIRI/F770W non-detections while the vast majority of galaxies at $z \gtrsim 8$ have MIRI/F770W detections. However, there are no prominent emission features entering or leaving the MIRI/F770W filter at $z \approx 8$, which makes it difficult to explain why this apparent redshift evolution is occurring. To explore whether the apparent redshift evolution is real, we compare the photometric redshift distributions of the detected and non-detected samples by performing Kolmogorov–Smirnov (K-S) and Anderson–Darling (A-D) tests. These are two-sided tests for the null hypothesis that the detected and non-detected samples are drawn from the same continuous distribution of photometric redshifts. We find that these tests indicate that the photometric redshift distributions of the detected and non-detected samples are statistically different from each other at the $\approx 2.5-2.6\sigma$ level (which corresponds to $p$-values of $p \approx 0.009-0.011$), although we caution that the sample sizes are relatively small ($N = 22$ and $N = 25$ for the detected and non-detected samples, respectively). One possible explanation for this behavior is that typical star-forming galaxies during the EoR have blue continua at rest-frame near-infrared wavelengths, resulting in an increased K-correction for MIRI/F770W at $z \approx 9$ compared to $z \approx 7$. Alternatively, an overestimation of photometric redshifts for the detected sample could alleviate the differences between the photometric redshift distributions of the detected and non-detected samples (see also Section~\ref{SectionThreeTwo}).

Figure~\ref{fig:Color_F410M_F444W} shows the observed color in NIRCam/F410M relative to NIRCam/F444W as a function of photometric redshift for the final photometric sample described in Section~\ref{SectionThreeOne}. The blue points represent galaxies with MIRI/F770W detections while the pink points represent galaxies with MIRI/F770W non-detections. Some of the brightest rest-optical nebular emission lines ($\mathrm{H}\beta$ and $\mathrm{[OIII]}\lambda\lambda 4960{,}5008$) are producing excess flux in the NIRCam/F410M filter relative to NIRCam/F444W at $z \lesssim 7.6$ and F444W relative to F410M at $z \gtrsim 7.6$. The presence of these bright rest-optical emission lines provides evidence that these galaxies are strongly star-forming and that their photometric redshifts are consistent with the emission-line excess that would be expected in the F410M$-$F444W color.

Table~\ref{tab:FinalSampleTable} provides a summary of the most relevant physical quantities for the final photometric sample of high-redshift galaxies with MIRI/F770W detections ($\mathrm{S/N} > 1.5$), as described in Section~\ref{SectionThreeOne}. In the order of their appearance, these quantities include: (1) identification number; (2) right ascension in degrees; (3) declination in degrees; (4) spectroscopic redshift, when available; (5) best-fit photometric redshift from \texttt{EAZY}, corresponding to the fit where the $\chi^{2}$ was minimized, along with the associated $1\sigma$ uncertainty; (6) summed probability of being at $z > 7$ from \texttt{EAZY}, assuming the uniform redshift prior $P(z) = \mathrm{exp}[-\chi^{2}(z)/2]$; (7) apparent magnitude in F277W, assuming the fiducial photometry described in Section~\ref{SectionTwo}, along with the associated $1\sigma$ uncertainty; (8) rest-UV absolute magnitude, as measured in Section~\ref{SectionFourOne}, along with the associated $1\sigma$ uncertainty; (9) rest-UV continuum slope, as measured in Section~\ref{SectionFourOne}, along with the associated $1\sigma$ uncertainty; and (10) original reference, when available. The identification numbers and coordinates are identical to the quantities reported in the JADES internal catalog and public data releases \citep[][]{Rieke:2023, D'Eugenio:2024, Eisenstein:2025}. The original reference was determined by searching SIMBAD with \texttt{astroquery} \citep[][]{Ginsburg:2019} and identifying sources from the literature that are within a radius of $1.0^{\prime\prime}$ around each of the galaxies that are part of the final photometric sample.

\subsection{Spectroscopic Redshifts}
\label{SectionThreeTwo}

We searched the available spectroscopic data in the MIRI/F770W parallel region to confirm the redshifts for both of the final photometric samples presented in Section~\ref{SectionThreeOne}. The available spectroscopic data includes JWST/NIRSpec multi-object spectroscopy (MOS) from JADES (PID: 1287, 5997; PI: N. L\"{u}tzgendorf, T. Looser) in addition to JWST/NIRCam wide field slitless spectroscopy (WFSS) from FRESCO (PID: 1895; PI: P. Oesch) and JADES (PID: 4540; PI: D. Eisenstein). A description of the JWST/NIRSpec MOS data reduction and analysis is provided in \citet{D'Eugenio:2025} and will be presented in more detail in a forthcoming paper from the JADES Collaboration (J. Scholtz et al. in preparation). A description of the JWST/NIRCam grism data reduction and analysis is provided in \citet{Helton:2024b} and will be presented in more detail in a forthcoming paper from the JADES Collaboration (F. Sun et al. in preparation). For galaxies that have both NIRSpec/MOS and NIRCam/WFSS data, we prioritize the NIRSpec/MOS data due to its increased sensitivity. Using these data, we spectroscopically confirm redshifts by detecting the $\left[\mathrm{OIII}\right]\lambda5008$ emission line at $\mathrm{S/N} > 3$ near the best-fit photometric redshift. As provided in Table~\ref{tab:FinalSampleTable}, there are $N = 17$ galaxies with spectroscopic redshifts in the final photometric sample of JWST/MIRI detections, but only $N = 4$ galaxies with spectroscopic redshifts in the final photometric sample of JWST/MIRI non-detections, as provided in Table~\ref{tab:FinalSampleTable_Appendix}. 

For the two final photometric samples presented in Section~\ref{SectionThreeOne}, we note that the photometric redshifts are systematically larger than the spectroscopic redshifts, which is a result that has already been discussed extensively in the literature \citep[e.g.,][]{ArrabalHaro:2023, Fujimoto:2024, Hainline:2024}. This result has typically been attributed to mismatches between the observed photometry and the galaxy templates used to estimate the photometric redshifts. These mismatches include differences in the treatment of bright rest-optical emission lines and damped Lyman-$\alpha$ absorption. The offsets we measure between the photometric and spectroscopic redshifts ($\langle \Delta z = z_{\mathrm{phot}} - z_{\mathrm{spec}} \rangle = 0.43 \pm 0.32$) are consistent with those reported in the literature ($\langle \Delta z = z_{\mathrm{phot}} - z_{\mathrm{spec}} \rangle \approx 0.3-0.5$). There are no catastrophic outliers in our final photometric samples, defined as objects satisfying $\lvert z_{\mathrm{phot}} - z_{\mathrm{spec}} \rvert / (1 + z_{\mathrm{spec}}) > 0.15$. For all of the subsequent analyses, we adopt the spectroscopic redshift when available, rather than the photometric redshift, unless otherwise stated.


\section{Inferring Physical Properties}
\label{SectionFour}

Using the data and observations presented in Section~\ref{SectionTwo} alongside the photometric and spectroscopic redshifts from Section~\ref{SectionThree}, we explore the SEDs for the high-redshift star-forming galaxies in our final photometric sample. The properties of the rest-UV photometry are derived and described in Section~\ref{SectionFourOne}, where we compare the properties between the JWST/MIRI detected and undetected samples to explore the types of galaxies that are selected and are omitted from the final photometric sample. The properties of the stellar populations are derived and described in Sections~\ref{SectionFourTwo} and \ref{SectionFourThree}, where we utilize multiple filter choices and SFH assumptions to explore the impact of these assumptions.

\begin{figure}
    \centering
    \includegraphics[width=1.0\linewidth]{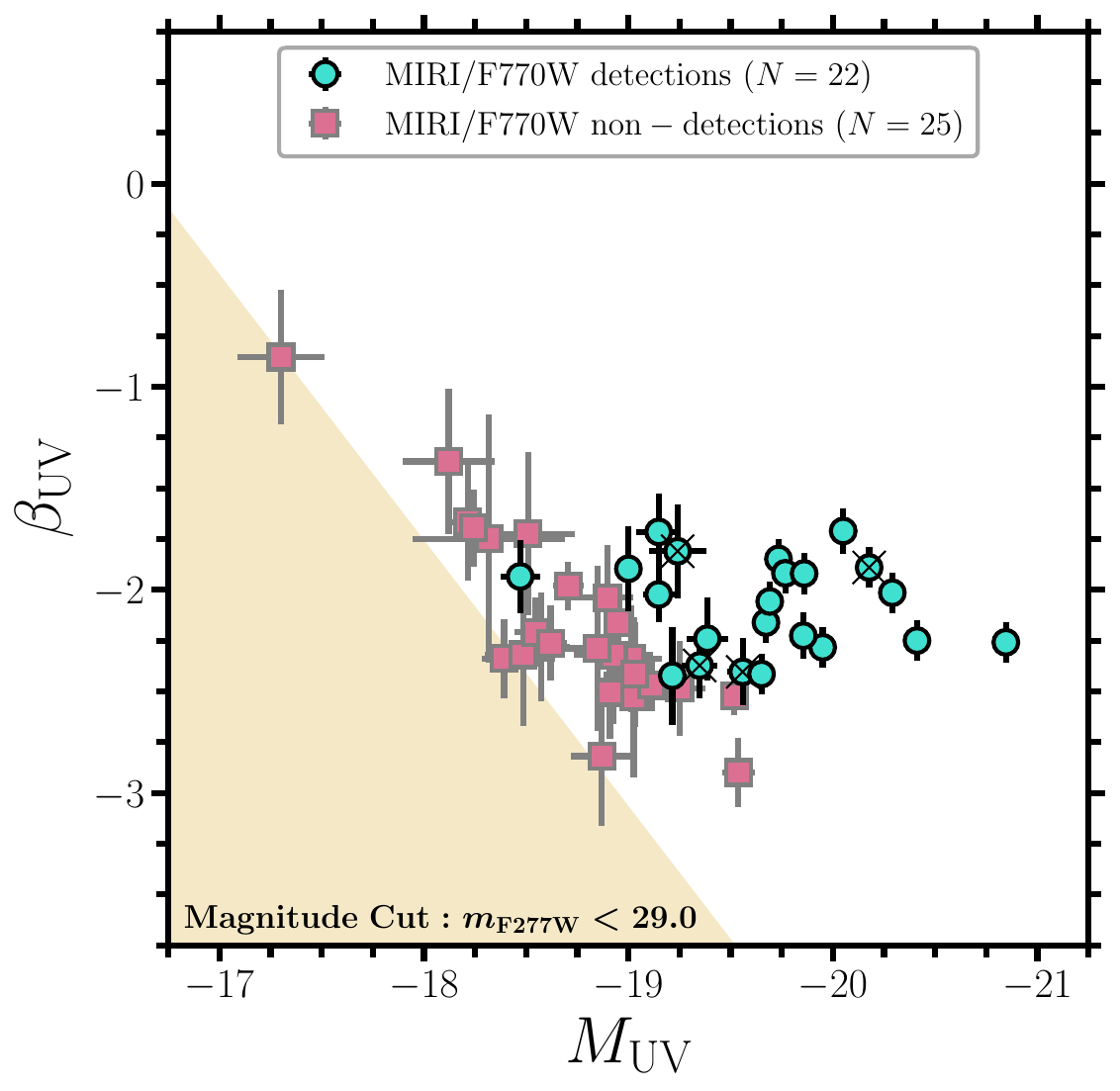}
    \caption{Rest-UV continuum slope versus absolute magnitude for the final photometric sample described in Section~\ref{SectionThreeOne}. The blue (pink) points represent galaxies with MIRI/F770W detections (non-detections) while the crosses represent atypical galaxies (see discussion in Section~\ref{SectionFourThree}). We should note that the adopted apparent magnitude cut ($m_{\mathrm{F277W}} < 29.0\ \mathrm{AB\ mag}$) is effectively a rest-UV absolute magnitude cut at fixed continuum slope for these redshifts. Therefore, the UV-faintest sources are also the UV-reddest. The yellow shaded region represents the area of parameter space that would not be selected by this adopted cut. \label{fig:RestUV_Properties}}
\end{figure}

\subsection{Properties of the Rest-UV Photometry}
\label{SectionFourOne}

The rest-UV absolute magnitudes ($M_{\mathrm{UV}}$) and continuum slopes ($\beta_{\mathrm{UV}}$) are derived following the methodology of \citet{Topping:2024} for each of the $N = 47$ galaxies contained within the final photometric sample described in Section~\ref{SectionThreeOne}. To quickly summarize, UV continuum slopes are obtained by fitting a simple power law to the observed wide-band photometry at $\lambda_{\mathrm{rest}} \approx 1200-2600$\AA\ \citep[$f_{\lambda} \propto \lambda^{\beta_{\mathrm{UV}}}$;][]{Calzetti:1994} while UV luminosities are obtained by measuring $\nu L_{\nu}$ at $\lambda_{\mathrm{rest}} = 1500$\AA\ using the same power-law fit as used for the continuum slopes. These properties are derived using the photometric measurements described in Section~\ref{SectionTwo}, while the redshift is fixed at either the spectroscopic redshift (if available) or the best-fit photometric redshift. A redshift-dependent set of filters is adopted in an attempt to avoid contamination from Lyman-$\alpha$ emission. The assumed filter set for objects in our final photometric sample includes F115W, F150W, and F200W for galaxies at $z < 7.4$ (F150W, F200W, and F277W at $z \geq 7.4$).

\begin{table*}
	\caption{A summary of the filter sets used for the derivation of the stellar population properties, as described in Section~\ref{SectionFourTwo}. All of the available medium- and wide-band filters from JWST/NIRCam are provided to illustrate the near completeness of the JOF filter set, which is only missing four medium-band filters and one wide-band filter.}
	\label{tab:FilterSetTable}
	\makebox[\textwidth]{
        \begin{tabular}{c||c|ccccccccc|ccccccccccc|c}
		\hline
		\hline
		Name (Number of Filters) & \rotatebox{90}{F814W } & \rotatebox{90}{F070W } & \rotatebox{90}{F090W } & \rotatebox{90}{F115W } & \rotatebox{90}{F140M } & \rotatebox{90}{F150W } & \rotatebox{90}{F162M } & \rotatebox{90}{F182M } & \rotatebox{90}{F200W } & \rotatebox{90}{F210M } & \rotatebox{90}{F250M } & \rotatebox{90}{F277W } & \rotatebox{90}{F300M } & \rotatebox{90}{F335M } & \rotatebox{90}{F356W } & \rotatebox{90}{F360M } & \rotatebox{90}{F410M } & \rotatebox{90}{F430M } & \rotatebox{90}{F444W } & \rotatebox{90}{F460M } & \rotatebox{90}{F480M } & \rotatebox{90}{F770W } \\
		\hline
            COSMOS-Web (6) & \textcolor{cyan}{\ding{52}} & \textcolor{magenta}{\ding{56}} & \textcolor{magenta}{\ding{56}} & \textcolor{cyan}{\ding{52}} & \textcolor{magenta}{\ding{56}} & \textcolor{cyan}{\ding{52}} & \textcolor{magenta}{\ding{56}} & \textcolor{magenta}{\ding{56}} & \textcolor{magenta}{\ding{56}} & \textcolor{magenta}{\ding{56}} & \textcolor{magenta}{\ding{56}} & \textcolor{cyan}{\ding{52}} & \textcolor{magenta}{\ding{56}} & \textcolor{magenta}{\ding{56}} & \textcolor{magenta}{\ding{56}} & \textcolor{magenta}{\ding{56}} & \textcolor{magenta}{\ding{56}} & \textcolor{magenta}{\ding{56}} & \textcolor{cyan}{\ding{52}} & \textcolor{magenta}{\ding{56}} & \textcolor{magenta}{\ding{56}} & \textcolor{cyan}{\ding{52}} \\
            CEERS (9) & \textcolor{cyan}{\ding{52}} & \textcolor{magenta}{\ding{56}} & \textcolor{magenta}{\ding{56}} & \textcolor{cyan}{\ding{52}} & \textcolor{magenta}{\ding{56}} & \textcolor{cyan}{\ding{52}} & \textcolor{magenta}{\ding{56}} & \textcolor{magenta}{\ding{56}} & \textcolor{cyan}{\ding{52}} & \textcolor{magenta}{\ding{56}} & \textcolor{magenta}{\ding{56}} & \textcolor{cyan}{\ding{52}} & \textcolor{magenta}{\ding{56}} & \textcolor{magenta}{\ding{56}} & \textcolor{cyan}{\ding{52}} & \textcolor{magenta}{\ding{56}} & \textcolor{cyan}{\ding{52}} & \textcolor{magenta}{\ding{56}} & \textcolor{cyan}{\ding{52}} & \textcolor{magenta}{\ding{56}} & \textcolor{magenta}{\ding{56}} & \textcolor{cyan}{\ding{52}} \\
            JADES (11) & \textcolor{cyan}{\ding{52}} & \textcolor{magenta}{\ding{56}} & \textcolor{cyan}{\ding{52}} & \textcolor{cyan}{\ding{52}} & \textcolor{magenta}{\ding{56}} & \textcolor{cyan}{\ding{52}} & \textcolor{magenta}{\ding{56}} & \textcolor{magenta}{\ding{56}} & \textcolor{cyan}{\ding{52}} & \textcolor{magenta}{\ding{56}} & \textcolor{magenta}{\ding{56}} & \textcolor{cyan}{\ding{52}} & \textcolor{magenta}{\ding{56}} & \textcolor{cyan}{\ding{52}} & \textcolor{cyan}{\ding{52}} & \textcolor{magenta}{\ding{56}} & \textcolor{cyan}{\ding{52}} & \textcolor{magenta}{\ding{56}} & \textcolor{cyan}{\ding{52}} & \textcolor{magenta}{\ding{56}} & \textcolor{magenta}{\ding{56}} & \textcolor{cyan}{\ding{52}} \\
            JOF (16) & \textcolor{cyan}{\ding{52}} & \textcolor{magenta}{\ding{56}} & \textcolor{cyan}{\ding{52}} & \textcolor{cyan}{\ding{52}} & \textcolor{magenta}{\ding{56}} & \textcolor{cyan}{\ding{52}} & \textcolor{cyan}{\ding{52}} & \textcolor{cyan}{\ding{52}} & \textcolor{cyan}{\ding{52}} & \textcolor{cyan}{\ding{52}} & \textcolor{cyan}{\ding{52}} & \textcolor{cyan}{\ding{52}} & \textcolor{cyan}{\ding{52}} & \textcolor{cyan}{\ding{52}} & \textcolor{cyan}{\ding{52}} & \textcolor{magenta}{\ding{56}} & \textcolor{cyan}{\ding{52}} & \textcolor{magenta}{\ding{56}} & \textcolor{cyan}{\ding{52}} & \textcolor{magenta}{\ding{56}} & \textcolor{magenta}{\ding{56}} & \textcolor{cyan}{\ding{52}} \\
            \hline
	\end{tabular}
        }
\end{table*}

In Figure~\ref{fig:RestUV_Properties}, we provide the rest-UV continuum slope versus absolute magnitude for the final photometric sample, as described in Section~\ref{SectionThreeOne}. The blue points represent galaxies with MIRI/F770W detections while the pink points represent galaxies with MIRI/F770W non-detections. Galaxies within the sample of detections (non-detections) typically have rest-UV absolute magnitudes $-19.0 \gtrsim M_{\mathrm{UV}} \gtrsim -21.0$ ($-17.5 \gtrsim M_{\mathrm{UV}} \gtrsim -19.5$). This simple comparison suggests that the properties of the rest-UV photometry are a key indicator of whether a star-forming galaxy will be detected with MIRI/F770W at $z \approx 8$. The implication is that the final photometric sample of MIRI/F770W detections is a biased subset of the full photometric sample, as to be expected for any flux-limited sample of galaxies. We caution the reader that our analysis and results may only be applicable for typical high-redshift star-forming galaxies (i.e., galaxies at $z \approx 8$ with $M_{\mathrm{UV}} \approx -20$ and $\beta_{\mathrm{UV}} \approx -2$).

Since the adopted apparent magnitude cut (Figure~\ref{fig:m_F277W_Distribution}) of $m_{\mathrm{F277W}} < 29.0\ \mathrm{AB\ mag}$ is effectively an absolute magnitude cut at fixed continuum slope, the UV-faintest sources are also the UV-reddest, which is unexpected. Typically, the UV-faintest sources are the UV-bluest \citep[e.g.,][]{Topping:2024}. The yellow shaded region represents the area of parameter space that would not be selected by the adopted apparent magnitude cut, if we assume the continuum is fit by a simple power law. We should mention that one of the objects in the final photometric sample of MIRI/F770W detections is noticeably fainter in the rest-UV (JADES$-$GS$-$ID$-$$180446$, or JADES$-$GS$+53.08626$$-$$27.85932$, at $z_{\mathrm{spec}} = 7.956$ with $M_{\mathrm{UV}} \approx -18.5$ and $\beta_{\mathrm{UV}} \approx -1.9$) when compared to the other objects in this sample, despite having a similar continuum slope. This may suggest additional contributions to the MIRI/F770W flux beyond what we would expect from the stellar and nebular emission alone, which could arise from dust emission and/or the presence of an active galactic nucleus (AGN).

\subsection{Stellar Population Modeling}
\label{SectionFourTwo}

We derive the properties of the stellar populations for each of the $N = 22$ galaxies that are part of the final photometric sample of MIRI/F770W detections following the methodology of \citet{Tacchella:2022} by utilizing the SED fitting code \texttt{Prospector} \citep[v1.2.0;][]{Johnson:2021} and using the photometric measurements described in Section~\ref{SectionTwo}. We sample the posterior distributions of the stellar population properties using the dynamic nested sampling code \texttt{dynesty} \citep[v1.2.3;][]{Speagle:2020}. We adopt the Flexible Stellar Population Synthesis code \citep[\texttt{FSPS};][]{Conroy:2009, Conroy:2010} which is accessed through the \texttt{python-FSPS} bindings \citep{Foreman-Mackey:2014}. We assume the Modules for Experiments in Stellar Astrophysics \citep[MESA;][]{Paxton:2011, Paxton:2013, Paxton:2015, Paxton:2018} stellar evolution package and the synthetic models from MESA Isochrones and Stellar Tracks \citep[MIST;][]{Dotter:2016, Choi:2016}, which includes the effects of stellar rotation. The IMF from \citet{Chabrier:2003} is used throughout with a lower bound of  $0.08\,M_{\odot}$ and an upper bound of $120\,M_{\odot}$.

IGM absorption is modeled after \citet{Madau:1995}, with one free parameter for the overall scaling of the IGM attenuation curve ($f_{\mathrm{IGM}}$, assuming a clipped normal prior with $\mathrm{min} = 0.0$, $\mathrm{max} = 2.0$, $\mu = 1.0$, and $\sigma = 0.3$). 

Dust attenuation is modeled after \citet{Charlot:2000} with two free parameters, one for the diffuse dust optical depth ($\tau_{\mathrm{dust},\,2}$, assuming a clipped normal prior with $\mathrm{min} = 0.0$, $\mathrm{max} = 4.0$, $\mu = 0.3$, and $\sigma = 1.0$) and another for the birth-cloud dust optical depth ($\tau_{\mathrm{dust},\,1}$, assuming a clipped normal prior with $\mathrm{min} = 0.0$, $\mathrm{max} = 2.0$, $\mu = 1.0$, and $\sigma = 0.3$). There is an additional free parameter for the power-law modifier to the shape of the \citet{Calzetti:2000} diffuse dust attenuation curve ($n$, assuming a uniform prior with $\mathrm{min} = -1.0$ and $\mathrm{max} = +0.4$), which is tied to the strength of the rest-UV bump and is modeled after \citet{Kriek:2013}. 

Nebular emission (both from emission lines and continuum) is self-consistently modeled after \citet{Byler:2017} using the photoionization code \texttt{Cloudy} \citep[][]{Ferland:2013} with two free parameters, one for the gas-phase metallicity ($\mathrm{log}_{10}[Z_{\mathrm{gas}}/Z_{\odot}]$, assuming a uniform prior with $\mathrm{min} = -2.0$ and $\mathrm{max} = +0.5$) and another for the ionization parameter ($\mathrm{log}_{10}[U]$, assuming a uniform prior with $\mathrm{min} = -4.0$ and $\mathrm{max} = -1.0$). 

Two additional free parameters are included for the total stellar mass formed ($\mathrm{log}_{10}[M_{\ast}/M_{\odot}]$, assuming a uniform prior with $\mathrm{min} = +5.0$ and $\mathrm{max} = +12.0$) and the stellar metallicity ($\mathrm{log}_{10}[Z_{\ast}/Z_{\odot}]$, assuming a uniform prior with $\mathrm{min} = -2.00$ and $\mathrm{max} = +0.19$). 

Redshift is also allowed to be a free parameter ($z$, assuming a clipped normal prior with $\mathrm{min} = 0.0$, $\mathrm{max} = 20.0$, $\mu = z_{\mathrm{a}}$, and $\sigma = \Delta z_{1}/2$) using the photometric redshift results from Section~\ref{SectionThreeOne} as informative priors. For objects with an available spectroscopic redshift, we additionally derive properties with redshift as a fixed parameter to explore the impact of allowing the redshift to be free. For these galaxies, we adopt the results with redshift as a fixed parameter for all of the subsequent analyses. We verified that fixing to the spectroscopic redshift does not change any of our results or conclusions. However, there is one galaxy where the inclusion of the spectroscopic redshift has a significant impact on the inferred properties and ultimately results in a much better fit when compared to using the photometric redshift (JADES$-$GS$-$ID$-$$174121$, or JADES$-$GS$+53.05567$$-$$27.86882$, at $z_{\mathrm{spec}} = 7.623$ with $M_{\mathrm{UV}} \approx -19.9$ and $\beta_{\mathrm{UV}} \approx -1.9$). The redshift of this galaxy places the nebular emission lines $\mathrm{H}\beta$ and $\mathrm{[OIII]}\lambda\lambda 4960{,}5008$ near the edge of the F410M transmission curve. As a result of this, these emission lines produce equal amounts of excess flux in F410M and F444W, resulting in an F410M$-$F444W color consistent with zero. Without the spectroscopic redshift, \texttt{Prospector} tries to explain this observed color with stellar continuum rather than emission-line contributions, which produces strong Balmer breaks in the models.

Finally, we assume four different models for the SFH, since we would like to understand how the derived stellar population properties depend on the assumed model, particularly for the different assumed filter sets. Two of these models are parametric (one is constant with respect to time and has $\mathrm{SFR}[t] \propto C$, and the other has the shape of a delayed-tau function with $\mathrm{SFR}[t] \propto [t - t_{0}] \,e^{-[t - t_{0}]/\tau}$) while the other two models are non-parametric \citep[one with the ``continuity'' prior, the other with the ``bursty'' prior; see also][]{Tacchella:2022}. Throughout this work, we assume the results from the parametric constant SFH model to be fiducial. These include all of the derived stellar population properties and rest-frame apparent magnitudes.

The parametric constant SFH (hereafter referred to as CSFH) model has one free parameter, corresponding to the galaxy age ($t_{0}/\mathrm{Myr}$, assuming a log-uniform prior with $\mathrm{min} = 1.0$, $\mathrm{max} = t_{\mathrm{univ}}/\mathrm{Myr}$, where $t_{\mathrm{univ}}$ is the age of the Universe measured with respect to the formation redshift $z_{\mathrm{form}} = 20$). The parametric delayed-tau SFH (hereafter referred to as DtSFH) model has two free parameters, corresponding to the galaxy age ($t_{0}/\mathrm{Myr}$, assuming a log-uniform prior with $\mathrm{min} = 1.0$, $\mathrm{max} = t_{\mathrm{univ}}/\mathrm{Myr}$) and the $e$-folding time ($\tau/\mathrm{Gyr}$, assuming a log-uniform prior with $\mathrm{min} = 0.001$, $\mathrm{max} = 30.0$). For these parametric SFHs, it is implicitly assumed that there is no star formation prior to the formation redshift $z_{\mathrm{form}} = 20$.

The non-parametric ``continuity'' prior SFH (hereafter referred to as ContSFH) and ``bursty continuity'' prior SFH (hereafter referred to as BurstySFH) models both have five free parameters, corresponding to the ratio of SFRs in adjacent time bins ($R_{\mathrm{SFR}}$, assuming Student's $t$-distribution prior with $\mu = 0.0, \sigma = 0.3$ for the ContSFH and $\mu = 0.0, \sigma = 1.0$ for the BurstySFH). These non-parametric SFHs assume that the SFH can be described as $N_{\mathrm{SFR}}$ distinct time bins, with the SFR remaining constant within each of these time bins, allowing for large flexibility in the shape of the SFHs since these are non-parametric with respect to time. The number of distinct time bins is fixed to be $N_{\mathrm{SFR}} = 6$. These time bins are calculated in units of lookback time, with the first three fixed at $0-3\ \mathrm{Myr}$, $3-10\ \mathrm{Myr}$, and $10-30\ \mathrm{Myr}$ while the last bin is fixed at $0.85\,t_{\mathrm{univ}}-t_{\mathrm{univ}}$, where $t_{\mathrm{univ}}$ is the age of the Universe measured with respect to the formation redshift $z_{\mathrm{form}} = 20$. The remaining time bins are divided evenly in logarithmic time between $30\ \mathrm{Myr}$ and $t_{\mathrm{univ}}$. For these non-parametric SFHs, it is implicitly assumed that the star formation rate is constant with respect to time and that there is no star formation prior to the formation redshift $z_{\mathrm{form}} = 20$.

One of the primary goals of this paper is to explore the impact of filter choice on the inferred properties of typical star-forming galaxies during the EoR. To accomplish this, we adopt four different filter sets when performing these \texttt{Prospector} fits with our four assumed models for the SFH, totaling $16$ different \texttt{Prospector} fits for each of the $N = 22$ galaxies that are part of the final photometric sample described in Section~\ref{SectionThreeOne}. These filter sets correspond to those commonly used in extragalactic observing programs with JWST and are summarized in Table~\ref{tab:FilterSetTable}, including filters from HST/ACS, JWST/NIRCam, and JWST/MIRI. The four different filter sets, in decreasing order of number of filters, including MIRI/F770W: the JOF \citep{Eisenstein:2025} set of $16$ filters, the JADES \citep{Eisenstein:2026} set of $11$ filters, the Cosmic Evolution Early Release Science Survey \citep[CEERS;][]{Finkelstein:2023} set of nine filters, and the COSMOS-Web \citep{Casey:2023} set of six filters. Table~\ref{tab:FilterSetTable} includes all of the available medium- and wide-band filters from JWST/NIRCam to illustrate the near completeness of the JOF filter set, which is only missing four medium bands and one wide band.

\subsection{Properties of the Stellar Populations}
\label{SectionFourThree}

\figsetstart
\figsetnum{5}
\graphicspath{{./}{files/ExampleSEDs/}}

\figsetgrpstart
\figsetgrpnum{5.1}
\figsetplot{SED_014647_ConstantPrior_JADES_freeRedshift.pdf}
\figsetgrpnote{\textit{Top panel}: Example of an SED for a typical galaxy (JADES$-$GS$-$ID$-$$14647$, or JADES$-$GS$+53.08646$$-$$27.88925$, at $z_{\mathrm{phot}} = 7.38 \pm 0.05$) from the final photometric sample of MIRI/F770W detections described in Section~\ref{SectionThreeOne}. The grey points represent the observed photometry assuming the JADES filter set. The medians for the  \texttt{Prospector} model photometry and spectroscopy are given by the squares and solid lines, respectively. The $68\%$ confidence interval of the \texttt{Prospector} models is shown by the shaded regions. \textit{Bottom panels}: Example of posterior distributions for the most relevant stellar population properties. From left to right: stellar mass, mass-weighted stellar age, sSFR averaged over the last $10$ Myr, and rest-frame EW of $\mathrm{[OIII]}+\mathrm{H}\beta$. The dashed lines represent the medians for the \texttt{Prospector} models while the dotted lines represent $68\%$ confidence intervals. \textit{Top and bottom panels}: The fiducial \texttt{Prospector} model is described in Section~\ref{SectionFourTwo} and assumes the constant SFH model. The blue (pink) squares, lines, and shaded regions represent results from fitting to the full JADES filter set, including (excluding) MIRI/F770W. For this typical galaxy, MIRI/F770W has little to no effect on the inferred SED and stellar population properties. This is true for the vast majority ($\approx 80\%$) of galaxies in the final photometric sample, suggesting that observations with JWST/MIRI are not needed to robustly determine the stellar population properties for typical star-forming galaxies during the EoR.}
\figsetgrpend

\figsetgrpstart
\figsetgrpnum{5.2}
\figsetplot{SED_025526_ConstantPrior_JADES_fixedRedshift.pdf}
\figsetgrpnote{\textit{Top panel}: Example of an SED for a typical galaxy (JADES$-$GS$-$ID$-$$25526$, or JADES$-$GS$+53.09942$$-$$27.88038$, at $z_{\mathrm{spec}} = 7.957$) from the final photometric sample of MIRI/F770W detections described in Section~\ref{SectionThreeOne}. The grey points represent the observed photometry assuming the JADES filter set. The medians for the  \texttt{Prospector} model photometry and spectroscopy are given by the squares and solid lines, respectively. The $68\%$ confidence interval of the \texttt{Prospector} models is shown by the shaded regions. \textit{Bottom panels}: Example of posterior distributions for the most relevant stellar population properties. From left to right: stellar mass, mass-weighted stellar age, sSFR averaged over the last $10$ Myr, and rest-frame EW of $\mathrm{[OIII]}+\mathrm{H}\beta$. The dashed lines represent the medians for the \texttt{Prospector} models while the dotted lines represent $68\%$ confidence intervals. \textit{Top and bottom panels}: The fiducial \texttt{Prospector} model is described in Section~\ref{SectionFourTwo} and assumes the constant SFH model. The blue (pink) squares, lines, and shaded regions represent results from fitting to the full JADES filter set, including (excluding) MIRI/F770W. For this typical galaxy, MIRI/F770W has little to no effect on the inferred SED and stellar population properties. This is true for the vast majority ($\approx 80\%$) of galaxies in the final photometric sample, suggesting that observations with JWST/MIRI are not needed to robustly determine the stellar population properties for typical star-forming galaxies during the EoR.}
\figsetgrpend

\figsetgrpstart
\figsetgrpnum{5.3}
\figsetplot{SED_027503_ConstantPrior_JADES_fixedRedshift.pdf}
\figsetgrpnote{\textit{Top panel}: Example of an SED for a typical galaxy (JADES$-$GS$-$ID$-$$27503$, or JADES$-$GS$+53.07581$$-$$27.87938$, at $z_{\mathrm{spec}} = 8.196$) from the final photometric sample of MIRI/F770W detections described in Section~\ref{SectionThreeOne}. The grey points represent the observed photometry assuming the JADES filter set. The medians for the  \texttt{Prospector} model photometry and spectroscopy are given by the squares and solid lines, respectively. The $68\%$ confidence interval of the \texttt{Prospector} models is shown by the shaded regions. \textit{Bottom panels}: Example of posterior distributions for the most relevant stellar population properties. From left to right: stellar mass, mass-weighted stellar age, sSFR averaged over the last $10$ Myr, and rest-frame EW of $\mathrm{[OIII]}+\mathrm{H}\beta$. The dashed lines represent the medians for the \texttt{Prospector} models while the dotted lines represent $68\%$ confidence intervals. \textit{Top and bottom panels}: The fiducial \texttt{Prospector} model is described in Section~\ref{SectionFourTwo} and assumes the constant SFH model. The blue (pink) squares, lines, and shaded regions represent results from fitting to the full JADES filter set, including (excluding) MIRI/F770W. For this typical galaxy, MIRI/F770W has little to no effect on the inferred SED and stellar population properties. This is true for the vast majority ($\approx 80\%$) of galaxies in the final photometric sample, suggesting that observations with JWST/MIRI are not needed to robustly determine the stellar population properties for typical star-forming galaxies during the EoR.}
\figsetgrpend

\figsetgrpstart
\figsetgrpnum{5.4}
\figsetplot{SED_030333_ConstantPrior_JADES_fixedRedshift.pdf}
\figsetgrpnote{\textit{Top panel}: Example of an SED for a typical galaxy (JADES$-$GS$-$ID$-$$30333$, or JADES$-$GS$+53.05373$$-$$27.87789$, at $z_{\mathrm{spec}} = 7.891$) from the final photometric sample of MIRI/F770W detections described in Section~\ref{SectionThreeOne}. The grey points represent the observed photometry assuming the JADES filter set. The medians for the  \texttt{Prospector} model photometry and spectroscopy are given by the squares and solid lines, respectively. The $68\%$ confidence interval of the \texttt{Prospector} models is shown by the shaded regions. \textit{Bottom panels}: Example of posterior distributions for the most relevant stellar population properties. From left to right: stellar mass, mass-weighted stellar age, sSFR averaged over the last $10$ Myr, and rest-frame EW of $\mathrm{[OIII]}+\mathrm{H}\beta$. The dashed lines represent the medians for the \texttt{Prospector} models while the dotted lines represent $68\%$ confidence intervals. \textit{Top and bottom panels}: The fiducial \texttt{Prospector} model is described in Section~\ref{SectionFourTwo} and assumes the constant SFH model. The blue (pink) squares, lines, and shaded regions represent results from fitting to the full JADES filter set, including (excluding) MIRI/F770W. For this typical galaxy, MIRI/F770W has little to no effect on the inferred SED and stellar population properties. This is true for the vast majority ($\approx 80\%$) of galaxies in the final photometric sample, suggesting that observations with JWST/MIRI are not needed to robustly determine the stellar population properties for typical star-forming galaxies during the EoR.}
\figsetgrpend

\figsetgrpstart
\figsetgrpnum{5.5}
\figsetplot{SED_037458_ConstantPrior_JADES_fixedRedshift.pdf}
\figsetgrpnote{\textit{Top panel}: Example of an SED for a typical galaxy (JADES$-$GS$-$ID$-$$37458$, or JADES$-$GS$+53.08932$$-$$27.87269$, at $z_{\mathrm{spec}} = 8.225$) from the final photometric sample of MIRI/F770W detections described in Section~\ref{SectionThreeOne}. The grey points represent the observed photometry assuming the JADES filter set. The medians for the  \texttt{Prospector} model photometry and spectroscopy are given by the squares and solid lines, respectively. The $68\%$ confidence interval of the \texttt{Prospector} models is shown by the shaded regions. \textit{Bottom panels}: Example of posterior distributions for the most relevant stellar population properties. From left to right: stellar mass, mass-weighted stellar age, sSFR averaged over the last $10$ Myr, and rest-frame EW of $\mathrm{[OIII]}+\mathrm{H}\beta$. The dashed lines represent the medians for the \texttt{Prospector} models while the dotted lines represent $68\%$ confidence intervals. \textit{Top and bottom panels}: The fiducial \texttt{Prospector} model is described in Section~\ref{SectionFourTwo} and assumes the constant SFH model. The blue (pink) squares, lines, and shaded regions represent results from fitting to the full JADES filter set, including (excluding) MIRI/F770W. For this typical galaxy, MIRI/F770W has little to no effect on the inferred SED and stellar population properties. This is true for the vast majority ($\approx 80\%$) of galaxies in the final photometric sample, suggesting that observations with JWST/MIRI are not needed to robustly determine the stellar population properties for typical star-forming galaxies during the EoR.}
\figsetgrpend

\figsetgrpstart
\figsetgrpnum{5.6}
\figsetplot{SED_057378_ConstantPrior_JADES_fixedRedshift.pdf}
\figsetgrpnote{\textit{Top panel}: Example of an SED for a typical galaxy (JADES$-$GS$-$ID$-$$57378$, or JADES$-$GS$+53.08650$$-$$27.85920$, at $z_{\mathrm{spec}} = 7.950$) from the final photometric sample of MIRI/F770W detections described in Section~\ref{SectionThreeOne}. The grey points represent the observed photometry assuming the JADES filter set. The medians for the  \texttt{Prospector} model photometry and spectroscopy are given by the squares and solid lines, respectively. The $68\%$ confidence interval of the \texttt{Prospector} models is shown by the shaded regions. \textit{Bottom panels}: Example of posterior distributions for the most relevant stellar population properties. From left to right: stellar mass, mass-weighted stellar age, sSFR averaged over the last $10$ Myr, and rest-frame EW of $\mathrm{[OIII]}+\mathrm{H}\beta$. The dashed lines represent the medians for the \texttt{Prospector} models while the dotted lines represent $68\%$ confidence intervals. \textit{Top and bottom panels}: The fiducial \texttt{Prospector} model is described in Section~\ref{SectionFourTwo} and assumes the constant SFH model. The blue (pink) squares, lines, and shaded regions represent results from fitting to the full JADES filter set, including (excluding) MIRI/F770W. For this typical galaxy, MIRI/F770W has little to no effect on the inferred SED and stellar population properties. This is true for the vast majority ($\approx 80\%$) of galaxies in the final photometric sample, suggesting that observations with JWST/MIRI are not needed to robustly determine the stellar population properties for typical star-forming galaxies during the EoR.}
\figsetgrpend

\figsetgrpstart
\figsetgrpnum{5.7}
\figsetplot{SED_164055_ConstantPrior_JADES_fixedRedshift.pdf}
\figsetgrpnote{\textit{Top panel}: Example of an SED for a typical galaxy (JADES$-$GS$-$ID$-$$164055$, or JADES$-$GS$+53.08168$$-$$27.88858$, at $z_{\mathrm{spec}} = 7.397$) from the final photometric sample of MIRI/F770W detections described in Section~\ref{SectionThreeOne}. The grey points represent the observed photometry assuming the JADES filter set. The medians for the  \texttt{Prospector} model photometry and spectroscopy are given by the squares and solid lines, respectively. The $68\%$ confidence interval of the \texttt{Prospector} models is shown by the shaded regions. \textit{Bottom panels}: Example of posterior distributions for the most relevant stellar population properties. From left to right: stellar mass, mass-weighted stellar age, sSFR averaged over the last $10$ Myr, and rest-frame EW of $\mathrm{[OIII]}+\mathrm{H}\beta$. The dashed lines represent the medians for the \texttt{Prospector} models while the dotted lines represent $68\%$ confidence intervals. \textit{Top and bottom panels}: The fiducial \texttt{Prospector} model is described in Section~\ref{SectionFourTwo} and assumes the constant SFH model. The blue (pink) squares, lines, and shaded regions represent results from fitting to the full JADES filter set, including (excluding) MIRI/F770W. For this typical galaxy, MIRI/F770W has little to no effect on the inferred SED and stellar population properties. This is true for the vast majority ($\approx 80\%$) of galaxies in the final photometric sample, suggesting that observations with JWST/MIRI are not needed to robustly determine the stellar population properties for typical star-forming galaxies during the EoR.}
\figsetgrpend

\figsetgrpstart
\figsetgrpnum{5.8}
\figsetplot{SED_165595_ConstantPrior_JADES_fixedRedshift.pdf}
\figsetgrpnote{\textit{Top panel}: Example of an SED for a typical galaxy (JADES$-$GS$-$ID$-$$165595$, or JADES$-$GS$+53.05830$$-$$27.88486$, at $z_{\mathrm{spec}} = 8.585$) from the final photometric sample of MIRI/F770W detections described in Section~\ref{SectionThreeOne}. The grey points represent the observed photometry assuming the JADES filter set. The medians for the  \texttt{Prospector} model photometry and spectroscopy are given by the squares and solid lines, respectively. The $68\%$ confidence interval of the \texttt{Prospector} models is shown by the shaded regions. \textit{Bottom panels}: Example of posterior distributions for the most relevant stellar population properties. From left to right: stellar mass, mass-weighted stellar age, sSFR averaged over the last $10$ Myr, and rest-frame EW of $\mathrm{[OIII]}+\mathrm{H}\beta$. The dashed lines represent the medians for the \texttt{Prospector} models while the dotted lines represent $68\%$ confidence intervals. \textit{Top and bottom panels}: The fiducial \texttt{Prospector} model is described in Section~\ref{SectionFourTwo} and assumes the constant SFH model. The blue (pink) squares, lines, and shaded regions represent results from fitting to the full JADES filter set, including (excluding) MIRI/F770W. For this typical galaxy, MIRI/F770W has little to no effect on the inferred SED and stellar population properties. This is true for the vast majority ($\approx 80\%$) of galaxies in the final photometric sample, suggesting that observations with JWST/MIRI are not needed to robustly determine the stellar population properties for typical star-forming galaxies during the EoR.}
\figsetgrpend

\figsetgrpstart
\figsetgrpnum{5.9}
\figsetplot{SED_173624_ConstantPrior_JADES_fixedRedshift.pdf}
\figsetgrpnote{\textit{Top panel}: Example of an SED for a typical galaxy (JADES$-$GS$-$ID$-$$173624$, or JADES$-$GS$+53.07688$$-$$27.86967$, at $z_{\mathrm{spec}} = 8.270$) from the final photometric sample of MIRI/F770W detections described in Section~\ref{SectionThreeOne}. The grey points represent the observed photometry assuming the JADES filter set. The medians for the  \texttt{Prospector} model photometry and spectroscopy are given by the squares and solid lines, respectively. The $68\%$ confidence interval of the \texttt{Prospector} models is shown by the shaded regions. \textit{Bottom panels}: Example of posterior distributions for the most relevant stellar population properties. From left to right: stellar mass, mass-weighted stellar age, sSFR averaged over the last $10$ Myr, and rest-frame EW of $\mathrm{[OIII]}+\mathrm{H}\beta$. The dashed lines represent the medians for the \texttt{Prospector} models while the dotted lines represent $68\%$ confidence intervals. \textit{Top and bottom panels}: The fiducial \texttt{Prospector} model is described in Section~\ref{SectionFourTwo} and assumes the constant SFH model. The blue (pink) squares, lines, and shaded regions represent results from fitting to the full JADES filter set, including (excluding) MIRI/F770W. For this typical galaxy, MIRI/F770W has little to no effect on the inferred SED and stellar population properties. This is true for the vast majority ($\approx 80\%$) of galaxies in the final photometric sample, suggesting that observations with JWST/MIRI are not needed to robustly determine the stellar population properties for typical star-forming galaxies during the EoR.}
\figsetgrpend

\figsetgrpstart
\figsetgrpnum{5.10}
\figsetplot{SED_173679_ConstantPrior_JADES_freeRedshift.pdf}
\figsetgrpnote{\textit{Top panel}: Example of an SED for a typical galaxy (JADES$-$GS$-$ID$-$$173679$, or JADES$-$GS$+53.07277$$-$$27.86929$, at $z_{\mathrm{phot}} = 8.08 \pm 0.19$) from the final photometric sample of MIRI/F770W detections described in Section~\ref{SectionThreeOne}. The grey points represent the observed photometry assuming the JADES filter set. The medians for the  \texttt{Prospector} model photometry and spectroscopy are given by the squares and solid lines, respectively. The $68\%$ confidence interval of the \texttt{Prospector} models is shown by the shaded regions. \textit{Bottom panels}: Example of posterior distributions for the most relevant stellar population properties. From left to right: stellar mass, mass-weighted stellar age, sSFR averaged over the last $10$ Myr, and rest-frame EW of $\mathrm{[OIII]}+\mathrm{H}\beta$. The dashed lines represent the medians for the \texttt{Prospector} models while the dotted lines represent $68\%$ confidence intervals. \textit{Top and bottom panels}: The fiducial \texttt{Prospector} model is described in Section~\ref{SectionFourTwo} and assumes the constant SFH model. The blue (pink) squares, lines, and shaded regions represent results from fitting to the full JADES filter set, including (excluding) MIRI/F770W. For this typical galaxy, MIRI/F770W has little to no effect on the inferred SED and stellar population properties. This is true for the vast majority ($\approx 80\%$) of galaxies in the final photometric sample, suggesting that observations with JWST/MIRI are not needed to robustly determine the stellar population properties for typical star-forming galaxies during the EoR.}
\figsetgrpend

\figsetgrpstart
\figsetgrpnum{5.11}
\figsetplot{SED_174121_ConstantPrior_JADES_fixedRedshift.pdf}
\figsetgrpnote{\textit{Top panel}: Example of an SED for a typical galaxy (JADES$-$GS$-$ID$-$$174121$, or JADES$-$GS$+53.05567$$-$$27.86882$, at $z_{\mathrm{spec}} = 7.623$) from the final photometric sample of MIRI/F770W detections described in Section~\ref{SectionThreeOne}. The grey points represent the observed photometry assuming the JADES filter set. The medians for the  \texttt{Prospector} model photometry and spectroscopy are given by the squares and solid lines, respectively. The $68\%$ confidence interval of the \texttt{Prospector} models is shown by the shaded regions. \textit{Bottom panels}: Example of posterior distributions for the most relevant stellar population properties. From left to right: stellar mass, mass-weighted stellar age, sSFR averaged over the last $10$ Myr, and rest-frame EW of $\mathrm{[OIII]}+\mathrm{H}\beta$. The dashed lines represent the medians for the \texttt{Prospector} models while the dotted lines represent $68\%$ confidence intervals. \textit{Top and bottom panels}: The fiducial \texttt{Prospector} model is described in Section~\ref{SectionFourTwo} and assumes the constant SFH model. The blue (pink) squares, lines, and shaded regions represent results from fitting to the full JADES filter set, including (excluding) MIRI/F770W. For this typical galaxy, MIRI/F770W has little to no effect on the inferred SED and stellar population properties. This is true for the vast majority ($\approx 80\%$) of galaxies in the final photometric sample, suggesting that observations with JWST/MIRI are not needed to robustly determine the stellar population properties for typical star-forming galaxies during the EoR.}
\figsetgrpend

\figsetgrpstart
\figsetgrpnum{5.12}
\figsetplot{SED_175729_ConstantPrior_JADES_fixedRedshift.pdf}
\figsetgrpnote{\textit{Top panel}: Example of an SED for a typical galaxy (JADES$-$GS$-$ID$-$$175729$, or JADES$-$GS$+53.06058$$-$$27.86603$, at $z_{\mathrm{spec}} = 7.883$) from the final photometric sample of MIRI/F770W detections described in Section~\ref{SectionThreeOne}. The grey points represent the observed photometry assuming the JADES filter set. The medians for the  \texttt{Prospector} model photometry and spectroscopy are given by the squares and solid lines, respectively. The $68\%$ confidence interval of the \texttt{Prospector} models is shown by the shaded regions. \textit{Bottom panels}: Example of posterior distributions for the most relevant stellar population properties. From left to right: stellar mass, mass-weighted stellar age, sSFR averaged over the last $10$ Myr, and rest-frame EW of $\mathrm{[OIII]}+\mathrm{H}\beta$. The dashed lines represent the medians for the \texttt{Prospector} models while the dotted lines represent $68\%$ confidence intervals. \textit{Top and bottom panels}: The fiducial \texttt{Prospector} model is described in Section~\ref{SectionFourTwo} and assumes the constant SFH model. The blue (pink) squares, lines, and shaded regions represent results from fitting to the full JADES filter set, including (excluding) MIRI/F770W. For this typical galaxy, MIRI/F770W has little to no effect on the inferred SED and stellar population properties. This is true for the vast majority ($\approx 80\%$) of galaxies in the final photometric sample, suggesting that observations with JWST/MIRI are not needed to robustly determine the stellar population properties for typical star-forming galaxies during the EoR.}
\figsetgrpend

\figsetgrpstart
\figsetgrpnum{5.13}
\figsetplot{SED_175837_ConstantPrior_JADES_fixedRedshift.pdf}
\figsetgrpnote{\textit{Top panel}: Example of an SED for a typical galaxy (JADES$-$GS$-$ID$-$$175837$, or JADES$-$GS$+53.06021$$-$$27.86572$, at $z_{\mathrm{spec}} = 7.884$) from the final photometric sample of MIRI/F770W detections described in Section~\ref{SectionThreeOne}. The grey points represent the observed photometry assuming the JADES filter set. The medians for the  \texttt{Prospector} model photometry and spectroscopy are given by the squares and solid lines, respectively. The $68\%$ confidence interval of the \texttt{Prospector} models is shown by the shaded regions. \textit{Bottom panels}: Example of posterior distributions for the most relevant stellar population properties. From left to right: stellar mass, mass-weighted stellar age, sSFR averaged over the last $10$ Myr, and rest-frame EW of $\mathrm{[OIII]}+\mathrm{H}\beta$. The dashed lines represent the medians for the \texttt{Prospector} models while the dotted lines represent $68\%$ confidence intervals. \textit{Top and bottom panels}: The fiducial \texttt{Prospector} model is described in Section~\ref{SectionFourTwo} and assumes the constant SFH model. The blue (pink) squares, lines, and shaded regions represent results from fitting to the full JADES filter set, including (excluding) MIRI/F770W. For this typical galaxy, MIRI/F770W has little to no effect on the inferred SED and stellar population properties. This is true for the vast majority ($\approx 80\%$) of galaxies in the final photometric sample, suggesting that observations with JWST/MIRI are not needed to robustly determine the stellar population properties for typical star-forming galaxies during the EoR.}
\figsetgrpend

\figsetgrpstart
\figsetgrpnum{5.14}
\figsetplot{SED_177322_ConstantPrior_JADES_fixedRedshift.pdf}
\figsetgrpnote{\textit{Top panel}: Example of an SED for a typical galaxy (JADES$-$GS$-$ID$-$$177322$, or JADES$-$GS$+53.06036$$-$$27.86355$, at $z_{\mathrm{spec}} = 7.885$) from the final photometric sample of MIRI/F770W detections described in Section~\ref{SectionThreeOne}. The grey points represent the observed photometry assuming the JADES filter set. The medians for the  \texttt{Prospector} model photometry and spectroscopy are given by the squares and solid lines, respectively. The $68\%$ confidence interval of the \texttt{Prospector} models is shown by the shaded regions. \textit{Bottom panels}: Example of posterior distributions for the most relevant stellar population properties. From left to right: stellar mass, mass-weighted stellar age, sSFR averaged over the last $10$ Myr, and rest-frame EW of $\mathrm{[OIII]}+\mathrm{H}\beta$. The dashed lines represent the medians for the \texttt{Prospector} models while the dotted lines represent $68\%$ confidence intervals. \textit{Top and bottom panels}: The fiducial \texttt{Prospector} model is described in Section~\ref{SectionFourTwo} and assumes the constant SFH model. The blue (pink) squares, lines, and shaded regions represent results from fitting to the full JADES filter set, including (excluding) MIRI/F770W. For this typical galaxy, MIRI/F770W has little to no effect on the inferred SED and stellar population properties. This is true for the vast majority ($\approx 80\%$) of galaxies in the final photometric sample, suggesting that observations with JWST/MIRI are not needed to robustly determine the stellar population properties for typical star-forming galaxies during the EoR.}
\figsetgrpend

\figsetgrpstart
\figsetgrpnum{5.15}
\figsetplot{SED_180446_ConstantPrior_JADES_fixedRedshift.pdf}
\figsetgrpnote{\textit{Top panel}: Example of an SED for a typical galaxy (JADES$-$GS$-$ID$-$$180446$, or JADES$-$GS$+53.08626$$-$$27.85932$, at $z_{\mathrm{spec}} = 7.956$) from the final photometric sample of MIRI/F770W detections described in Section~\ref{SectionThreeOne}. The grey points represent the observed photometry assuming the JADES filter set. The medians for the  \texttt{Prospector} model photometry and spectroscopy are given by the squares and solid lines, respectively. The $68\%$ confidence interval of the \texttt{Prospector} models is shown by the shaded regions. \textit{Bottom panels}: Example of posterior distributions for the most relevant stellar population properties. From left to right: stellar mass, mass-weighted stellar age, sSFR averaged over the last $10$ Myr, and rest-frame EW of $\mathrm{[OIII]}+\mathrm{H}\beta$. The dashed lines represent the medians for the \texttt{Prospector} models while the dotted lines represent $68\%$ confidence intervals. \textit{Top and bottom panels}: The fiducial \texttt{Prospector} model is described in Section~\ref{SectionFourTwo} and assumes the constant SFH model. The blue (pink) squares, lines, and shaded regions represent results from fitting to the full JADES filter set, including (excluding) MIRI/F770W. For this typical galaxy, MIRI/F770W has little to no effect on the inferred SED and stellar population properties. This is true for the vast majority ($\approx 80\%$) of galaxies in the final photometric sample, suggesting that observations with JWST/MIRI are not needed to robustly determine the stellar population properties for typical star-forming galaxies during the EoR.}
\figsetgrpend

\figsetgrpstart
\figsetgrpnum{5.16}
\figsetplot{SED_300287_ConstantPrior_JADES_freeRedshift.pdf}
\figsetgrpnote{\textit{Top panel}: Example of an SED for a typical galaxy (JADES$-$GS$-$ID$-$$300287$, or JADES$-$GS$+53.08812$$-$$27.90817$, at $z_{\mathrm{phot}} = 8.08 \pm 0.10$) from the final photometric sample of MIRI/F770W detections described in Section~\ref{SectionThreeOne}. The grey points represent the observed photometry assuming the JADES filter set. The medians for the  \texttt{Prospector} model photometry and spectroscopy are given by the squares and solid lines, respectively. The $68\%$ confidence interval of the \texttt{Prospector} models is shown by the shaded regions. \textit{Bottom panels}: Example of posterior distributions for the most relevant stellar population properties. From left to right: stellar mass, mass-weighted stellar age, sSFR averaged over the last $10$ Myr, and rest-frame EW of $\mathrm{[OIII]}+\mathrm{H}\beta$. The dashed lines represent the medians for the \texttt{Prospector} models while the dotted lines represent $68\%$ confidence intervals. \textit{Top and bottom panels}: The fiducial \texttt{Prospector} model is described in Section~\ref{SectionFourTwo} and assumes the constant SFH model. The blue (pink) squares, lines, and shaded regions represent results from fitting to the full JADES filter set, including (excluding) MIRI/F770W. For this typical galaxy, MIRI/F770W has little to no effect on the inferred SED and stellar population properties. This is true for the vast majority ($\approx 80\%$) of galaxies in the final photometric sample, suggesting that observations with JWST/MIRI are not needed to robustly determine the stellar population properties for typical star-forming galaxies during the EoR.}
\figsetgrpend

\figsetgrpstart
\figsetgrpnum{5.17}
\figsetplot{SED_300391_ConstantPrior_JADES_freeRedshift.pdf}
\figsetgrpnote{\textit{Top panel}: Example of an SED for a typical galaxy (JADES$-$GS$-$ID$-$$300391$, or JADES$-$GS$+53.08513$$-$$27.90636$, at $z_{\mathrm{phot}} = 8.29 \pm 0.08$) from the final photometric sample of MIRI/F770W detections described in Section~\ref{SectionThreeOne}. The grey points represent the observed photometry assuming the JADES filter set. The medians for the  \texttt{Prospector} model photometry and spectroscopy are given by the squares and solid lines, respectively. The $68\%$ confidence interval of the \texttt{Prospector} models is shown by the shaded regions. \textit{Bottom panels}: Example of posterior distributions for the most relevant stellar population properties. From left to right: stellar mass, mass-weighted stellar age, sSFR averaged over the last $10$ Myr, and rest-frame EW of $\mathrm{[OIII]}+\mathrm{H}\beta$. The dashed lines represent the medians for the \texttt{Prospector} models while the dotted lines represent $68\%$ confidence intervals. \textit{Top and bottom panels}: The fiducial \texttt{Prospector} model is described in Section~\ref{SectionFourTwo} and assumes the constant SFH model. The blue (pink) squares, lines, and shaded regions represent results from fitting to the full JADES filter set, including (excluding) MIRI/F770W. For this typical galaxy, MIRI/F770W has little to no effect on the inferred SED and stellar population properties. This is true for the vast majority ($\approx 80\%$) of galaxies in the final photometric sample, suggesting that observations with JWST/MIRI are not needed to robustly determine the stellar population properties for typical star-forming galaxies during the EoR.}
\figsetgrpend

\figsetgrpstart
\figsetgrpnum{5.18}
\figsetplot{SED_380203_ConstantPrior_JADES_freeRedshift.pdf}
\figsetgrpnote{\textit{Top panel}: Example of an SED for a typical galaxy (JADES$-$GS$-$ID$-$$380203$, or JADES$-$GS$+53.08172$$-$$27.89881$, at $z_{\mathrm{phot}} = 8.03 \pm 0.26$) from the final photometric sample of MIRI/F770W detections described in Section~\ref{SectionThreeOne}. The grey points represent the observed photometry assuming the JADES filter set. The medians for the  \texttt{Prospector} model photometry and spectroscopy are given by the squares and solid lines, respectively. The $68\%$ confidence interval of the \texttt{Prospector} models is shown by the shaded regions. \textit{Bottom panels}: Example of posterior distributions for the most relevant stellar population properties. From left to right: stellar mass, mass-weighted stellar age, sSFR averaged over the last $10$ Myr, and rest-frame EW of $\mathrm{[OIII]}+\mathrm{H}\beta$. The dashed lines represent the medians for the \texttt{Prospector} models while the dotted lines represent $68\%$ confidence intervals. \textit{Top and bottom panels}: The fiducial \texttt{Prospector} model is described in Section~\ref{SectionFourTwo} and assumes the constant SFH model. The blue (pink) squares, lines, and shaded regions represent results from fitting to the full JADES filter set, including (excluding) MIRI/F770W. For this typical galaxy, MIRI/F770W has little to no effect on the inferred SED and stellar population properties. This is true for the vast majority ($\approx 80\%$) of galaxies in the final photometric sample, suggesting that observations with JWST/MIRI are not needed to robustly determine the stellar population properties for typical star-forming galaxies during the EoR.}
\figsetgrpend

\figsetend

\begin{figure*}
    \centering
    \figurenum{5}
    \includegraphics[width=1.0\linewidth]{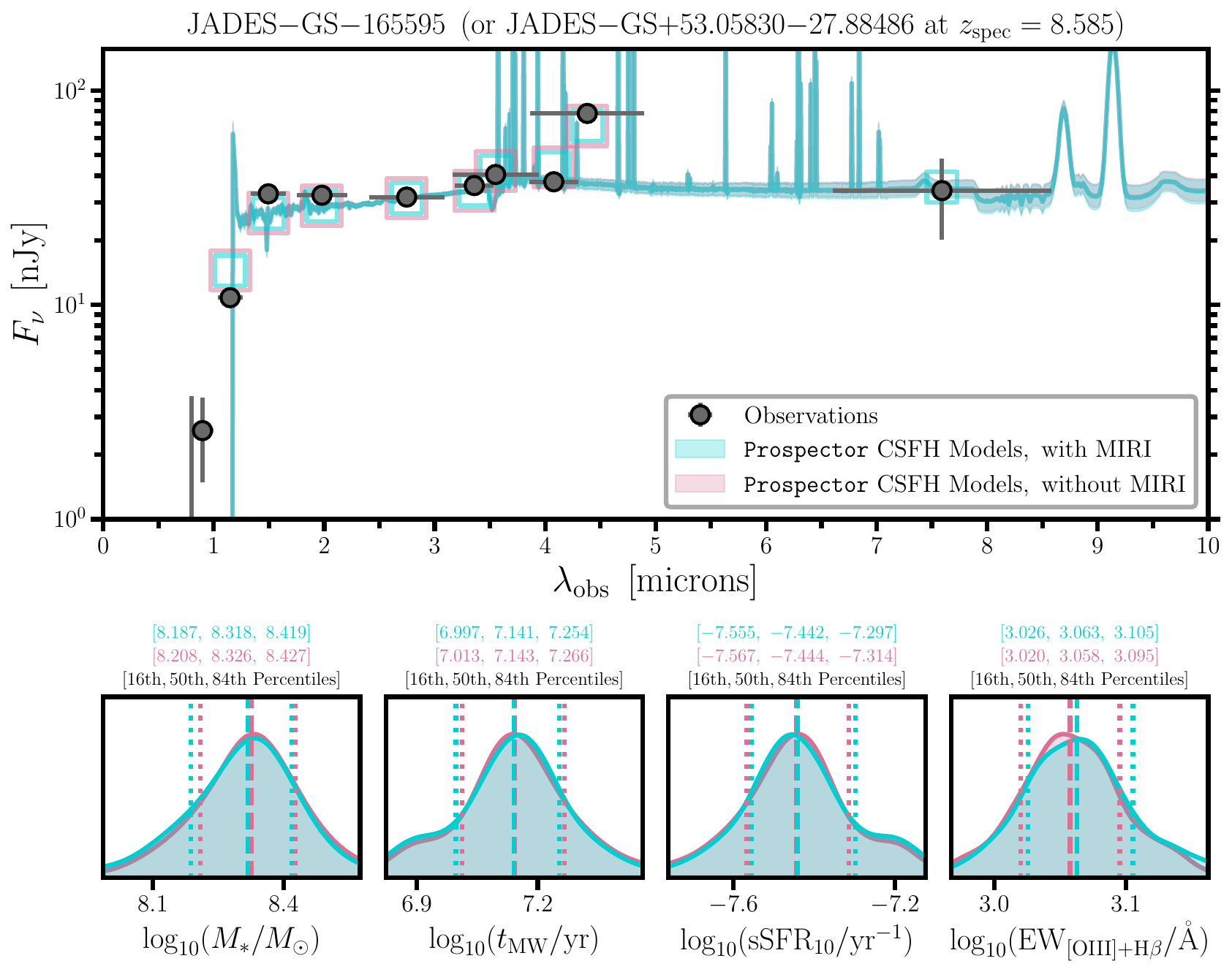}
    \caption{\textit{Top panel}: Example of an SED for a typical galaxy (JADES$-$GS$-$ID$-$$165595$, or JADES$-$GS$+53.05830$$-$$27.88486$, at $z_{\mathrm{spec}} = 8.585$) from the final photometric sample of MIRI/F770W detections described in Section~\ref{SectionThreeOne}. The grey points represent the observed photometry assuming the JADES filter set. The medians for the  \texttt{Prospector} model photometry and spectroscopy are given by the squares and solid lines, respectively. The $68\%$ confidence interval of the \texttt{Prospector} models is shown by the shaded regions. \textit{Bottom panels}: Example of posterior distributions for the most relevant stellar population properties. From left to right: stellar mass, mass-weighted stellar age, sSFR averaged over the last $10$ Myr, and rest-frame EW of $\mathrm{[OIII]}+\mathrm{H}\beta$. The dashed lines represent the medians for the \texttt{Prospector} models while the dotted lines represent $68\%$ confidence intervals. \textit{Top and bottom panels}: The fiducial \texttt{Prospector} model is described in Section~\ref{SectionFourTwo} and assumes the constant SFH model. The blue (pink) squares, lines, and shaded regions represent results from fitting to the full JADES filter set, including (excluding) MIRI/F770W. For this typical galaxy, MIRI/F770W has little to no effect on the inferred SED and stellar population properties. This is true for the vast majority ($\approx 80\%$) of galaxies in the final photometric sample, suggesting that observations with JWST/MIRI are not needed to robustly determine the stellar population properties for typical star-forming galaxies during the EoR. The complete figure set of these typical galaxies ($18$ figures) is available in the online journal. \label{figset:ExampleSEDs_1}}
\end{figure*}

\figsetstart
\figsetnum{6}
\graphicspath{{./}{files/ExampleSEDs/}}

\figsetgrpstart
\figsetgrpnum{6.1}
\figsetplot{SED_021468_ConstantPrior_JADES_fixedRedshift.pdf}
\figsetgrpnote{Similar to Figure~\ref{figset:ExampleSEDs_1}, but for an atypical galaxy (JADES$-$GS$-$ID$-$$21468$, or JADES$-$GS$+53.10108$$-$$27.88310$, at $z_{\mathrm{spec}} = 8.808$) from the final photometric sample of MIRI/F770W detections described in Section~\ref{SectionThreeOne}. For this atypical galaxy, MIRI/F770W has a significant effect on the inferred SED and stellar population properties, suggesting roughly two-thirds the amount of flux at $7.7$ microns than one would expect from fitting the JWST/NIRCam data alone (which can be interpreted as evidence for younger stellar populations and/or small amounts of dust). This is true for a small fraction ($\approx 10\%$) of galaxies in the final photometric sample, suggesting that observations with JWST/MIRI can reveal anomalously blue sources with interesting physical properties.}
\figsetgrpend

\figsetgrpstart
\figsetgrpnum{6.2}
\figsetplot{SED_066293_ConstantPrior_JADES_fixedRedshift.pdf}
\figsetgrpnote{Similar to Figure~\ref{figset:ExampleSEDs_1}, but for an atypical galaxy (JADES$-$GS$-$ID$-$$66293$, or JADES$-$GS$+53.04601$$-$$27.85399$, at $z_{\mathrm{spec}} = 8.065$) from the final photometric sample of MIRI/F770W detections described in Section~\ref{SectionThreeOne}. For this atypical galaxy, MIRI/F770W has a significant effect on the inferred SED and stellar population properties, suggesting roughly four times the amount of flux at $7.7$ microns than one would expect from fitting the JWST/NIRCam data alone (which can be interpreted as evidence for older stellar populations, large amounts of dust, and/or significant AGN contributions). This is true for a small fraction ($\approx 10\%$) of galaxies in the final photometric sample, suggesting that observations with JWST/MIRI can reveal anomalously red sources with interesting physical properties.}
\figsetgrpend

\figsetgrpstart
\figsetgrpnum{6.3}
\figsetplot{SED_174693_ConstantPrior_JADES_fixedRedshift.pdf}
\figsetgrpnote{Similar to Figure~\ref{figset:ExampleSEDs_1}, but for an atypical galaxy (JADES$-$GS$-$ID$-$$174693$, or JADES$-$GS$+53.06058$$-$$27.86795$, at $z_{\mathrm{spec}} = 7.882$) from the final photometric sample of MIRI/F770W detections described in Section~\ref{SectionThreeOne}. For this atypical galaxy, MIRI/F770W has a significant effect on the inferred SED and stellar population properties, suggesting roughly two-thirds the amount of flux at $7.7$ microns than one would expect from fitting the JWST/NIRCam data alone (which can be interpreted as evidence for younger stellar populations and/or small amounts of dust). This is true for a small fraction ($\approx 10\%$) of galaxies in the final photometric sample, suggesting that observations with JWST/MIRI can reveal anomalously blue sources with interesting physical properties.}
\figsetgrpend

\figsetgrpstart
\figsetgrpnum{6.4}
\figsetplot{SED_179485_ConstantPrior_JADES_fixedRedshift.pdf}
\figsetgrpnote{Similar to Figure~\ref{figset:ExampleSEDs_1}, but for an atypical galaxy (JADES$-$GS$-$ID$-$$179485$, or JADES$-$GS$+53.08738$$-$$27.86031$, at $z_{\mathrm{spec}} = 7.955$) from the final photometric sample of MIRI/F770W detections described in Section~\ref{SectionThreeOne}. For this atypical galaxy, MIRI/F770W has a significant effect on the inferred SED and stellar population properties, suggesting roughly double the amount of flux at $7.7$ microns than one would expect from fitting the JWST/NIRCam data alone (which can be interpreted as evidence for older stellar populations, large amounts of dust, and/or significant AGN contributions). This is true for a small fraction ($\approx 10\%$) of galaxies in the final photometric sample, suggesting that observations with JWST/MIRI can reveal anomalously red sources with interesting physical properties.}
\figsetgrpend

\figsetend

\begin{figure*}
    \centering
    \figurenum{6}
    \includegraphics[width=1.0\linewidth]{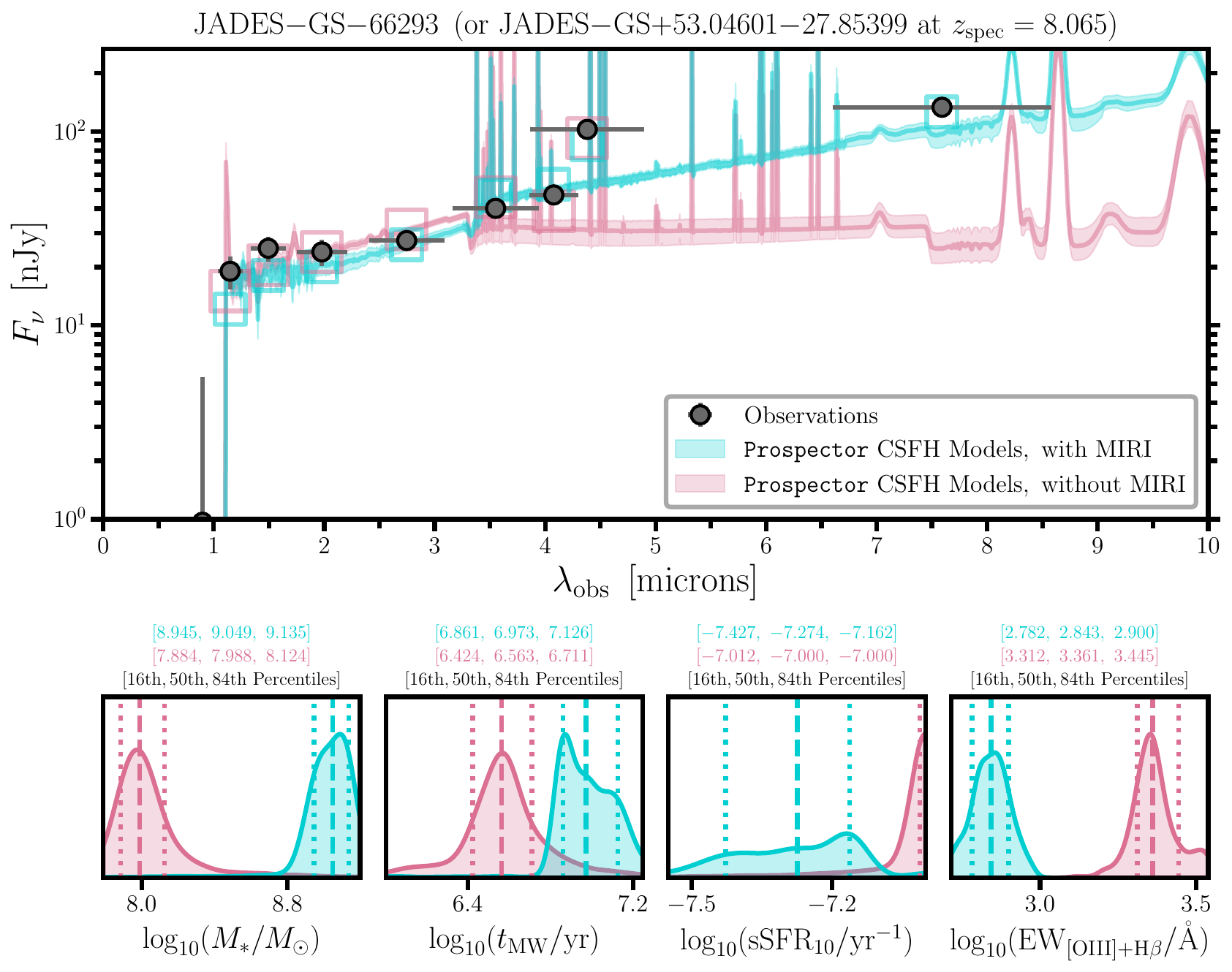}
    \caption{Similar to Figure~\ref{figset:ExampleSEDs_1}, but for an atypical galaxy (JADES$-$GS$-$ID$-$$66293$, or JADES$-$GS$+53.04601$$-$$27.85399$, at $z_{\mathrm{spec}} = 8.065$) from the final photometric sample of MIRI/F770W detections described in Section~\ref{SectionThreeOne}. For this atypical galaxy, MIRI/F770W has a significant effect on the inferred SED and stellar population properties, suggesting roughly four times the amount of flux at $7.7$ microns than one would expect from fitting the JWST/NIRCam data alone (which can be interpreted as evidence for older stellar populations, large amounts of dust, and/or significant AGN contributions). This is true for a small fraction ($\approx 10\%$) of galaxies in the final photometric sample, suggesting that observations with JWST/MIRI can reveal anomalously red sources with interesting physical properties. The complete figure set of these atypical galaxies (four figures) is available in the online journal, which also includes two anomalously blue sources. \label{figset:ExampleSEDs_2}}
\end{figure*}

\setcounter{figure}{6}

\graphicspath{{./}{files/}}

Figure~\ref{figset:ExampleSEDs_1} illustrates an example of an SED for one of the typical galaxies from the final photometric sample (JADES$-$GS$-$ID$-$$165595$ at $z_{\mathrm{spec}} = 8.585$). This object has a rest-UV absolute magnitude $M_{\mathrm{UV}} \approx -19.7$ and a rest-UV continuum slope $\beta_{\mathrm{UV}} \approx -2.1$. In the upper panel, the observed photometry is represented by the gray points, assuming the JADES filter set. The medians of the photometry and spectroscopy from the \texttt{Prospector} models are given by the squares in the upper panel and the solid lines in the lower panels, assuming the fiducial CSFH model. The $68\%$ confidence interval is shown by the shaded regions in the upper panel but by the dotted lines in the lower panels. 

In the lower panels of Figure~\ref{figset:ExampleSEDs_1}, there are examples of posterior distributions for some of the most relevant stellar population properties. From left to right, these properties are: stellar mass ($M_{\ast}$), mass-weighted stellar age ($t_{\mathrm{MW}}$), specific star formation rate ($\mathrm{sSFR}_{10}$) averaged over the last $10$ Myr of lookback time, and rest-frame equivalent width ($\mathrm{EW}_{\mathrm{[OIII]}+\mathrm{H}\beta}$) of the rest-optical emission lines $\mathrm{[OIII]}+\mathrm{H}\beta$. The medians of these posterior distributions are given by the dashed lines. The median derived stellar mass $M_{\ast} \approx 10^{8.3}\ M_{\odot}$ is roughly one-hundredth the characteristic mass for the stellar mass function at $z \approx 8$ \citep[e.g.,][]{Navarro-Carrera:2024} while the median derived star formation rate $\mathrm{SFR} \approx 10\ M_{\odot}/\mathrm{yr}$ is consistent with expectations from the star-forming main sequence observed at these redshifts \citep[e.g.,][]{Popesso:2023}. Furthermore, the median derived stellar age $t_{\mathrm{MW}} \approx 10\ \mathrm{Myr}$ and the median derived rest-frame optical equivalent width $\mathrm{EW}_{\mathrm{[OIII]}+\mathrm{H}\beta} \approx 10^{3}\ \mathrm{\mathring{A}}$ are consistent with the values measured for some of the youngest galaxies observed at $z \approx 8$ \citep[e.g.,][]{Endsley:2024}. These median derived quantities are representative of the typical values that we derive for each of the $N = 22$ galaxies that are part of the final photometric sample of MIRI/F770W detections.

In both the upper and lower panels, the results from the \texttt{Prospector} models assuming the full JADES filter set (including MIRI/F770W) are represented by the blue squares, lines, and shaded regions. Likewise, those results assuming the partial JADES filter set (excluding MIRI/F770W) are represented by the pink squares, lines, and shaded regions. It is evident from Figure~\ref{figset:ExampleSEDs_1} that for this typical galaxy, the inclusion of MIRI/F770W has little to no observable effect on the inferred SED and stellar population properties since the medians and $68\%$ confidence intervals of the \texttt{Prospector} models are entirely overlapping. This result is true for the vast majority ($\approx 80\%$) of galaxies in the final photometric sample ($N = 18$). This seems to suggest that observations with JWST/MIRI are not needed to robustly determine the stellar population properties for typical star-forming galaxies during the EoR (i.e., star-forming galaxies at $z \approx 8$ with $M_{\mathrm{UV}} \approx -20$ and $\beta_{\mathrm{UV}} \approx -2$).

Similar to Figure~\ref{figset:ExampleSEDs_1}, Figure~\ref{figset:ExampleSEDs_2} illustrates an example of an atypical SED for one of the galaxies from the final photometric sample (JADES$-$GS$-$ID$-$$66293$). This object has a rest-UV absolute magnitude $M_{\mathrm{UV}} \approx -19.2$ and a rest-UV continuum slope $\beta_{\mathrm{UV}} \approx -1.8$. We are defining atypical galaxies as those that have an observed MIRI/F770W flux which deviates by more than $1\sigma$ when compared to the predicted value from the \texttt{Prospector} fits to the JADES filter set (excluding MIRI/F770W). Unlike the example of the typical galaxy in Figure~\ref{figset:ExampleSEDs_1}, the inclusion of MIRI/F770W has a significant effect on the inferred SED and stellar population properties for this source. This result is true for a small fraction ($\approx 20\%$) of galaxies in the final photometric sample ($N = 4$). This seems to suggest that observations with JWST/MIRI can reveal anomalously red sources with interesting properties during the EoR. One of the other three examples of galaxies in which MIRI/F770W has a significant impact on the inferred properties is quite similar to the example atypical galaxy JADES$-$GS$-$ID$-$$66293$ shown in Figure~\ref{figset:ExampleSEDs_2}. This galaxy is JADES$-$GS$-$ID$-$$179485$ with a rest-UV absolute magnitude $M_{\mathrm{UV}} \approx -20.2$ and a rest-UV continuum slope $\beta_{\mathrm{UV}} \approx -1.9$. These are some of the reddest galaxies in the final photometric sample.

The other two galaxies where the inclusion of the MIRI/F770W data point has a significant effect on the inferred properties are JADES$-$GS$-$ID$-$$21468$ and JADES$-$GS$-$ID$-$$174693$. These two galaxies are similar to one another, with rest-UV absolute magnitudes and continuum slopes of $M_{\mathrm{UV}} \approx -19.5$ and $\beta_{\mathrm{UV}} \approx -2.3$. They are some of the bluest galaxies in the final photometric sample of MIRI/F770W detections described in Section~\ref{SectionThreeOne}, much bluer in the rest-UV than the previous two atypical galaxies. Unlike those previous two galaxies, the observed MIRI/F770W photometry for JADES$-$GS$-$ID$-$$21468$ and JADES$-$GS$-$ID$-$$174693$ is fainter than the predicted value from fitting the HST/ACS and JWST/NIRCam photometry alone. This seems to suggest that observations with JWST/MIRI can also reveal anomalously blue sources during the EoR, similar to the results found in \citet{Papovich:2023} at $z = 4-9$, where they fit with a base set of seven filters from HST/ACS, HST/WFC3, and Spitzer/IRAC (plus MIRI/F560W and MIRI/F770W).

It is interesting to note that two of the bluest galaxies in our sample (which have rest-UV continuum slopes $\beta_{\mathrm{UV}} \approx -2.3$) are also the two galaxies where the observed MIRI/F770W photometry is marginally fainter than the predicted value from fitting the HST/ACS and JWST/NIRCam photometry alone. On the other hand, two of the reddest galaxies in our sample (which have rest-UV continuum slopes $\beta_{\mathrm{UV}} \approx -1.8$) are also the two galaxies in which the observed MIRI/F770W photometry is significantly brighter than the predicted value. Future population studies with larger samples of high-redshift star-forming galaxies that have JWST/MIRI detections will be necessary to determine whether it is typical for the bluest and reddest objects in the rest-UV to exhibit this sort of behavior at $z \approx 8$.


\section{Results \& Discussion}
\label{SectionFive}

\begin{figure*}
    \centering
    \includegraphics[width=1.0\linewidth]{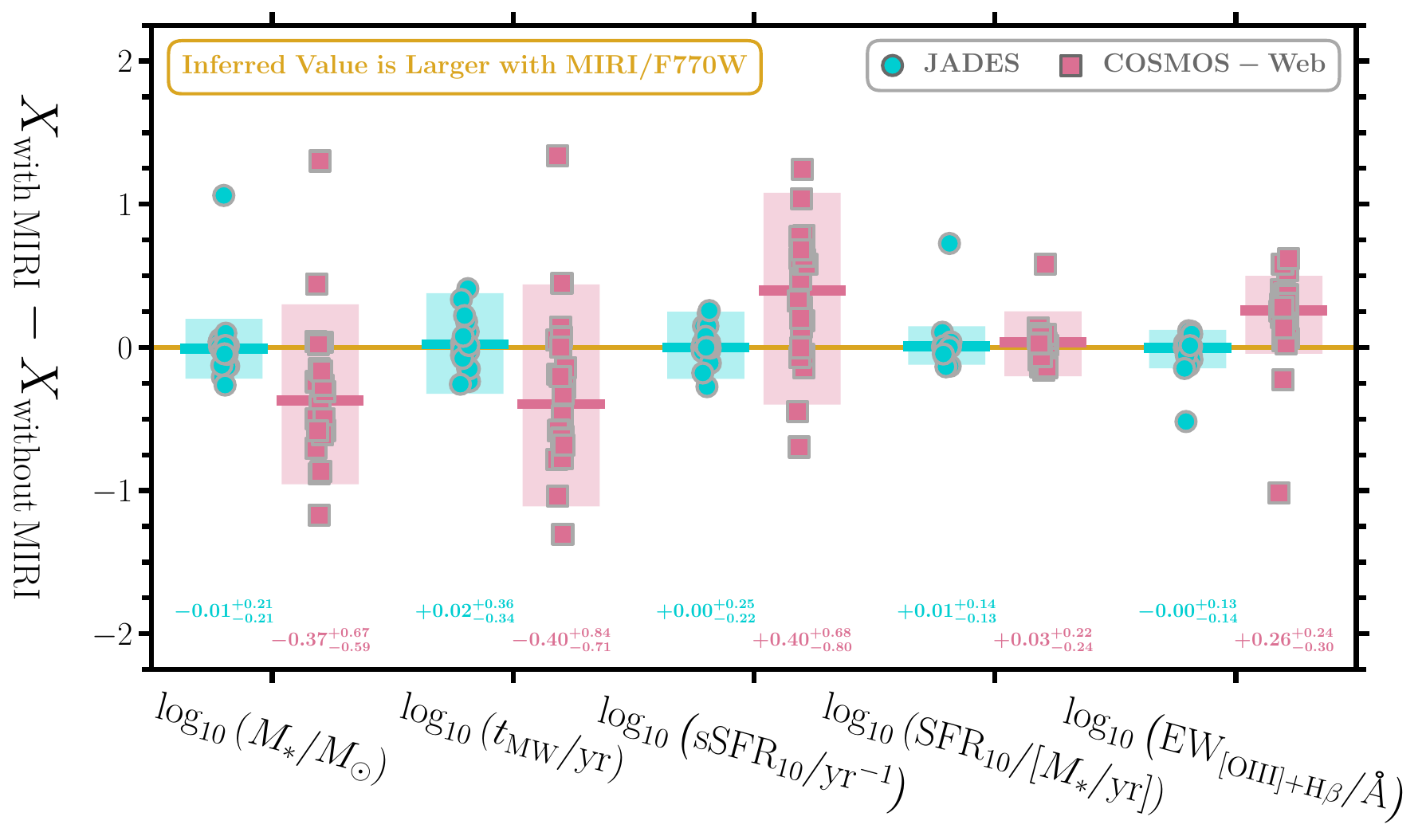}
    \caption{Comparison of derived stellar population properties for galaxies from the final photometric sample of MIRI/F770W detections described in Section~\ref{SectionThreeOne}. Values are reported as the difference between derived properties when including ($X_{\mathrm{with\,MIRI}}$) and excluding ($X_{\mathrm{without\,MIRI}}$) MIRI/F770W, such that positive (negative) values imply larger (smaller) inferred values when including this data point. These are derived using the fiducial \texttt{Prospector} model as described in Section~\ref{SectionFourOne} and assume the constant SFH model. The median values for each of the galaxies are given by the circles or squares. The median and 68\% confidence interval of the full sample are shown by the lines and shaded regions. These values for the full sample are also reported at the bottom of the figure. The blue circles (pink squares), lines, and shaded regions represent results from fitting to the JADES (COSMOS-Web) filter set. It is evident that the inclusion of MIRI/F770W has a smaller impact on the derived properties when using the JADES filter set (not systematically offset and smaller scatter) compared to COSMOS-Web (systematically offset and larger scatter). This suggests that observations with JWST/MIRI are not needed to robustly determine the stellar population properties for typical star-forming galaxies during the EoR for filter sets with sufficient JWST/NIRCam photometric coverage. \label{fig:Summary_StellarPopulations}}
\end{figure*}

\begin{figure*}
    \centering
    \includegraphics[width=1.0\linewidth]{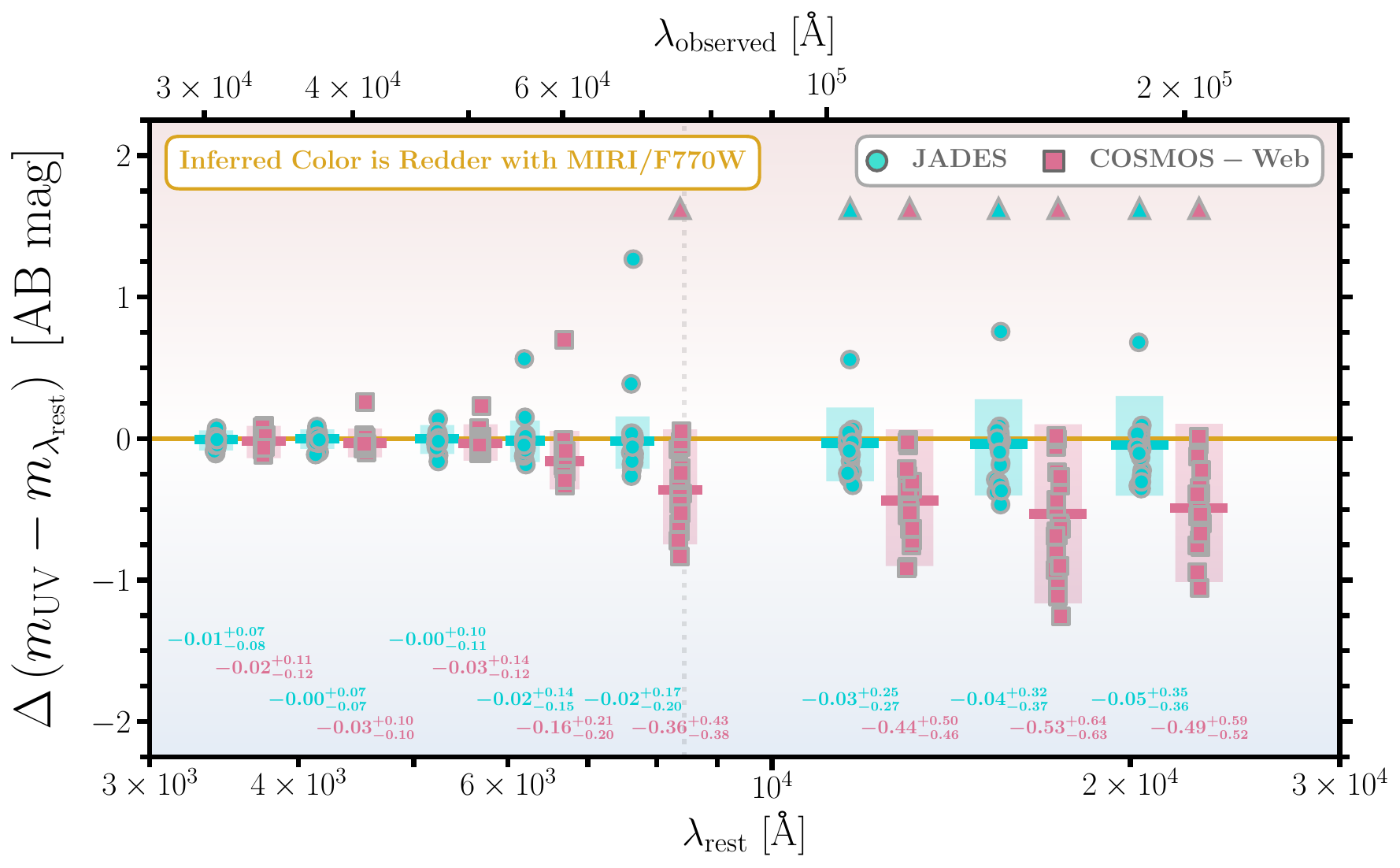}
    \caption{Similar to Figure~\ref{fig:Summary_StellarPopulations}, but comparing the derived rest-frame colors for galaxies from the final photometric sample of MIRI/F770W detections described in Section~\ref{SectionThreeOne}. Colors are defined relative to the measured apparent magnitude at rest-frame $1500\ \mathrm{\AA}$. Values are reported as the difference between derived properties when including and excluding MIRI/F770W. Positive (negative) values imply redder (bluer) inferred colors when including this data point. Triangles represent limits for points that fall outside the plotted range of values. From left to right: $U$-band, $B$-band, $V$-band, $R$-band, $I$-band, $J$-band, $H$-band, and $K$-band. The faint, dotted gray line indicates the effective wavelength of MIRI/F770W. Observed wavelengths are reported at $z = 8$. The inclusion of MIRI/F770W has a smaller impact on the derived colors when using the JADES filter set compared to COSMOS-Web, suggesting that JWST/MIRI is not needed to robustly determine the rest-frame colors out to $\lambda_{\mathrm{rest}} \approx 2\ \mu\mathrm{m}$ for typical star-forming galaxies during the EoR for filter sets with sufficient JWST/NIRCam photometric coverage. \label{fig:Summary_RestFrameColors}}
\end{figure*}

Using the physical properties inferred in Section~\ref{SectionFour}, we perform comparisons of the high-redshift star-forming galaxies within our final photometric sample to explore the impact of filter choice and SFH assumption. Comparisons of the inferred stellar population properties are presented in Section~\ref{SectionFiveOne} while the inferred rest-frame colors are presented in Section~\ref{SectionFiveTwo}. These comparisons are made with four different filter sets and four different assumed SFHs (see Section~\ref{SectionFour}). The four filter sets are JOF ($N_{\mathrm{filter}} = 16$), JADES ($N_{\mathrm{filter}} = 11$), CEERS ($N_{\mathrm{filter}} = 9$), and COSMOS-Web ($N_{\mathrm{filter}} = 6$). The four assumed SFHs are parametric constant (CSFH), parametric delayed-tau (DtSFH), non-parametric ``continuity'' (ContSFH), and non-parametric ``bursty continuity'' (BurstySFH). We should emphasize that we do not simulate or explore the impact of differences in observational depth for the different filter sets. This is an important point to consider in relation to the extreme depth of the observations used here ($5\sigma$ limiting magnitudes of $m \approx 28.1\ \mathrm{AB\ mag}$ for MIRI/F770W, assuming circular apertures with diameters of $0.7^{\prime\prime}$). Thus, we are only changing the adopted set of filters, while using the same photometric measurements in each case. Succinct summaries of our results are provided in Tables~\ref{tab:StellarPopulations} and \ref{tab:RestFrameColors} (see also Figures~\ref{fig:Summary_StellarPopulations} and \ref{fig:Summary_RestFrameColors}). Our results demonstrate that the JOF, JADES, and CEERS filter sets perform significantly better than COSMOS-Web in the absence of MIRI/F770W data. Similarly, the parametric DtSFH model performs better than the parametric CSFH model in addition to both of the non-parametric models (ContSFH and BurstySFH). Finally, spectroscopic redshifts decrease the scatter in the derived properties for the final photometric sample of detections.

\subsection{Comparison of the Stellar Populations}
\label{SectionFiveOne}

In Figure~\ref{fig:Summary_StellarPopulations}, the median values for each of the galaxies are given by the circles or squares. After accounting for the uncertainties on the values for individual galaxies, the median and 68\% confidence interval of the full sample are shown by the lines and shaded regions. Additionally, the medians and 68\% confidence intervals of the full sample are reported at the bottom of the figure, while these quantities are presented for the four different filter sets and the four different assumed SFHs in Table~\ref{tab:StellarPopulations}. From left to right, the derived stellar population properties are: stellar mass ($M_{\ast}$), mass-weighted stellar age ($t_{\mathrm{MW}}$), specific star formation rate ($\mathrm{sSFR}_{10}$) averaged over the last $10$ Myr of lookback time, star formation rate ($\mathrm{SFR}_{10}$) averaged over the same timescale, and rest-frame equivalent width ($\mathrm{EW}_{\mathrm{[OIII]}+\mathrm{H}\beta}$) of the rest-optical emission lines $\mathrm{[OIII]}+\mathrm{H}\beta$.

\begin{table*}
	\caption{A comparison of the derived stellar population properties, as described in Section~\ref{SectionFiveOne}. Values are reported as the difference between derived properties when including and excluding the MIRI/F770W data point. Thus, positive values imply larger inferred values when including this data point, while negative values imply smaller inferred values. The median and 68\% confidence interval of the full sample are reported, after accounting for the uncertainties on the values for individual galaxies.}
	\label{tab:StellarPopulations}
	\makebox[\textwidth]{
        \hspace{+1pt}
        \begin{tabular}{c|c||c|c|c|c|c}
		\hline
		\hline
            & & \multicolumn{5}{c}{Derived Stellar Population Property} \\
		\hline
		& Filter Set & $\mathrm{log}_{10} ( M_{\ast}/M_{\odot} )$ & $\mathrm{log}_{10} ( t_\mathrm{MW}/\mathrm{yr} )$ & $\mathrm{log}_{10} ( \mathrm{SFR}_{10}/[M_{\ast}/\mathrm{yr}] )$ & $\mathrm{log}_{10} ( \mathrm{sSFR}_{10}/\mathrm{yr}^{-1} )$ & $\mathrm{log}_{10} ( \mathrm{EW}_{\mathrm{[OIII]}+\mathrm{H}\beta}/\mathrm{\AA} )$ \\
		\hline
            \multirow{4}{*}{\rotatebox{90}{CSFH }} & COSMOS-Web & $-0.37^{+0.67}_{-0.59}$ & $-0.40^{+0.84}_{-0.71}$ & $+0.40^{+0.68}_{-0.80}$ & $+0.03^{+0.22}_{-0.24}$ & $+0.26^{+0.24}_{-0.30}$ \\
            & CEERS & $-0.02^{+0.23}_{-0.22}$ & $+0.02^{+0.38}_{-0.37}$ & $+0.00^{+0.26}_{-0.22}$ & $+0.00^{+0.15}_{-0.14}$ & $+0.01^{+0.13}_{-0.14}$ \\
            & JADES & $-0.01^{+0.21}_{-0.21}$ & $+0.02^{+0.36}_{-0.34}$ & $+0.00^{+0.25}_{-0.22}$ & $+0.01^{+0.14}_{-0.13}$ & $-0.00^{+0.13}_{-0.14}$ \\
            & JOF & $+0.01^{+0.20}_{-0.21}$ & $+0.04^{+0.40}_{-0.33}$ & $+0.00^{+0.21}_{-0.24}$ & $+0.01^{+0.14}_{-0.12}$ & $-0.01^{+0.12}_{-0.14}$ \\
            \hline
            \multirow{4}{*}{\rotatebox{90}{DtSFH }} & COSMOS-Web & $-0.34^{+0.53}_{-0.47}$ & $-0.29^{+0.66}_{-0.60}$ & $+0.28^{+0.56}_{-0.59}$ & $-0.00^{+0.20}_{-0.21}$ & $+0.24^{+0.21}_{-0.27}$ \\
            & CEERS & $+0.00^{+0.22}_{-0.21}$ & $-0.00^{+0.46}_{-0.43}$ & $+0.00^{+0.25}_{-0.25}$ & $+0.01^{+0.15}_{-0.12}$ & $+0.02^{+0.12}_{-0.13}$ \\
            & JADES & $-0.02^{+0.21}_{-0.19}$ & $-0.04^{+0.36}_{-0.37}$ & $+0.01^{+0.26}_{-0.23}$ & $+0.00^{+0.14}_{-0.13}$ & $+0.02^{+0.12}_{-0.13}$ \\
            & JOF & $+0.01^{+0.21}_{-0.21}$ & $+0.01^{+0.39}_{-0.34}$ & $+0.00^{+0.23}_{-0.24}$ & $+0.01^{+0.13}_{-0.12}$ & $+0.01^{+0.11}_{-0.14}$ \\
            \hline
            \multirow{4}{*}{\rotatebox{90}{ContSFH }} & COSMOS-Web & $-0.22^{+0.43}_{-0.57}$ & $-0.03^{+0.14}_{-0.20}$ & $+0.46^{+0.83}_{-1.09}$ & $+0.16^{+0.52}_{-0.67}$ & $+0.40^{+0.61}_{-0.70}$ \\
            & CEERS & $+0.02^{+0.46}_{-0.40}$ & $-0.01^{+0.27}_{-0.32}$ & $-0.06^{+0.51}_{-0.54}$ & $-0.00^{+0.32}_{-0.32}$ & $-0.01^{+0.18}_{-0.27}$ \\
            & JADES & $+0.09^{+0.34}_{-0.38}$ & $+0.00^{+0.62}_{-0.32}$ & $-0.07^{+0.49}_{-0.44}$ & $+0.01^{+0.27}_{-0.24}$ & $-0.03^{+0.19}_{-0.16}$ \\
            & JOF & $-0.05^{+0.52}_{-0.40}$ & $-0.00^{+0.25}_{-0.29}$ & $+0.11^{+0.45}_{-0.43}$ & $+0.08^{+0.30}_{-0.20}$ & $+0.02^{+0.21}_{-0.22}$ \\
            \hline
            \multirow{4}{*}{\rotatebox{90}{BurstySFH }} & COSMOS-Web & $-0.28^{+0.62}_{-1.06}$ & $+0.05^{+0.44}_{-0.80}$ & $+0.56^{+1.81}_{-1.57}$ & $+0.16^{+1.22}_{-1.16}$ & $+0.61^{+1.93}_{-1.18}$ \\
            & CEERS & $+0.03^{+0.32}_{-0.31}$ & $+0.02^{+0.68}_{-0.69}$ & $-0.03^{+0.41}_{-0.36}$ & $+0.03^{+0.30}_{-0.27}$ & $-0.02^{+0.16}_{-0.20}$ \\
            & JADES & $-0.07^{+0.27}_{-0.42}$ & $-0.12^{+0.69}_{-0.68}$ & $+0.10^{+0.45}_{-0.39}$ & $+0.01^{+0.30}_{-0.34}$ & $+0.03^{+0.21}_{-0.22}$ \\
            & JOF & $-0.04^{+0.59}_{-0.41}$ & $-0.12^{+0.71}_{-0.78}$ & $+0.07^{+0.55}_{-0.60}$ & $+0.04^{+0.31}_{-0.21}$ & $+0.02^{+0.20}_{-0.32}$ \\
            \hline
	\end{tabular}
        }
\end{table*}

The results from the \texttt{Prospector} models assuming the JADES filter set are represented by the blue circles, lines, and shaded regions. Those results assuming the COSMOS-Web filter set are represented by the pink squares, lines, and shaded regions. When assuming the JADES filter set, it is evident from Figure~\ref{fig:Summary_StellarPopulations} that the inclusion of MIRI/F770W has little to no observable effect on all of the inferred stellar population properties for the vast majority ($\approx 80\%$) of galaxies in the final photometric sample ($N = 18$). These typical galaxies have stellar masses that are offset by $\Delta M_{\ast} \lesssim 0.2\ \mathrm{dex}$, mass-weighted stellar ages by $\Delta t_{\mathrm{MW}} \lesssim 0.3\ \mathrm{dex}$, specific star formation rates by $\Delta \mathrm{sSFR}_{10} \lesssim 0.2\ \mathrm{dex}$, star formation rates by $\Delta \mathrm{SFR}_{10} \lesssim 0.1\ \mathrm{dex}$, and rest-optical equivalent widths by $\Delta \mathrm{EW} \lesssim 0.3\ \mathrm{dex}$. For these galaxies, the mass-weighted stellar age is the property that is subject to the most scatter, while the star formation rate and rest-frame equivalent widths are the properties that are subject to the least scatter.

The remaining galaxies ($N = 4$), which represent the minority ($\approx 20\%$) of the final photometric sample, are those where the inclusion of MIRI/F770W has a significant effect on the inferred SED and/or stellar population properties. Two of these atypical galaxies are anomalously red and characterized by relatively red rest-UV continuum slopes ($\beta_{\mathrm{UV}} \approx -1.8$), as discussed in Section~\ref{SectionFourThree}. These red galaxies have stellar masses that are increased by $\Delta M_{\ast} \approx 0.1-1.1\ \mathrm{dex}$, mass-weighted stellar ages that are increased by $\Delta t_{\mathrm{MW}} \approx 0.0-0.4\ \mathrm{dex}$, specific star formation rates that are decreased by $\Delta \mathrm{sSFR}_{10} \approx 0.0-0.4\ \mathrm{dex}$, star formation rates that are increased by $\Delta \mathrm{SFR}_{10} \approx 0.1-1.1\ \mathrm{dex}$, and rest-optical equivalent widths that are decreased by $\Delta \mathrm{EW} \approx 0.1-0.6\ \mathrm{dex}$.

The other two atypical galaxies are anomalously blue and characterized by relatively blue rest-UV continuum slopes ($\beta_{\mathrm{UV}} \approx -2.3$), as discussed in Section~\ref{SectionFourThree}. These blue galaxies have stellar masses that are decreased by $\Delta M_{\ast} \approx 0.1-0.2\ \mathrm{dex}$, mass-weighted stellar ages that are decreased by $\Delta t_{\mathrm{MW}} \approx 0.1-0.2\ \mathrm{dex}$, specific star formation rates that are increased by $\Delta \mathrm{sSFR}_{10} \approx 0.1-0.2\ \mathrm{dex}$, star formation rates that are increased by $\Delta \mathrm{SFR}_{10} \approx 0.1-0.2\ \mathrm{dex}$, and rest-optical equivalent widths that are increased by $\Delta \mathrm{EW} \approx 0.1\ \mathrm{dex}$.

However, when assuming the COSMOS-Web filter set, it is evident from Figure~\ref{fig:Summary_StellarPopulations} and Table~\ref{tab:StellarPopulations} that the inclusion of MIRI/F770W has a significant effect on the inferred stellar population properties for the vast majority of galaxies in the final photometric sample. Once again, the mass-weighted stellar age is the property that is subject to the most scatter, as shown by the confidence intervals, while the star formation rate is the property that is subject to the least scatter. The stellar masses and mass-weighted stellar ages are typically smaller with MIRI/F770W, corresponding to median offsets of $-0.37_{-0.59}^{+0.67}\ \mathrm{dex}$ and $-0.40_{-0.71}^{+0.84}\ \mathrm{dex}$, respectively. The specific star formation rates and rest-optical equivalent widths are typically larger with MIRI/F770W, corresponding to median offsets of $+0.40_{-0.80}^{+0.68}\ \mathrm{dex}$ and $+0.26_{-0.30}^{+0.24}\ \mathrm{dex}$, respectively.  Finally, the star formation rates are roughly the same with MIRI/F770W, corresponding to a median offset of $+0.03_{-0.24}^{+0.22}\ \mathrm{dex}$. These results suggest that galaxies are modeled as typically less massive and younger, with larger specific star formation rates and rest-optical equivalent widths, when including the MIRI/F770W data point and assuming the COSMOS-Web filter set. For comparison, when assuming the JADES filter set, the median offsets are: $-0.01_{-0.21}^{+0.21}\ \mathrm{dex}$ for stellar masses, $-0.02_{-0.34}^{+0.36}\ \mathrm{dex}$ for stellar ages, $+0.00_{-0.22}^{+0.25}\ \mathrm{dex}$ for specific star formation rates, $+0.01_{-0.13}^{+0.14}\ \mathrm{dex}$ for star formation rates, and $-0.00_{-0.14}^{+0.13}\ \mathrm{dex}$ for rest-optical equivalent widths.

Taken together, the results provided in Figure~\ref{fig:Summary_StellarPopulations} and Table~\ref{tab:StellarPopulations} demonstrate that observations with JWST/MIRI are not needed to robustly determine the stellar population properties for typical star-forming galaxies at $z \approx 8$, so long as the filter set has sufficient photometric coverage with JWST/NIRCam (e.g., CEERS, JADES, and JOF as described in Table~\ref{tab:FilterSetTable}). This statement is true for the vast majority ($\approx 80\%$) of galaxies in the final photometric sample. In this case, ``sufficient'' photometric coverage refers to $8-14$ filters with JWST/NIRCam, including $1-7$ medium-bands. However, in the absence of sufficient photometric coverage with JWST/NIRCam (e.g., COSMOS-Web), observations with JWST/MIRI are needed to robustly determine the stellar population properties for typical star-forming galaxies at $z \approx 8$, consistent with the results presented in \citet{Papovich:2023}.

The primary reason that including MIRI/F770W has a significant impact on the derived properties of typical star-forming galaxies at $z \approx 8$ when using the COSMOS-Web filter set, but not when using CEERS, JADES, and JOF, is that a limited filter set is unable to provide a robust measurement of the rest-frame optical continuum. This occurs because the only JWST/NIRCam filter in COSMOS-Web at rest-frame optical wavelengths (i.e., F444W) is a wide-band filter contaminated with strong emission lines (i.e., $\mathrm{H}\beta$ and $\mathrm{[OIII]}\lambda\lambda 4960{,}5008$) at these redshifts. Since it is outside the scope of this work, we do not test for the unverified potential of other filter combinations in producing comparable results.

We speculate that adding any filter to the COSMOS-Web filter set which is capable of distinguishing between contributions from continuum and emission lines should be sufficient to significantly improve constraints on their relative contributions. This additional filter could be any of the JWST/MIRI filters that are free of strong emission lines (e.g., F770W, or any filter except for F560W, which contains $\mathrm{H}\alpha$ at $z \approx 8$). Alternatively, this filter could be any of the JWST/NIRCam medium-band filters at $\lambda_{\mathrm{obs}} \approx 4-5\ \mu\mathrm{m}$ (e.g., F410M) as demonstrated by the CEERS, JADES, and JOF filter sets. However, the JWST/MIRI filters have some advantages when compared to JWST/NIRCam. Most notably, the longer wavelength coverage of JWST/MIRI provides more sensitivity to older stellar populations, diffuse dust attenuation, and contributions from an AGN. For the vast majority ($\approx 90\%$) of our sample, JWST/MIRI does not reveal old stars, dust, or an AGN. However, we argue that this is expected given we are studying a sample of relatively young, star-forming galaxies at $z \approx 8$.

Another interesting, and potentially surprising, result is that the medium-band filters at $\lambda_{\mathrm{obs}} \approx 1.5-3.5\ \mu\mathrm{m}$ included in the JOF filter set offer little improvement in determining the stellar population properties when compared to JADES and CEERS. For $z \approx 8$ \mbox{galaxies}, the medium-band filters at $\lambda_{\mathrm{obs}} \lesssim 3\ \mu\mathrm{m}$ do not contain much information beyond what is provided by the wide-band filters at these wavelengths (i.e., the rest-frame UV continuum shape). While the filters at $\lambda_{\mathrm{obs}} \approx 3.0-3.5\ \mu\mathrm{m}$ provide information about the strength of the Balmer continuum break, this feature is expected to be weak for the relatively young, star-forming galaxies within the final photometric sample of MIRI/F770W detections.

\subsection{Comparison of the Rest-Frame Colors}
\label{SectionFiveTwo}

Similar to Figure~\ref{fig:Summary_StellarPopulations}, Figure~\ref{fig:Summary_RestFrameColors} provides a comparison of the derived rest-frame colors for galaxies from the final photometric sample of MIRI/F770W detections described in Section~\ref{SectionThreeOne}. The reported colors are defined relative to the measured apparent magnitude at rest-frame $1500\ \mathrm{\AA}$, which corresponds to the rest-UV. The reported values are the difference between derived properties from SED fitting that includes and excludes the MIRI/F770W data point. This means that positive values imply redder inferred colors when including this additional data point, while negative values imply bluer inferred colors. These are derived using the fiducial \texttt{Prospector} model as described in Section~\ref{SectionFourOne} and assume the parametric constant SFH model.

\begin{table*}
	\caption{A comparison of the derived rest-frame colors, as described in Section~\ref{SectionFiveTwo}. Values are reported as the difference between derived properties when including and excluding the MIRI/F770W data point. Thus, positive values imply larger inferred values when including this data point, while negative values imply smaller inferred values. The median and 68\% confidence interval of the full sample are reported, after accounting for the uncertainties on the values for individual galaxies.}
	\label{tab:RestFrameColors}
	\makebox[\textwidth]{
        \hspace{+1pt}
        \begin{tabular}{c|c||c|c|c|c|c|c|c|c}
		\hline
		\hline
            & & \multicolumn{8}{c}{Derived Rest-Frame Color} \\
		\hline
		& Filter Set & $\mathrm{UV} - U$ & $\mathrm{UV} - B$ & $\mathrm{UV} - V$ & $\mathrm{UV} - R$ & $\mathrm{UV} - I$ & $\mathrm{UV} - J$ & $\mathrm{UV} - H$ & $\mathrm{UV} - K$ \\
		\hline
            \multirow{4}{*}{\rotatebox{90}{CSFH }} & COSMOS-Web & $-0.02^{+0.11}_{-0.12}$ & $-0.03^{+0.10}_{-0.10}$ & $-0.03^{+0.14}_{-0.12}$ & $-0.16^{+0.21}_{-0.20}$ & $-0.36^{+0.43}_{-0.38}$ &$-0.44^{+0.50}_{-0.46}$ & $-0.53^{+0.64}_{-0.63}$ & $-0.49^{+0.59}_{-0.52}$ \\
            & CEERS & $-0.01^{+0.07}_{-0.09}$ & $+0.00^{+0.07}_{-0.07}$ & $+0.01^{+0.11}_{-0.10}$ & $-0.02^{+0.17}_{-0.18}$ & $-0.02^{+0.17}_{-0.18}$ &$-0.04^{+0.26}_{-0.26}$ & $-0.04^{+0.32}_{-0.33}$ & $-0.05^{+0.38}_{-0.37}$ \\
            & JADES & $-0.01^{+0.07}_{-0.08}$ & $-0.00^{+0.07}_{-0.07}$ & $-0.00^{+0.10}_{-0.11}$ & $-0.02^{+0.14}_{-0.15}$ & $-0.02^{+0.17}_{-0.20}$ &$-0.03^{+0.25}_{-0.27}$ & $-0.04^{+0.32}_{-0.37}$ & $-0.05^{+0.35}_{-0.36}$ \\
            & JOF & $-0.01^{+0.06}_{-0.09}$ & $+0.00^{+0.06}_{-0.07}$ & $+0.01^{+0.09}_{-0.10}$ & $-0.01^{+0.14}_{-0.18}$ & $+0.00^{+0.17}_{-0.18}$ &$-0.01^{+0.24}_{-0.26}$ & $+0.01^{+0.30}_{-0.34}$ & $-0.01^{+0.35}_{-0.36}$ \\
            \hline
            \multirow{4}{*}{\rotatebox{90}{DtSFH }} & COSMOS-Web & $+0.00^{+0.11}_{-0.11}$ & $-0.01^{+0.11}_{-0.10}$ & $-0.01^{+0.13}_{-0.12}$ & $-0.12^{+0.23}_{-0.21}$ & $-0.33^{+0.41}_{-0.38}$ &$-0.38^{+0.50}_{-0.47}$ & $-0.52^{+0.66}_{-0.62}$ & $-0.43^{+0.59}_{-0.56}$ \\
            & CEERS & $+0.00^{+0.07}_{-0.08}$ & $+0.01^{+0.09}_{-0.08}$ & $+0.01^{+0.13}_{-0.11}$ & $+0.01^{+0.18}_{-0.16}$ & $-0.00^{+0.18}_{-0.17}$ &$+0.00^{+0.26}_{-0.25}$ & $-0.00^{+0.33}_{-0.32}$ & $+0.01^{+0.37}_{-0.33}$ \\
            & JADES & $+0.01^{+0.07}_{-0.06}$ & $+0.01^{+0.08}_{-0.07}$ & $+0.02^{+0.12}_{-0.10}$ & $+0.02^{+0.18}_{-0.14}$ & $-0.01^{+0.18}_{-0.16}$ &$-0.00^{+0.29}_{-0.23}$ & $-0.01^{+0.36}_{-0.29}$ & $-0.00^{+0.40}_{-0.32}$ \\
            & JOF & $-0.01^{+0.06}_{-0.08}$ & $+0.00^{+0.06}_{-0.06}$ & $+0.01^{+0.10}_{-0.09}$ & $+0.00^{+0.14}_{-0.14}$ & $+0.00^{+0.18}_{-0.17}$ &$+0.00^{+0.26}_{-0.24}$ & $+0.01^{+0.32}_{-0.31}$ & $+0.01^{+0.35}_{-0.33}$ \\
            \hline
            \multirow{4}{*}{\rotatebox{90}{ContSFH }} & COSMOS-Web & $+0.01^{+0.08}_{-0.09}$ & $-0.05^{+0.10}_{-0.10}$ & $-0.08^{+0.13}_{-0.12}$ & $-0.16^{+0.19}_{-0.17}$ & $-0.35^{+0.34}_{-0.34}$ &$-0.40^{+0.40}_{-0.43}$ & $-0.51^{+0.52}_{-0.54}$ & $-0.44^{+0.48}_{-0.52}$ \\
            & CEERS & $-0.02^{+0.10}_{-0.14}$ & $-0.01^{+0.09}_{-0.14}$ & $-0.02^{+0.12}_{-0.17}$ & $-0.02^{+0.25}_{-0.22}$ & $-0.03^{+0.24}_{-0.22}$ &$-0.05^{+0.36}_{-0.33}$ & $-0.06^{+0.48}_{-0.38}$ & $-0.07^{+0.52}_{-0.41}$ \\
            & JADES & $-0.00^{+0.11}_{-0.09}$ & $+0.02^{+0.10}_{-0.10}$ & $+0.03^{+0.14}_{-0.16}$ & $+0.03^{+0.17}_{-0.21}$ & $+0.03^{+0.19}_{-0.19}$ &$+0.04^{+0.30}_{-0.27}$ & $+0.06^{+0.39}_{-0.36}$ & $+0.06^{+0.40}_{-0.36}$ \\
            & JOF & $+0.02^{+0.13}_{-0.10}$ & $+0.01^{+0.11}_{-0.09}$ & $+0.01^{+0.17}_{-0.13}$ & $+0.03^{+0.22}_{-0.14}$ & $+0.01^{+0.23}_{-0.24}$ &$+0.04^{+0.33}_{-0.26}$ & $+0.03^{+0.44}_{-0.34}$ & $+0.08^{+0.42}_{-0.33}$ \\
            \hline
            \multirow{4}{*}{\rotatebox{90}{BurstySFH }} & COSMOS-Web & $+0.05^{+0.11}_{-0.10}$ & $-0.01^{+0.12}_{-0.13}$ & $-0.08^{+0.15}_{-0.13}$ & $-0.19^{+0.23}_{-0.23}$ & $-0.47^{+0.40}_{-0.52}$ &$-0.62^{+0.52}_{-0.62}$ & $-0.83^{+0.69}_{-0.82}$ & $-0.73^{+0.64}_{-0.81}$ \\
            & CEERS & $-0.00^{+0.09}_{-0.09}$ & $+0.01^{+0.10}_{-0.09}$ & $-0.00^{+0.14}_{-0.14}$ & $-0.01^{+0.20}_{-0.19}$ & $+0.01^{+0.17}_{-0.22}$ &$-0.00^{+0.28}_{-0.32}$ & $+0.01^{+0.34}_{-0.43}$ & $+0.00^{+0.43}_{-0.42}$ \\
            & JADES & $+0.00^{+0.10}_{-0.11}$ & $+0.01^{+0.13}_{-0.10}$ & $+0.02^{+0.20}_{-0.14}$ & $-0.02^{+0.23}_{-0.26}$ & $-0.03^{+0.23}_{-0.24}$ &$-0.04^{+0.38}_{-0.41}$ & $-0.05^{+0.47}_{-0.55}$ & $-0.07^{+0.58}_{-0.53}$ \\
            & JOF & $+0.00^{+0.09}_{-0.11}$ & $+0.00^{+0.12}_{-0.09}$ & $+0.01^{+0.16}_{-0.15}$ & $+0.03^{+0.21}_{-0.17}$ & $-0.04^{+0.25}_{-0.27}$ &$-0.04^{+0.40}_{-0.30}$ & $-0.08^{+0.55}_{-0.45}$ & $-0.02^{+0.51}_{-0.38}$ \\
            \hline
	\end{tabular}
        }
\end{table*}

In Figure~\ref{fig:Summary_RestFrameColors}, the median values for each of the galaxies are given by the circles or squares. After accounting for the uncertainties on the values for individual galaxies, the median and 68\% confidence interval of the full sample are shown by the lines and shaded regions. Additionally, the medians and 68\% confidence intervals are reported at the bottom of the figure, while these quantities are presented for the four different filter sets and the four different assumed SFHs in Table~\ref{tab:RestFrameColors}. From left to right, relative to the apparent magnitude at rest-frame $1500\ \mathrm{\AA}$, the derived colors are for the following photometric filters: $U$-band, $B$-band, $V$-band, $R$-band, $I$-band, $J$-band, $H$-band, and $K$-band. The faint, dotted gray line indicates the effective wavelength of MIRI/F770W ($7.7\ \mu\mathrm{m}$) at $z = 8$.

The results from the \texttt{Prospector} models assuming the JADES filter set are represented by the blue circles, lines, and shaded regions. Those results assuming the COSMOS-Web filter set are represented by the pink squares, lines, and shaded regions. Triangles represent limits for points that fall outside the plotted range of values. When assuming the JADES filter set, it is evident from Figure~\ref{fig:Summary_RestFrameColors} that the inclusion of MIRI/F770W has little to no observable effect on the inferred rest-frame colors for the vast majority ($\approx 80\%$) of galaxies in the final photometric sample ($N = 18$). For these galaxies, the scatter in the color increases as the filters move toward redder wavelengths.

However, when assuming the COSMOS-Web filter set, it is evident from Figure~\ref{fig:Summary_RestFrameColors} and Table~\ref{tab:RestFrameColors} that the inclusion of MIRI/F770W has a significant effect on the inferred rest-frame colors at the reddest wavelengths (i.e., $R$-band, $I$-band, $J$-band, $H$-band, and $K$-band) for the vast majority of galaxies in the final photometric sample. At the bluest wavelengths (i.e., $U$-band, $B$-band, and $V$-band), the inclusion of MIRI/F770W does not have a significant effect on the inferred rest-frame colors. At the reddest wavelengths, the median offsets are: $-0.16_{-0.20}^{+0.21}\ \mathrm{AB\ mag}$ for $R$-band, $-0.36_{-0.38}^{+0.43}\ \mathrm{AB\ mag}$ for $I$-band, $-0.44_{-0.46}^{+0.50}\ \mathrm{AB\ mag}$ for $J$-band, $-0.53_{-0.63}^{+0.64}\ \mathrm{AB\ mag}$ for $H$-band, and $-0.49_{-0.52}^{+0.59}\ \mathrm{AB\ mag}$ for $K$-band. These results suggest that galaxies are typically bluer at the reddest wavelengths when including the MIRI/F770W data point and assuming the COSMOS-Web filter set. For comparison, when assuming the JADES filter set, the median offsets are: $-0.02_{-0.15}^{+0.14}\ \mathrm{AB\ mag}$ for $R$-band, $-0.02_{-0.20}^{+0.17}\ \mathrm{AB\ mag}$ for $I$-band, $-0.03_{-0.27}^{+0.25}\ \mathrm{AB\ mag}$ for $J$-band, $-0.04_{-0.37}^{+0.32}\ \mathrm{AB\ mag}$ for $H$-band, and $-0.05_{-0.36}^{+0.35}\ \mathrm{AB\ mag}$ for $K$-band.

Similar to Section~\ref{SectionFiveOne}, the results presented in Figure~\ref{fig:Summary_RestFrameColors} and Table~\ref{tab:RestFrameColors} demonstrate that observations with JWST/MIRI are not needed to robustly determine the rest-frame colors for typical star-forming galaxies at $z \approx 8$. This statement is true for the vast majority ($\approx 80\%$) of galaxies in the final photometric sample for filter sets that have sufficient photometric coverage with JWST/NIRCam. However, in the absence of sufficient photometric coverage with JWST/NIRCam, observations with JWST/MIRI are needed to robustly determine the rest-frame colors for typical star-forming galaxies at $z \approx 8$. These results are once again consistent with the results presented in \citet{Papovich:2023}.

\subsection{Comparison of Star Formation Histories}
\label{SectionFiveThree}

Although the comparisons of the derived stellar population properties and rest-frame colors presented in Sections~\ref{SectionFiveOne} and \ref{SectionFiveTwo} are assuming the parametric constant SFH model, we have also provided these comparisons for three other SFH models: parametric delayed-tau, non-parametric ``continuity,'' and non-parametric ``bursty continuity.'' Tables~\ref{tab:StellarPopulations} and \ref{tab:RestFrameColors} present these comparisons for all four of the SFH models. As a reminder, the values provided in these tables are reported as the difference between derived properties and colors when including and excluding the MIRI/F770W data point, such that positive (negative) values imply larger (smaller) inferred values. To explore which of these SFH models provide the most consistent properties and colors when excluding MIRI/F770W, we compare the offsets and scatters around these offsets when including MIRI/F770W.

We will first investigate the derived stellar population properties from Table~\ref{tab:StellarPopulations}. In general, for the COSMOS-Web filter set, we find that the delayed-tau SFH model provides the smallest offsets and scatters around these offsets for stellar mass, star formation rate, specific star formation rate, and rest-optical equivalent width. Stellar ages are highly uncertain for the COSMOS-Web filter set with the delayed-tau SFH model, and it is actually the ``continuity'' SFH model that provides the smallest offsets and scatters around these offsets as a result of being the most stringent prior. For the CEERS, JADES, and JOF filter sets, we find that the constant and delayed-tau SFH models provide the smallest offsets and scatters around these offsets for stellar mass, star formation rate, specific star formation rate, and rest-optical equivalent width. Once again, stellar ages are highly uncertain for the constant and delayed-tau SFH models, and it is the ``continuity'' SFH model that provides the smallest offsets and scatters around these offsets as a result of being the most stringent prior.

We will now investigate the derived rest-frame colors from Table~\ref{tab:RestFrameColors}. At the bluest wavelengths, all of the SFH models for all of the assumed filter sets produce nearly identical results, since JWST/NIRCam is providing sufficient coverage of these wavelengths. At the reddest wavelengths, the different SFH models and filter sets begin to diverge from one another. As a reminder, MIRI/F770W yields coverage of the rest-frame $I$-band at $z \approx 8$. In general, for the COSMOS-Web filter set, we find that the delayed-tau SFH model provides the smallest offsets and scatters around these offsets for the colors at $\lambda_{\mathrm{rest}} \gtrsim 0.5\ \mu\mathrm{m}$ (corresponding to $V$-band and redder). For the CEERS, JADES, and JOF filter sets, we find that all of the SFH models provide similar offsets and scatters around these offsets for the colors across all of the wavelengths that we consider.

Taken together, the results described in Tables~\ref{tab:StellarPopulations} and \ref{tab:RestFrameColors} suggest that the parametric delayed-tau SFH model provides the most consistent stellar population properties and rest-frame colors when excluding MIRI/F770W for all four of the filter sets considered here. By ``most consistent,'' we mean that this SFH model provides the smallest offsets and scatters around these offsets when including MIRI/F770W. The parametric constant SFH model also provides consistent stellar population properties, so long as the filter set has sufficient photometric coverage with JWST/NIRCam (e.g., CEERS, JADES, and JOF as described in Table~\ref{tab:FilterSetTable}). We should caution that the inferred stellar ages are subject to extremely large scatter for nearly all of the SFH models and filter sets, aside from the non-parametric ``continuity'' SFH model, which includes the most stringent prior. Finally, the rest-frame colors are most sensitive to the assumed SFH model in the absence of sufficient photometric coverage with JWST/NIRCam (e.g., COSMOS-Web).

The primary reason that the delayed-tau SFH model performs better than the other SFH models for the CEERS, JADES, and JOF filter sets is that the delayed-tau SFH model does not implicitly assume that the star formation rate is constant with respect to time, whereas the other SFH models do make this assumption. Another reason is that the delayed-tau SFH model is simpler and has fewer free parameters when compared to the non-parametric SFH models. As previously discussed, observations and theory have both demonstrated that on average, SFHs are increasing and becoming more bursty with cosmic time at $z > 3$. This increase in the SFH directly follows the mass accretion rate of dark matter halos \citep[e.g.,][]{Tacchella:2018}. The delayed-tau SFH model naturally captures this cosmologically motivated growth pattern while remaining mathematically simple enough to be constrained by photometry alone. The balance between physical realism and model simplicity makes it particularly well suited for studying high-redshift galaxies, where observations are limited and the available cosmic time is short. The delayed-tau SFH model helps mitigate some of the degeneracies and uncertainties that are inherent among the apparent properties of the stars, nebular gas, and dust for galaxies during the EoR. However, it is important to acknowledge that significant systematic uncertainties in the SED modeling remain, such as assumptions about the IMF and stellar populations, which are not currently captured by the Bayesian framework.

As already mentioned, the inferred stellar population properties are all consistent within the associated uncertainties for the CEERS, JADES, and JOF filter sets. However, these properties are not all consistent with one another for the four different SFH assumptions. This is best demonstrated by the table in Appendix~\ref{AppendixB}, which provides the inferred stellar population properties for the different SFH assumptions. To quickly summarize, the two non-parametric SFHs (``continuity'' and ``bursty continuity'') generally infer larger stellar masses, mass-weighted ages, and specific star-formation rates when compared to the two parametric SFHs (constant and delayed-tau). As discussed in \citet[][]{Endsley:2024}, this is because the non-parametric models impose a mild preference for significant star formation extending back to the formation redshift $z_{\mathrm{form}} = 20$ (corresponding to a lookback time of $t_{\mathrm{form}} \approx 450\ \mathrm{Myr}$ for a galaxy at $z \approx 8$) by penalizing strong variability in the SFH.

\subsection{Impact of Photometric Redshift Uncertainty}
\label{SectionFiveFour}

As described in Section~\ref{SectionFourTwo}, for objects with spectroscopic redshifts, we derive properties using two types of physical models to explore the impact of the photometric redshift uncertainty. One of these models has redshift as a fixed parameter, while the other has redshift as a free parameter. As already mentioned in Section~\ref{SectionFourTwo}, there is only one galaxy (JADES$-$GS$-$ID$-$$174121$) where the inclusion of the spectroscopic redshift has a significant impact on the inferred properties and ultimately results in a much better fit when compared to using the photometric redshift. This is caused by misplacement of rest-frame optical emission lines ($\mathrm{H}\beta$ and $\mathrm{[OIII]}\lambda\lambda 4960{,}5008$) falling near the edge of the F410M transmission curve. Furthermore, we verified that allowing redshift to be a free parameter does not change any of the results or conclusions that have already been presented in Section~\ref{SectionFive}. The only notable difference is that the scatters in the median offsets increase for most of the inferred stellar population properties and rest-frame colors when redshift is allowed to be free. However, we note the median offsets themselves remain unchanged.

When assuming the JADES filter set, the scatters in the median offsets increase by $\approx 25\%$ for stellar masses, stellar ages, specific star formation rates, and rest-optical equivalent widths. The scatter for star formation rates remains unchanged. When assuming the COSMOS-Web filter set, the scatters in the median offsets increase by $\approx 5\%$ for stellar masses and specific star formation rates while they increase by $\approx 10-15\%$ for stellar ages. The scatters for star formation rates and rest-optical equivalent widths remains unchanged. Similarly, when assuming the JADES filter set, the scatters in the median offsets increase by $\approx 20\%$ for $R$-band and $I$-band while they increase by $\approx 10\%$ for $J$-band, $H$-band, and $K$-band. The scatters at bluer wavelengths remains unchanged. When assuming the COSMOS-Web filter set, the scatters in the median offsets increase by $\approx 20\%$ for $I$-band, $J$-band, and $H$-band. The scatters for the remaining filters remains unchanged.

Regardless of the available redshifts, we find that observations with JWST/MIRI are not needed to robustly determine the stellar population properties and rest-frame colors for typical star-forming galaxies at $z \approx 8$. However, fixing the redshift to the spectroscopic value decreases the JWST/MIRI's impact by decreasing the scatters in the derived properties. The reason for the decreased scatters in galaxies with spectroscopic redshifts is the decreased dimensionality in the physical modeling, since one of the free parameters has been removed.

\subsection{Other Physical Properties}
\label{SectionFiveFive}

We also compare the inferred properties of the nebular gas and dust for the four different filter sets and four different assumed SFHs. We find median offsets equal to zero for ionization parameter, diffuse dust optical depth, birth-cloud dust optical depth, and the power-law index of the dust attenuation curve for all of the filter sets (including COSMOS-Web) and assumed SFHs.


\section{Summary \& Conclusions}
\label{SectionSix}

For star-forming galaxies during the Epoch of Reionization (EoR; at $z > 6$), the properties of the stars, nebular gas, and dust are unfortunately degenerate with one another when they are derived from a limited set of observations. These degeneracies are most prevalent in the absence of photometry (or spectroscopy) with sufficient resolution and rest-frame spectral coverage. Using some of the deepest existing imaging with JWST/NIRCam and JWST/MIRI from the JWST Advanced Deep Extragalactic Survey \citep[JADES;][]{Eisenstein:2026}, we explore ways to break these degeneracies at $z \approx 8$. Key to this study is the imaging from JWST/MIRI at $7.7\ \mu\mathrm{m}$, which provides spectral coverage of the rest-frame $I$-band at the observed redshifts.  Our findings can be summarized as follows.
\begin{enumerate}
    \topsep 0pt
    \parsep 0pt
    \itemsep 0pt
    \item Observations from HST/ACS, JWST/NIRCam, and JWST/MIRI were used to select a flux-limited ($m_{\mathrm{F277W}} < 29.0$) sample of $N = 47$ high-redshift star-forming galaxies at $7 < z_{\mathrm{phot}} \leq 9$ using photometry spanning $\lambda_{\mathrm{obs}} = 0.8-8.9\ \mu\mathrm{m}$. This sample includes $N = 22$ ($N = 25$) galaxies with (in)significant detections of MIRI/F770W, where we adopted a signal-to-noise ratio of $\mathrm{S/N} = 1.5$ to separate the detections and non-detections.
    \item The properties of the rest-UV photometry were inferred for the flux-limited sample. We found that selecting a sample of MIRI/F770W detections at $z \approx 8$ is similar to selecting a sample of galaxies based on their rest-UV properties ($M_{\mathrm{UV}} \lesssim -19$).
    \item The properties of the stellar populations and rest-frame colors were inferred for the flux-limited sample of MIRI/F770W detections using four different filter sets and four different assumed SFHs. By evaluating these quantities both with and without the $7.7\ \mu\mathrm{m}$ data point, we found that dense spectral coverage with JWST/NIRCam ($8-14$ filters, including $1-7$ medium-bands) can compensate for lacking the rest-frame $I$-band coverage for the vast majority ($\approx 80\%$) of our sample. This suggests that observations with JWST/MIRI are not needed to robustly determine these physical properties for typical star-forming galaxies during the EoR (i.e., galaxies at $z \approx 8$ with $M_{\mathrm{UV}} \approx -20$ and $\beta_{\mathrm{UV}} \approx -2$). For the remaining minority ($\approx 20\%$) of our sample, our results suggest that observations with JWST/MIRI can reveal anomalous sources that are redder or bluer than expected from HST/ACS and JWST/NIRCam alone.
    \item However, in the absence of sufficient spectral coverage with JWST/NIRCam, observations with JWST/MIRI are needed to robustly determine the properties of the stellar populations and rest-frame colors, albeit with larger scatter than using denser sampling with JWST/NIRCam. In this case of insufficient spectral coverage with JWST/NIRCam but including JWST/MIRI, the inferred properties of high-redshift star-forming galaxies are typically: less massive and younger, with larger specific star formation rates and rest-optical equivalent widths, in addition to being bluer at both rest-frame optical and near-infrared wavelengths.
    \item Finally, regardless of the adopted filter set, the properties of the stellar populations and rest-frame colors are most consistently determined by assuming a star formation history with the parametric delayed-tau prior, at least when compared to the following priors: parametric constant, non-parametric ``continuity,'' and non-parametric ``bursty continuity.'' This assumption provided the smallest offsets and scatters around these offsets when including JWST/MIRI.
\end{enumerate}

Our findings have important implications for future surveys of high-redshift galaxies with JWST. We found that dense spectral coverage with JWST/NIRCam can provide robust constraints on the stellar population properties and rest-frame colors for the vast majority of star-forming galaxies at $z \approx 8$, without requiring observations from JWST/MIRI. Given the limited field of view and sensitivity of JWST/MIRI when compared to JWST/NIRCam, this greatly expands the potential sample of high-redshift galaxies that can be studied in detail. For follow-up observations of typical galaxies during the EoR, we recommend prioritizing multiband observations with JWST/NIRCam using both medium- and wide-band filters as opposed to observations with JWST/MIRI. However, JWST/MIRI remains valuable for identifying the minority of galaxies with anomalous properties, such as galaxies that are unusually red \citep[see also, e.g.,][]{Rinaldi:2025}. Future work should study these anomalously red galaxies in more detail, since they potentially have older stellar populations, large amounts of dust, and/or significant contributions from an \mbox{obscured} active galactic nucleus (AGN).

Our work also suggests that future studies of high-redshift galaxies should utilize the parametric delayed-tau prior for their star formation history assumption since this model provides a physically motivated compromise between overly simplistic models (i.e., parametric constant) and those that are unnecessarily complex (i.e., non-parametric ``continuity'' and ``bursty continuity''). The preference for the delayed-tau star formation history provides key insights into the physics governing early galaxy formation and evolution, while setting the stage for more detailed investigations of galaxies in the first few hundred million years after the Big Bang.


\section*{Acknowledgments}

We thank the anonymous referee for providing constructive feedback that helped strengthen this paper. We also thank Daniel Stark, Ryan Endsley, Robert Kennicutt, and David Sand for valuable conversations that improved this paper. This work is based on observations made with the NASA/ESA/CSA James Webb Space Telescope (JWST). These data were obtained from the Mikulski Archive for Space Telescopes (MAST) at the Space Telescope Science Institute (STScI), which is operated by the Association of Universities for Research in Astronomy, Inc., under NASA contract \mbox{NAS 5-03127} for JWST. These observations are associated with program \#1180, 1181, 1210, 1286, and 1895. The specific observations analyzed here can be accessed via \dataset[DOI: 10.17909/z2gw-mk31]{https://doi.org/10.17909/z2gw-mk31} \citep{JADES_DR2_DOI} and \dataset[DOI: 10.17909/T91019]{https://doi.org/10.17909/T91019} \citep{HLF_DOI}. This work is also based on observations made with the NASA/ESA Hubble Space Telescope (HST). These data were obtained from STScI, which is operated by the Association of Universities for Research in Astronomy, Inc., under NASA contract NAS5–26555 for HST. Additionally, this work made use of the {\it lux} supercomputer at UC Santa Cruz, which is funded by NSF MRI grant AST 1828315, as well as the High Performance Computing (HPC) resources at the University of Arizona, which is funded by the Office of Research Discovery and Innovation (ORDI), Chief Information Officer (CIO), and University Information Technology Services (UITS).

We respectfully acknowledge that the University of Arizona is on the land and territories of Indigenous peoples. Today, Arizona is home to 22 federally recognized tribes, with Tucson being home to the O’odham and the Yaqui. The University of Arizona strives to build sustainable relationships with sovereign Native Nations and Indigenous communities through education offerings, partnerships, and community service.

J.M.H., G.H.R., Z.J., M.J.R., B.D.J., B.R., E.E., D.J.E., and C.N.A.W acknowledge support from the JWST Near Infrared Camera (NIRCam) Science Team Lead, NAS5-02015, from NASA Goddard Space Flight Center to the University of Arizona. J.M.H. and B.R. also acknowledge support from JWST Program \#3215. S.A. acknowledges support from the JWST Mid-Infrared Instrument (MIRI) Science Team Lead, 80NSSC18K0555, from NASA Goddard Space Flight Center to the University of Arizona. S.T. acknowledges support by the Royal Society Research Grant G125142. L.W. acknowledges support from the National Science Foundation Graduate Research Fellowship under grant No. DGE-2137419. W.M.B. gratefully acknowledges support from DARK via the DARK fellowship; this work was supported by a research grant (VIL54489) from Villum Fonden. K.B. acknowledges support from the Australian Research Council Centre of Excellence for All Sky Astrophysics in 3 Dimensions (ASTRO 3D), through project number CE170100013. A.J.B. and J.C. acknowledge funding from the ``FirstGalaxies'' Advanced Grant from the European Research Council (ERC) under the European Union’s Horizon 2020 research and innovation program (grant agreement No. 789056). S.C. acknowledges support by European Union’s HE ERC Starting Grant No. 101040227 - WINGS. E.C.-L. acknowledges support of an STFC Webb Fellowship (ST/W001438/1). D.J.E. is also supported as a Simons Investigator. R.H. acknowledges support from the Johns Hopkins University, Institute for Data Intensive Engineering and Science (IDIES). R.M. and J.W. acknowledge support by the Science and Technology Facilities Council (STFC), by the ERC through Advanced Grant 695671 ``QUENCH,'' and by the UKRI Frontier Research grant RISEandFALL. R.M. also acknowledges funding from a research professorship from the Royal Society. P.G.P.-G. acknowledges support from grant PID2022-139567NB-I00 funded by Spanish Ministerio de Ciencia e Innovaci\'{o}n MCIN/AEI/10.13039/501100011033, FEDER, UE. C.C.W. acknowledges support from NOIRLab, which is managed by the Association of Universities for Research in Astronomy (AURA) under a cooperative agreement with the National Science Foundation.

\facilities{HST (ACS), JWST (MIRI and NIRCam)}

\software{\texttt{AstroPy} \citep[][]{Astropy:2013, Astropy:2018}, \texttt{astroquery} \citep[][]{Ginsburg:2019}, \texttt{Cloudy} \citep{Byler:2017}, \texttt{corner} \citep[][]{Foreman-Mackey:2016}, \texttt{dynesty} \citep[][]{Speagle:2020}, \texttt{FSPS} \citep[][]{Conroy:2009, Conroy:2010}, \texttt{Matplotlib} \citep[][]{Matplotlib:2007}, \texttt{NumPy} \citep[][]{NumPy:2011, NumPy:2020}, \texttt{Pandas} \citep[][]{Pandas:2022}, \texttt{photutils} \citep[][]{Bradley:2022}, \texttt{Prospector} \citep[][]{Johnson:2021}, \texttt{python-FSPS} \citep[][]{Foreman-Mackey:2014}, \texttt{SciPy} \citep[][]{SciPy:2020}, \texttt{seaborn} \citep[][]{Waskom:2021}, \texttt{TinyTim} \citep[][]{Krist:2011}, \texttt{WebbPSF} \citep[][]{Perrin:2014}}

\bibliographystyle{aasjournal}
\bibliography{main}{}

\begin{thebibliography}{}
\expandafter\ifx\csname natexlab\endcsname\relax\def\natexlab#1{#1}\fi
\providecommand{\url}[1]{\href{#1}{#1}}
\providecommand{\dodoi}[1]{doi:~\href{http://doi.org/#1}{\nolinkurl{#1}}}
\providecommand{\doeprint}[1]{\href{http://ascl.net/#1}{\nolinkurl{http://ascl.net/#1}}}
\providecommand{\doarXiv}[1]{\href{https://arxiv.org/abs/#1}{\nolinkurl{https://arxiv.org/abs/#1}}}

\bibitem[{{Alberts} {et~al.}(2024{\natexlab{a}}){Alberts}, {Williams}, {Helton}, {Suess}, {Ji}, {Shivaei}, {Lyu}, {Rieke}, {Baker}, {Bonaventura}, {Bunker}, {Carniani}, {Charlot}, {Curtis-Lake}, {D'Eugenio}, {Eisenstein}, {de Graaff}, {Hainline}, {Hausen}, {Johnson}, {Maiolino}, {Parlanti}, {Rieke}, {Robertson}, {Sun}, {Tacchella}, {Willmer}, \& {Willott}}]{Alberts:2024a}
{Alberts}, S., {Williams}, C.~C., {Helton}, J.~M., {et~al.} 2024{\natexlab{a}}, \apj, 975, 85, \dodoi{10.3847/1538-4357/ad66cc}

\bibitem[{{Alberts} {et~al.}(2024{\natexlab{b}}){Alberts}, {Lyu}, {Shivaei}, {Rieke}, {P{\'e}rez-Gonz{\'a}lez}, {Bonaventura}, {Zhu}, {Helton}, {Ji}, {Morrison}, {Robertson}, {Stone}, {Sun}, {Williams}, \& {Willmer}}]{Alberts:2024b}
{Alberts}, S., {Lyu}, J., {Shivaei}, I., {et~al.} 2024{\natexlab{b}}, \apj, 976, 224, \dodoi{10.3847/1538-4357/ad7396}

\bibitem[{{Arrabal Haro} {et~al.}(2023){Arrabal Haro}, {Dickinson}, {Finkelstein}, {Kartaltepe}, {Donnan}, {Burgarella}, {Carnall}, {Cullen}, {Dunlop}, {Fern{\'a}ndez}, {Fujimoto}, {Jung}, {Krips}, {Larson}, {Papovich}, {P{\'e}rez-Gonz{\'a}lez}, {Amor{\'\i}n}, {Bagley}, {Buat}, {Casey}, {Chworowsky}, {Cohen}, {Ferguson}, {Giavalisco}, {Huertas-Company}, {Hutchison}, {Kocevski}, {Koekemoer}, {Lucas}, {McLeod}, {McLure}, {Pirzkal}, {Seill{\'e}}, {Trump}, {Weiner}, {Wilkins}, \& {Zavala}}]{ArrabalHaro:2023}
{Arrabal Haro}, P., {Dickinson}, M., {Finkelstein}, S.~L., {et~al.} 2023, \nat, 622, 707, \dodoi{10.1038/s41586-023-06521-7}

\bibitem[{{Astropy Collaboration} {et~al.}(2013){Astropy Collaboration}, {Robitaille}, {Tollerud}, {Greenfield}, {Droettboom}, {Bray}, {Aldcroft}, {Davis}, {Ginsburg}, {Price-Whelan}, {Kerzendorf}, {Conley}, {Crighton}, {Barbary}, {Muna}, {Ferguson}, {Grollier}, {Parikh}, {Nair}, {Unther}, {Deil}, {Woillez}, {Conseil}, {Kramer}, {Turner}, {Singer}, {Fox}, {Weaver}, {Zabalza}, {Edwards}, {Azalee Bostroem}, {Burke}, {Casey}, {Crawford}, {Dencheva}, {Ely}, {Jenness}, {Labrie}, {Lim}, {Pierfederici}, {Pontzen}, {Ptak}, {Refsdal}, {Servillat}, \& {Streicher}}]{Astropy:2013}
{Astropy Collaboration}, {Robitaille}, T.~P., {Tollerud}, E.~J., {et~al.} 2013, \aap, 558, A33, \dodoi{10.1051/0004-6361/201322068}

\bibitem[{{Astropy Collaboration} {et~al.}(2018){Astropy Collaboration}, {Price-Whelan}, {Sip{\H{o}}cz}, {G{\"u}nther}, {Lim}, {Crawford}, {Conseil}, {Shupe}, {Craig}, {Dencheva}, {Ginsburg}, {VanderPlas}, {Bradley}, {P{\'e}rez-Su{\'a}rez}, {de Val-Borro}, {Aldcroft}, {Cruz}, {Robitaille}, {Tollerud}, {Ardelean}, {Babej}, {Bach}, {Bachetti}, {Bakanov}, {Bamford}, {Barentsen}, {Barmby}, {Baumbach}, {Berry}, {Biscani}, {Boquien}, {Bostroem}, {Bouma}, {Brammer}, {Bray}, {Breytenbach}, {Buddelmeijer}, {Burke}, {Calderone}, {Cano Rodr{\'\i}guez}, {Cara}, {Cardoso}, {Cheedella}, {Copin}, {Corrales}, {Crichton}, {D'Avella}, {Deil}, {Depagne}, {Dietrich}, {Donath}, {Droettboom}, {Earl}, {Erben}, {Fabbro}, {Ferreira}, {Finethy}, {Fox}, {Garrison}, {Gibbons}, {Goldstein}, {Gommers}, {Greco}, {Greenfield}, {Groener}, {Grollier}, {Hagen}, {Hirst}, {Homeier}, {Horton}, {Hosseinzadeh}, {Hu}, {Hunkeler}, {Ivezi{\'c}}, {Jain}, {Jenness}, {Kanarek}, {Kendrew}, {Kern}, {Kerzendorf}, {Khvalko}, {King}, {Kirkby}, {Kulkarni},
  {Kumar}, {Lee}, {Lenz}, {Littlefair}, {Ma}, {Macleod}, {Mastropietro}, {McCully}, {Montagnac}, {Morris}, {Mueller}, {Mumford}, {Muna}, {Murphy}, {Nelson}, {Nguyen}, {Ninan}, {N{\"o}the}, {Ogaz}, {Oh}, {Parejko}, {Parley}, {Pascual}, {Patil}, {Patil}, {Plunkett}, {Prochaska}, {Rastogi}, {Reddy Janga}, {Sabater}, {Sakurikar}, {Seifert}, {Sherbert}, {Sherwood-Taylor}, {Shih}, {Sick}, {Silbiger}, {Singanamalla}, {Singer}, {Sladen}, {Sooley}, {Sornarajah}, {Streicher}, {Teuben}, {Thomas}, {Tremblay}, {Turner}, {Terr{\'o}n}, {van Kerkwijk}, {de la Vega}, {Watkins}, {Weaver}, {Whitmore}, {Woillez}, {Zabalza}, \& {Astropy Contributors}}]{Astropy:2018}
{Astropy Collaboration}, {Price-Whelan}, A.~M., {Sip{\H{o}}cz}, B.~M., {et~al.} 2018, \aj, 156, 123, \dodoi{10.3847/1538-3881/aabc4f}

\bibitem[{{Bagley} {et~al.}(2024){Bagley}, {Finkelstein}, {Rojas-Ruiz}, {Diekmann}, {Finkelstein}, {Song}, {Papovich}, {Somerville}, {Baronchelli}, \& {Dai}}]{Bagley:2024}
{Bagley}, M.~B., {Finkelstein}, S.~L., {Rojas-Ruiz}, S., {et~al.} 2024, \apj, 961, 209, \dodoi{10.3847/1538-4357/ad09dc}

\bibitem[{{Beckwith} {et~al.}(2006){Beckwith}, {Stiavelli}, {Koekemoer}, {Caldwell}, {Ferguson}, {Hook}, {Lucas}, {Bergeron}, {Corbin}, {Jogee}, {Panagia}, {Robberto}, {Royle}, {Somerville}, \& {Sosey}}]{Beckwith:2006}
{Beckwith}, S. V.~W., {Stiavelli}, M., {Koekemoer}, A.~M., {et~al.} 2006, \aj, 132, 1729, \dodoi{10.1086/507302}

\bibitem[{{Bouwens} {et~al.}(2008){Bouwens}, {Illingworth}, {Franx}, \& {Ford}}]{Bouwens:2008}
{Bouwens}, R.~J., {Illingworth}, G.~D., {Franx}, M., \& {Ford}, H. 2008, \apj, 686, 230, \dodoi{10.1086/590103}

\bibitem[{{Bouwens} {et~al.}(2010){Bouwens}, {Illingworth}, {Oesch}, {Stiavelli}, {van Dokkum}, {Trenti}, {Magee}, {Labb{\'e}}, {Franx}, {Carollo}, \& {Gonzalez}}]{Bouwens:2010}
{Bouwens}, R.~J., {Illingworth}, G.~D., {Oesch}, P.~A., {et~al.} 2010, \apjl, 709, L133, \dodoi{10.1088/2041-8205/709/2/L133}

\bibitem[{{Bouwens} {et~al.}(2015){Bouwens}, {Illingworth}, {Oesch}, {Trenti}, {Labb{\'e}}, {Bradley}, {Carollo}, {van Dokkum}, {Gonzalez}, {Holwerda}, {Franx}, {Spitler}, {Smit}, \& {Magee}}]{Bouwens:2015}
---. 2015, \apj, 803, 34, \dodoi{10.1088/0004-637X/803/1/34}

\bibitem[{{Bouwens} {et~al.}(2021){Bouwens}, {Oesch}, {Stefanon}, {Illingworth}, {Labb{\'e}}, {Reddy}, {Atek}, {Montes}, {Naidu}, {Nanayakkara}, {Nelson}, \& {Wilkins}}]{Bouwens:2021}
{Bouwens}, R.~J., {Oesch}, P.~A., {Stefanon}, M., {et~al.} 2021, \aj, 162, 47, \dodoi{10.3847/1538-3881/abf83e}

\bibitem[{{Boyett} {et~al.}(2024){Boyett}, {Bunker}, {Curtis-Lake}, {Chevallard}, {Cameron}, {Jones}, {Saxena}, {Charlot}, {Curti}, {Wallace}, {Arribas}, {Carniani}, {Willott}, {Alberts}, {Eisenstein}, {Hainline}, {Hausen}, {Johnson}, {Rieke}, {Robertson}, {Stark}, {Tacchella}, {Williams}, {Chen}, {Egami}, {Endsley}, {Kumari}, {Laseter}, {Looser}, {Maseda}, {Scholtz}, {Shivaei}, {Simmonds}, {Smit}, {{\"U}bler}, \& {Witstok}}]{Boyett:2024}
{Boyett}, K., {Bunker}, A.~J., {Curtis-Lake}, E., {et~al.} 2024, \mnras, 535, 1796, \dodoi{10.1093/mnras/stae2430}

\bibitem[{{Bradley} {et~al.}(2022){Bradley}, {Sip{\H{o}}cz}, {Robitaille}, {Tollerud}, {Vin{\'\i}cius}, {Deil}, {Barbary}, {Wilson}, {Busko}, {Donath}, {G{\"u}nther}, {Cara}, {Lim}, {Me{\ss}linger}, {Conseil}, {Bostroem}, {Droettboom}, {Bray}, {Andersen Bratholm}, {Barentsen}, {Craig}, {Rathi}, {Pascual}, {Perren}, {Georgiev}, {De Val-Borro}, {Kerzendorf}, {Bach}, {Quint}, \& {Souchereau}}]{Bradley:2022}
{Bradley}, L., {Sip{\H{o}}cz}, B., {Robitaille}, T., {et~al.} 2022, {astropy/photutils: 1.5.0}, 1.5.0, Zenodo,  Zenodo, \dodoi{10.5281/zenodo.6825092}

\bibitem[{{Brammer} {et~al.}(2008){Brammer}, {van Dokkum}, \& {Coppi}}]{Brammer:2008}
{Brammer}, G.~B., {van Dokkum}, P.~G., \& {Coppi}, P. 2008, \apj, 686, 1503, \dodoi{10.1086/591786}

\bibitem[{{Bunker} {et~al.}(2010){Bunker}, {Wilkins}, {Ellis}, {Stark}, {Lorenzoni}, {Chiu}, {Lacy}, {Jarvis}, \& {Hickey}}]{Bunker:2010}
{Bunker}, A.~J., {Wilkins}, S., {Ellis}, R.~S., {et~al.} 2010, \mnras, 409, 855, \dodoi{10.1111/j.1365-2966.2010.17350.x}

\bibitem[{{Byler} {et~al.}(2017){Byler}, {Dalcanton}, {Conroy}, \& {Johnson}}]{Byler:2017}
{Byler}, N., {Dalcanton}, J.~J., {Conroy}, C., \& {Johnson}, B.~D. 2017, \apj, 840, 44, \dodoi{10.3847/1538-4357/aa6c66}

\bibitem[{{Calzetti} {et~al.}(2000){Calzetti}, {Armus}, {Bohlin}, {Kinney}, {Koornneef}, \& {Storchi-Bergmann}}]{Calzetti:2000}
{Calzetti}, D., {Armus}, L., {Bohlin}, R.~C., {et~al.} 2000, \apj, 533, 682, \dodoi{10.1086/308692}

\bibitem[{{Calzetti} {et~al.}(1994){Calzetti}, {Kinney}, \& {Storchi-Bergmann}}]{Calzetti:1994}
{Calzetti}, D., {Kinney}, A.~L., \& {Storchi-Bergmann}, T. 1994, \apj, 429, 582, \dodoi{10.1086/174346}

\bibitem[{{Casey} {et~al.}(2023){Casey}, {Kartaltepe}, {Drakos}, {Franco}, {Harish}, {Paquereau}, {Ilbert}, {Rose}, {Cox}, {Nightingale}, {Robertson}, {Silverman}, {Koekemoer}, {Massey}, {McCracken}, {Rhodes}, {Akins}, {Allen}, {Amvrosiadis}, {Arango-Toro}, {Bagley}, {Bongiorno}, {Capak}, {Champagne}, {Chartab}, {Ch{\'a}vez Ortiz}, {Chworowsky}, {Cooke}, {Cooper}, {Darvish}, {Ding}, {Faisst}, {Finkelstein}, {Fujimoto}, {Gentile}, {Gillman}, {Gould}, {Gozaliasl}, {Hayward}, {He}, {Hemmati}, {Hirschmann}, {Jahnke}, {Jin}, {Khostovan}, {Kokorev}, {Lambrides}, {Laigle}, {Larson}, {Leung}, {Liu}, {Liaudat}, {Long}, {Magdis}, {Mahler}, {Mainieri}, {Manning}, {Maraston}, {Martin}, {McCleary}, {McKinney}, {McPartland}, {Mobasher}, {Pattnaik}, {Renzini}, {Rich}, {Sanders}, {Sattari}, {Scognamiglio}, {Scoville}, {Sheth}, {Shuntov}, {Sparre}, {Suzuki}, {Talia}, {Toft}, {Trakhtenbrot}, {Urry}, {Valentino}, {Vanderhoof}, {Vardoulaki}, {Weaver}, {Whitaker}, {Wilkins}, {Yang}, \& {Zavala}}]{Casey:2023}
{Casey}, C.~M., {Kartaltepe}, J.~S., {Drakos}, N.~E., {et~al.} 2023, \apj, 954, 31, \dodoi{10.3847/1538-4357/acc2bc}

\bibitem[{{Chabrier}(2003)}]{Chabrier:2003}
{Chabrier}, G. 2003, \pasp, 115, 763, \dodoi{10.1086/376392}

\bibitem[{{Charlot} \& {Fall}(2000)}]{Charlot:2000}
{Charlot}, S., \& {Fall}, S.~M. 2000, \apj, 539, 718, \dodoi{10.1086/309250}

\bibitem[{{Choi} {et~al.}(2016){Choi}, {Dotter}, {Conroy}, {Cantiello}, {Paxton}, \& {Johnson}}]{Choi:2016}
{Choi}, J., {Dotter}, A., {Conroy}, C., {et~al.} 2016, \apj, 823, 102, \dodoi{10.3847/0004-637X/823/2/102}

\bibitem[{{Conroy} \& {Gunn}(2010)}]{Conroy:2010}
{Conroy}, C., \& {Gunn}, J.~E. 2010, \apj, 712, 833, \dodoi{10.1088/0004-637X/712/2/833}

\bibitem[{{Conroy} {et~al.}(2009){Conroy}, {Gunn}, \& {White}}]{Conroy:2009}
{Conroy}, C., {Gunn}, J.~E., \& {White}, M. 2009, \apj, 699, 486, \dodoi{10.1088/0004-637X/699/1/486}

\bibitem[{{D'Eugenio} {et~al.}(2024){D'Eugenio}, {Maiolino}, {Carniani}, {Chevallard}, {Curtis-Lake}, {Witstok}, {Charlot}, {Baker}, {Arribas}, {Boyett}, {Bunker}, {Curti}, {Eisenstein}, {Hainline}, {Ji}, {Johnson}, {Kumari}, {Looser}, {Nakajima}, {Nelson}, {Rieke}, {Robertson}, {Scholtz}, {Smit}, {Sun}, {Venturi}, {Tacchella}, {{\"U}bler}, {Willmer}, \& {Willott}}]{D'Eugenio:2024}
{D'Eugenio}, F., {Maiolino}, R., {Carniani}, S., {et~al.} 2024, \aap, 689, A152, \dodoi{10.1051/0004-6361/202348636}

\bibitem[{{D'Eugenio} {et~al.}(2025){D'Eugenio}, {Cameron}, {Scholtz}, {Carniani}, {Willott}, {Curtis-Lake}, {Bunker}, {Parlanti}, {Maiolino}, {Willmer}, {Jakobsen}, {Robertson}, {Johnson}, {Tacchella}, {Cargile}, {Rawle}, {Arribas}, {Chevallard}, {Curti}, {Egami}, {Eisenstein}, {Kumari}, {Looser}, {Rieke}, {Rodr{\'\i}guez Del Pino}, {Saxena}, {{\"U}bler}, {Venturi}, {Witstok}, {Baker}, {Bhatawdekar}, {Bonaventura}, {Boyett}, {Charlot}, {Danhaive}, {Hainline}, {Hausen}, {Helton}, {Ji}, {Ji}, {Jones}, {Juod{\v{z}}balis}, {Maseda}, {P{\'e}rez-Gonz{\'a}lez}, {Perna}, {Pusk{\'a}s}, {Shivaei}, {Silcock}, {Simmonds}, {Smit}, {Sun}, {Villanueva}, {Williams}, \& {Zhu}}]{D'Eugenio:2025}
{D'Eugenio}, F., {Cameron}, A.~J., {Scholtz}, J., {et~al.} 2025, \apjs, 277, 4, \dodoi{10.3847/1538-4365/ada148}

\bibitem[{{Dotter}(2016)}]{Dotter:2016}
{Dotter}, A. 2016, \apjs, 222, 8, \dodoi{10.3847/0067-0049/222/1/8}

\bibitem[{{Duncan} {et~al.}(2014){Duncan}, {Conselice}, {Mortlock}, {Hartley}, {Guo}, {Ferguson}, {Dav{\'e}}, {Lu}, {Ownsworth}, {Ashby}, {Dekel}, {Dickinson}, {Faber}, {Giavalisco}, {Grogin}, {Kocevski}, {Koekemoer}, {Somerville}, \& {White}}]{Duncan:2014}
{Duncan}, K., {Conselice}, C.~J., {Mortlock}, A., {et~al.} 2014, \mnras, 444, 2960, \dodoi{10.1093/mnras/stu1622}

\bibitem[{{Eisenstein} {et~al.}(2025){Eisenstein}, {Johnson}, {Robertson}, {Tacchella}, {Hainline}, {Jakobsen}, {Maiolino}, {Bonaventura}, {Bunker}, {Cameron}, {Cargile}, {Curtis-Lake}, {Hausen}, {Pusk{\'a}s}, {Rieke}, {Sun}, {Willmer}, {Willott}, {Alberts}, {Arribas}, {Baker}, {Baum}, {Bhatawdekar}, {Carniani}, {Charlot}, {Chen}, {Chevallard}, {Curti}, {DeCoursey}, {D'Eugenio}, {de Graaff}, {Egami}, {Helton}, {Ji}, {Jones}, {Kumari}, {L{\"u}tzgendorf}, {Laseter}, {Looser}, {Lyu}, {Maseda}, {Nelson}, {Parlanti}, {Rauscher}, {Rawle}, {Rieke}, {Rix}, {Rujopakarn}, {Sandles}, {Saxena}, {Scholtz}, {Sharpe}, {Shivaei}, {Simmonds}, {Smit}, {Topping}, {{\"U}bler}, {Venturi}, {Williams}, {Witstok}, \& {Woodrum}}]{Eisenstein:2025}
{Eisenstein}, D.~J., {Johnson}, B.~D., {Robertson}, B., {et~al.} 2025, \apjs, 281, 50, \dodoi{10.3847/1538-4365/ae1137}

\bibitem[{{Eisenstein} {et~al.}(2026){Eisenstein}, {Willott}, {Alberts}, {Arribas}, {Bonaventura}, {Bunker}, {Cameron}, {Carniani}, {Charlot}, {Curtis-Lake}, {D'Eugenio}, {Ferruit}, {Giardino}, {Hainline}, {Hausen}, {Jakobsen}, {Johnson}, {Maiolino}, {Rauscher}, {Rieke}, {Rieke}, {Rix}, {Robertson}, {Stark}, {Tacchella}, {Williams}, {Willmer}, {Baker}, {Baum}, {Bhatawdekar}, {Boyett}, {Chen}, {Chevallard}, {Circosta}, {Curti}, {Danhaive}, {DeCoursey}, {Endsley}, {de Graaff}, {Dressler}, {Egami}, {Helton}, {Hviding}, {Ji}, {Jones}, {Kumari}, {L{\"u}tzgendorf}, {Laseter}, {Looser}, {Lyu}, {Maseda}, {Nelson}, {Parlanti}, {Perna}, {Pusk{\'a}s}, {Rawle}, {Rodr{\'\i}guez Del Pino}, {Rujopakarn}, {Sandles}, {Saxena}, {Scholtz}, {Sharpe}, {Shivaei}, {Silcock}, {Simmonds}, {Skarbinski}, {Smit}, {Stone}, {Suess}, {Sun}, {Tang}, {Topping}, {{\"U}bler}, {Villanueva}, {Wallace}, {Whitler}, {Witstok}, \& {Woodrum}}]{Eisenstein:2026}
{Eisenstein}, D.~J., {Willott}, C., {Alberts}, S., {et~al.} 2026, \apjs, 283, 6, \dodoi{10.3847/1538-4365/ae3163}

\bibitem[{{Endsley} {et~al.}(2023){Endsley}, {Stark}, {Whitler}, {Topping}, {Chen}, {Plat}, {Chisholm}, \& {Charlot}}]{Endsley:2023}
{Endsley}, R., {Stark}, D.~P., {Whitler}, L., {et~al.} 2023, \mnras, 524, 2312, \dodoi{10.1093/mnras/stad1919}

\bibitem[{{Endsley} {et~al.}(2024){Endsley}, {Stark}, {Whitler}, {Topping}, {Johnson}, {Robertson}, {Tacchella}, {Alberts}, {Baker}, {Bhatawdekar}, {Boyett}, {Bunker}, {Cameron}, {Carniani}, {Charlot}, {Chen}, {Chevallard}, {Curtis-Lake}, {Danhaive}, {Egami}, {Eisenstein}, {Hainline}, {Helton}, {Ji}, {Looser}, {Maiolino}, {Nelson}, {Pusk{\'a}s}, {Rieke}, {Rieke}, {Rix}, {Sandles}, {Saxena}, {Simmonds}, {Smit}, {Sun}, {Williams}, {Willmer}, {Willott}, \& {Witstok}}]{Endsley:2024}
---. 2024, \mnras, 533, 1111, \dodoi{10.1093/mnras/stae1857}

\bibitem[{{Ferland} {et~al.}(2013){Ferland}, {Porter}, {van Hoof}, {Williams}, {Abel}, {Lykins}, {Shaw}, {Henney}, \& {Stancil}}]{Ferland:2013}
{Ferland}, G.~J., {Porter}, R.~L., {van Hoof}, P.~A.~M., {et~al.} 2013, \rmxaa, 49, 137, \dodoi{10.48550/arXiv.1302.4485}

\bibitem[{{Finkelstein} {et~al.}(2022){Finkelstein}, {Bagley}, {Song}, {Larson}, {Papovich}, {Dickinson}, {Finkelstein}, {Koekemoer}, {Pirzkal}, {Somerville}, {Yung}, {Behroozi}, {Ferguson}, {Giavalisco}, {Grogin}, {Hathi}, {Hutchison}, {Jung}, {Kocevski}, {Kawinwanichakij}, {Rojas-Ruiz}, {Ryan}, {Snyder}, \& {Tacchella}}]{Finkelstein:2022}
{Finkelstein}, S.~L., {Bagley}, M., {Song}, M., {et~al.} 2022, \apj, 928, 52, \dodoi{10.3847/1538-4357/ac3aed}

\bibitem[{{Finkelstein} {et~al.}(2023){Finkelstein}, {Bagley}, {Ferguson}, {Wilkins}, {Kartaltepe}, {Papovich}, {Yung}, {Arrabal Haro}, {Behroozi}, {Dickinson}, {Kocevski}, {Koekemoer}, {Larson}, {Le Bail}, {Morales}, {P{\'e}rez-Gonz{\'a}lez}, {Burgarella}, {Dav{\'e}}, {Hirschmann}, {Somerville}, {Wuyts}, {Bromm}, {Casey}, {Fontana}, {Fujimoto}, {Gardner}, {Giavalisco}, {Grazian}, {Grogin}, {Hathi}, {Hutchison}, {Jha}, {Jogee}, {Kewley}, {Kirkpatrick}, {Long}, {Lotz}, {Pentericci}, {Pierel}, {Pirzkal}, {Ravindranath}, {Ryan}, {Trump}, {Yang}, {Bhatawdekar}, {Bisigello}, {Buat}, {Calabr{\`o}}, {Castellano}, {Cleri}, {Cooper}, {Croton}, {Daddi}, {Dekel}, {Elbaz}, {Franco}, {Gawiser}, {Holwerda}, {Huertas-Company}, {Jaskot}, {Leung}, {Lucas}, {Mobasher}, {Pandya}, {Tacchella}, {Weiner}, \& {Zavala}}]{Finkelstein:2023}
{Finkelstein}, S.~L., {Bagley}, M.~B., {Ferguson}, H.~C., {et~al.} 2023, \apjl, 946, L13, \dodoi{10.3847/2041-8213/acade4}

\bibitem[{{Foreman-Mackey}(2016)}]{Foreman-Mackey:2016}
{Foreman-Mackey}, D. 2016, The Journal of Open Source Software, 1, 24, \dodoi{10.21105/joss.00024}

\bibitem[{{Foreman-Mackey} {et~al.}(2014){Foreman-Mackey}, {Sick}, \& {Johnson}}]{Foreman-Mackey:2014}
{Foreman-Mackey}, D., {Sick}, J., \& {Johnson}, B. 2014, {python-fsps: Python bindings to FSPS (v0.1.1)}, v0.1.1, Zenodo,  Zenodo, \dodoi{10.5281/zenodo.12157}

\bibitem[{{Fujimoto} {et~al.}(2024){Fujimoto}, {Wang}, {Weaver}, {Kokorev}, {Atek}, {Bezanson}, {Labbe}, {Brammer}, {Greene}, {Chemerynska}, {Dayal}, {de Graaff}, {Furtak}, {Oesch}, {Setton}, {Price}, {Miller}, {Williams}, {Whitaker}, {Zitrin}, {Cutler}, {Leja}, {Pan}, {Coe}, {van Dokkum}, {Feldmann}, {Fudamoto}, {Goulding}, {Khullar}, {Marchesini}, {Maseda}, {Nanayakkara}, {Nelson}, {Smit}, {Stefanon}, \& {Weibel}}]{Fujimoto:2024}
{Fujimoto}, S., {Wang}, B., {Weaver}, J.~R., {et~al.} 2024, \apj, 977, 250, \dodoi{10.3847/1538-4357/ad9027}

\bibitem[{{Gaia Collaboration} {et~al.}(2023){Gaia Collaboration}, {Vallenari}, {Brown}, {Prusti}, {de Bruijne}, {Arenou}, {Babusiaux}, {Biermann}, {Creevey}, {Ducourant}, {Evans}, {Eyer}, {Guerra}, {Hutton}, {Jordi}, {Klioner}, {Lammers}, {Lindegren}, {Luri}, {Mignard}, {Panem}, {Pourbaix}, {Randich}, {Sartoretti}, {Soubiran}, {Tanga}, {Walton}, {Bailer-Jones}, {Bastian}, {Drimmel}, {Jansen}, {Katz}, {Lattanzi}, {van Leeuwen}, {Bakker}, {Cacciari}, {Casta{\~n}eda}, {De Angeli}, {Fabricius}, {Fouesneau}, {Fr{\'e}mat}, {Galluccio}, {Guerrier}, {Heiter}, {Masana}, {Messineo}, {Mowlavi}, {Nicolas}, {Nienartowicz}, {Pailler}, {Panuzzo}, {Riclet}, {Roux}, {Seabroke}, {Sordo}, {Th{\'e}venin}, {Gracia-Abril}, {Portell}, {Teyssier}, {Altmann}, {Andrae}, {Audard}, {Bellas-Velidis}, {Benson}, {Berthier}, {Blomme}, {Burgess}, {Busonero}, {Busso}, {C{\'a}novas}, {Carry}, {Cellino}, {Cheek}, {Clementini}, {Damerdji}, {Davidson}, {de Teodoro}, {Nu{\~n}ez Campos}, {Delchambre}, {Dell'Oro}, {Esquej},
  {Fern{\'a}ndez-Hern{\'a}ndez}, {Fraile}, {Garabato}, {Garc{\'\i}a-Lario}, {Gosset}, {Haigron}, {Halbwachs}, {Hambly}, {Harrison}, {Hern{\'a}ndez}, {Hestroffer}, {Hodgkin}, {Holl}, {Jan{\ss}en}, {Jevardat de Fombelle}, {Jordan}, {Krone-Martins}, {Lanzafame}, {L{\"o}ffler}, {Marchal}, {Marrese}, {Moitinho}, {Muinonen}, {Osborne}, {Pancino}, {Pauwels}, {Recio-Blanco}, {Reyl{\'e}}, {Riello}, {Rimoldini}, {Roegiers}, {Rybizki}, {Sarro}, {Siopis}, {Smith}, {Sozzetti}, {Utrilla}, {van Leeuwen}, {Abbas}, {{\'A}brah{\'a}m}, {Abreu Aramburu}, {Aerts}, {Aguado}, {Ajaj}, {Aldea-Montero}, {Altavilla}, {{\'A}lvarez}, {Alves}, {Anders}, {Anderson}, {Anglada Varela}, {Antoja}, {Baines}, {Baker}, {Balaguer-N{\'u}{\~n}ez}, {Balbinot}, {Balog}, {Barache}, {Barbato}, {Barros}, {Barstow}, {Bartolom{\'e}}, {Bassilana}, {Bauchet}, {Becciani}, {Bellazzini}, {Berihuete}, {Bernet}, {Bertone}, {Bianchi}, {Binnenfeld}, {Blanco-Cuaresma}, {Blazere}, {Boch}, {Bombrun}, {Bossini}, {Bouquillon}, {Bragaglia}, {Bramante}, {Breedt},
  {Bressan}, {Brouillet}, {Brugaletta}, {Bucciarelli}, {Burlacu}, {Butkevich}, {Buzzi}, {Caffau}, {Cancelliere}, {Cantat-Gaudin}, {Carballo}, {Carlucci}, {Carnerero}, {Carrasco}, {Casamiquela}, {Castellani}, {Castro-Ginard}, {Chaoul}, {Charlot}, {Chemin}, {Chiaramida}, {Chiavassa}, {Chornay}, {Comoretto}, {Contursi}, {Cooper}, {Cornez}, {Cowell}, {Crifo}, {Cropper}, {Crosta}, {Crowley}, {Dafonte}, {Dapergolas}, {David}, {David}, {de Laverny}, {De Luise}, {De March}, {De Ridder}, {de Souza}, {de Torres}, {del Peloso}, {del Pozo}, {Delbo}, {Delgado}, {Delisle}, {Demouchy}, {Dharmawardena}, {Di Matteo}, {Diakite}, {Diener}, {Distefano}, {Dolding}, {Edvardsson}, {Enke}, {Fabre}, {Fabrizio}, {Faigler}, {Fedorets}, {Fernique}, {Fienga}, {Figueras}, {Fournier}, {Fouron}, {Fragkoudi}, {Gai}, {Garcia-Gutierrez}, {Garcia-Reinaldos}, {Garc{\'\i}a-Torres}, {Garofalo}, {Gavel}, {Gavras}, {Gerlach}, {Geyer}, {Giacobbe}, {Gilmore}, {Girona}, {Giuffrida}, {Gomel}, {Gomez}, {Gonz{\'a}lez-N{\'u}{\~n}ez},
  {Gonz{\'a}lez-Santamar{\'\i}a}, {Gonz{\'a}lez-Vidal}, {Granvik}, {Guillout}, {Guiraud}, {Guti{\'e}rrez-S{\'a}nchez}, {Guy}, {Hatzidimitriou}, {Hauser}, {Haywood}, {Helmer}, {Helmi}, {Sarmiento}, {Hidalgo}, {Hilger}, {H{\l}adczuk}, {Hobbs}, {Holland}, {Huckle}, {Jardine}, {Jasniewicz}, {Jean-Antoine Piccolo}, {Jim{\'e}nez-Arranz}, {Jorissen}, {Juaristi Campillo}, {Julbe}, {Karbevska}, {Kervella}, {Khanna}, {Kontizas}, {Kordopatis}, {Korn}, {K{\'o}sp{\'a}l}, {Kostrzewa-Rutkowska}, {Kruszy{\'n}ska}, {Kun}, {Laizeau}, {Lambert}, {Lanza}, {Lasne}, {Le Campion}, {Lebreton}, {Lebzelter}, {Leccia}, {Leclerc}, {Lecoeur-Taibi}, {Liao}, {Licata}, {Lindstr{\o}m}, {Lister}, {Livanou}, {Lobel}, {Lorca}, {Loup}, {Madrero Pardo}, {Magdaleno Romeo}, {Managau}, {Mann}, {Manteiga}, {Marchant}, {Marconi}, {Marcos}, {Marcos Santos}, {Mar{\'\i}n Pina}, {Marinoni}, {Marocco}, {Marshall}, {Martin Polo}, {Mart{\'\i}n-Fleitas}, {Marton}, {Mary}, {Masip}, {Massari}, {Mastrobuono-Battisti}, {Mazeh}, {McMillan}, {Messina}, {Michalik},
  {Millar}, {Mints}, {Molina}, {Molinaro}, {Moln{\'a}r}, {Monari}, {Mongui{\'o}}, {Montegriffo}, {Montero}, {Mor}, {Mora}, {Morbidelli}, {Morel}, {Morris}, {Muraveva}, {Murphy}, {Musella}, {Nagy}, {Noval}, {Oca{\~n}a}, {Ogden}, {Ordenovic}, {Osinde}, {Pagani}, {Pagano}, {Palaversa}, {Palicio}, {Pallas-Quintela}, {Panahi}, {Payne-Wardenaar}, {Pe{\~n}alosa Esteller}, {Penttil{\"a}}, {Pichon}, {Piersimoni}, {Pineau}, {Plachy}, {Plum}, {Poggio}, {Pr{\v{s}}a}, {Pulone}, {Racero}, {Ragaini}, {Rainer}, {Raiteri}, {Rambaux}, {Ramos}, {Ramos-Lerate}, {Re Fiorentin}, {Regibo}, {Richards}, {Rios Diaz}, {Ripepi}, {Riva}, {Rix}, {Rixon}, {Robichon}, {Robin}, {Robin}, {Roelens}, {Rogues}, {Rohrbasser}, {Romero-G{\'o}mez}, {Rowell}, {Royer}, {Ruz Mieres}, {Rybicki}, {Sadowski}, {S{\'a}ez N{\'u}{\~n}ez}, {Sagrist{\`a} Sell{\'e}s}, {Sahlmann}, {Salguero}, {Samaras}, {Sanchez Gimenez}, {Sanna}, {Santove{\~n}a}, {Sarasso}, {Schultheis}, {Sciacca}, {Segol}, {Segovia}, {S{\'e}gransan}, {Semeux}, {Shahaf}, {Siddiqui}, {Siebert},
  {Siltala}, {Silvelo}, {Slezak}, {Slezak}, {Smart}, {Snaith}, {Solano}, {Solitro}, {Souami}, {Souchay}, {Spagna}, {Spina}, {Spoto}, {Steele}, {Steidelm{\"u}ller}, {Stephenson}, {S{\"u}veges}, {Surdej}, {Szabados}, {Szegedi-Elek}, {Taris}, {Taylor}, {Teixeira}, {Tolomei}, {Tonello}, {Torra}, {Torra}, {Torralba Elipe}, {Trabucchi}, {Tsounis}, {Turon}, {Ulla}, {Unger}, {Vaillant}, {van Dillen}, {van Reeven}, {Vanel}, {Vecchiato}, {Viala}, {Vicente}, {Voutsinas}, {Weiler}, {Wevers}, {Wyrzykowski}, {Yoldas}, {Yvard}, {Zhao}, {Zorec}, {Zucker}, \& {Zwitter}}]{GaiaDR3}
{Gaia Collaboration}, {Vallenari}, A., {Brown}, A.~G.~A., {et~al.} 2023, \aap, 674, A1, \dodoi{10.1051/0004-6361/202243940}

\bibitem[{{Giavalisco} {et~al.}(2004){Giavalisco}, {Ferguson}, {Koekemoer}, {Dickinson}, {Alexander}, {Bauer}, {Bergeron}, {Biagetti}, {Brandt}, {Casertano}, {Cesarsky}, {Chatzichristou}, {Conselice}, {Cristiani}, {Da Costa}, {Dahlen}, {de Mello}, {Eisenhardt}, {Erben}, {Fall}, {Fassnacht}, {Fosbury}, {Fruchter}, {Gardner}, {Grogin}, {Hook}, {Hornschemeier}, {Idzi}, {Jogee}, {Kretchmer}, {Laidler}, {Lee}, {Livio}, {Lucas}, {Madau}, {Mobasher}, {Moustakas}, {Nonino}, {Padovani}, {Papovich}, {Park}, {Ravindranath}, {Renzini}, {Richardson}, {Riess}, {Rosati}, {Schirmer}, {Schreier}, {Somerville}, {Spinrad}, {Stern}, {Stiavelli}, {Strolger}, {Urry}, {Vandame}, {Williams}, \& {Wolf}}]{Giavalisco:2004}
{Giavalisco}, M., {Ferguson}, H.~C., {Koekemoer}, A.~M., {et~al.} 2004, \apjl, 600, L93, \dodoi{10.1086/379232}

\bibitem[{{Ginsburg} {et~al.}(2019){Ginsburg}, {Sip{\H{o}}cz}, {Brasseur}, {Cowperthwaite}, {Craig}, {Deil}, {Guillochon}, {Guzman}, {Liedtke}, {Lian Lim}, {Lockhart}, {Mommert}, {Morris}, {Norman}, {Parikh}, {Persson}, {Robitaille}, {Segovia}, {Singer}, {Tollerud}, {de Val-Borro}, {Valtchanov}, {Woillez}, {Astroquery Collaboration}, \& {a subset of astropy Collaboration}}]{Ginsburg:2019}
{Ginsburg}, A., {Sip{\H{o}}cz}, B.~M., {Brasseur}, C.~E., {et~al.} 2019, \aj, 157, 98, \dodoi{10.3847/1538-3881/aafc33}

\bibitem[{{Grazian} {et~al.}(2012){Grazian}, {Castellano}, {Fontana}, {Pentericci}, {Dunlop}, {McLure}, {Koekemoer}, {Dickinson}, {Faber}, {Ferguson}, {Galametz}, {Giavalisco}, {Grogin}, {Hathi}, {Kocevski}, {Lai}, {Newman}, \& {Vanzella}}]{Grazian:2012}
{Grazian}, A., {Castellano}, M., {Fontana}, A., {et~al.} 2012, \aap, 547, A51, \dodoi{10.1051/0004-6361/201219669}

\bibitem[{{Grazian} {et~al.}(2015){Grazian}, {Fontana}, {Santini}, {Dunlop}, {Ferguson}, {Castellano}, {Amorin}, {Ashby}, {Barro}, {Behroozi}, {Boutsia}, {Caputi}, {Chary}, {Dekel}, {Dickinson}, {Faber}, {Fazio}, {Finkelstein}, {Galametz}, {Giallongo}, {Giavalisco}, {Grogin}, {Guo}, {Kocevski}, {Koekemoer}, {Koo}, {Lee}, {Lu}, {Merlin}, {Mobasher}, {Nonino}, {Papovich}, {Paris}, {Pentericci}, {Reddy}, {Renzini}, {Salmon}, {Salvato}, {Sommariva}, {Song}, \& {Vanzella}}]{Grazian:2015}
{Grazian}, A., {Fontana}, A., {Santini}, P., {et~al.} 2015, \aap, 575, A96, \dodoi{10.1051/0004-6361/201424750}

\bibitem[{{Hainline} {et~al.}(2024){Hainline}, {Johnson}, {Robertson}, {Tacchella}, {Helton}, {Sun}, {Eisenstein}, {Simmonds}, {Topping}, {Whitler}, {Willmer}, {Rieke}, {Suess}, {Hviding}, {Cameron}, {Alberts}, {Baker}, {Baum}, {Bhatawdekar}, {Bonaventura}, {Boyett}, {Bunker}, {Carniani}, {Charlot}, {Chevallard}, {Chen}, {Curti}, {Curtis-Lake}, {D'Eugenio}, {Egami}, {Endsley}, {Hausen}, {Ji}, {Looser}, {Lyu}, {Maiolino}, {Nelson}, {Pusk{\'a}s}, {Rawle}, {Sandles}, {Saxena}, {Smit}, {Stark}, {Williams}, {Willott}, \& {Witstok}}]{Hainline:2024}
{Hainline}, K.~N., {Johnson}, B.~D., {Robertson}, B., {et~al.} 2024, \apj, 964, 71, \dodoi{10.3847/1538-4357/ad1ee4}

\bibitem[{{Harris} {et~al.}(2020){Harris}, {Millman}, {van der Walt}, {Gommers}, {Virtanen}, {Cournapeau}, {Wieser}, {Taylor}, {Berg}, {Smith}, {Kern}, {Picus}, {Hoyer}, {van Kerkwijk}, {Brett}, {Haldane}, {del R{\'\i}o}, {Wiebe}, {Peterson}, {G{\'e}rard-Marchant}, {Sheppard}, {Reddy}, {Weckesser}, {Abbasi}, {Gohlke}, \& {Oliphant}}]{NumPy:2020}
{Harris}, C.~R., {Millman}, K.~J., {van der Walt}, S.~J., {et~al.} 2020, \nat, 585, 357, \dodoi{10.1038/s41586-020-2649-2}

\bibitem[{{Helton} {et~al.}(2024){Helton}, {Sun}, {Woodrum}, {Hainline}, {Willmer}, {Rieke}, {Rieke}, {Alberts}, {Eisenstein}, {Tacchella}, {Robertson}, {Johnson}, {Baker}, {Bhatawdekar}, {Bunker}, {Chen}, {Egami}, {Ji}, {Maiolino}, {Willott}, \& {Witstok}}]{Helton:2024b}
{Helton}, J.~M., {Sun}, F., {Woodrum}, C., {et~al.} 2024, \apj, 974, 41, \dodoi{10.3847/1538-4357/ad6867}

\bibitem[{{Helton} {et~al.}(2025){Helton}, {Rieke}, {Alberts}, {Wu}, {Eisenstein}, {Hainline}, {Carniani}, {Ji}, {Baker}, {Bhatawdekar}, {Bunker}, {Cargile}, {Charlot}, {Chevallard}, {D'Eugenio}, {Egami}, {Johnson}, {Jones}, {Lyu}, {Maiolino}, {P{\'e}rez-Gonz{\'a}lez}, {Rieke}, {Robertson}, {Saxena}, {Scholtz}, {Shivaei}, {Sun}, {Tacchella}, {Whitler}, {Williams}, {Willmer}, {Willott}, {Witstok}, \& {Zhu}}]{Helton:2025}
{Helton}, J.~M., {Rieke}, G.~H., {Alberts}, S., {et~al.} 2025, Nature Astronomy, \dodoi{10.1038/s41550-025-02503-z}

\bibitem[{{Hunter}(2007)}]{Matplotlib:2007}
{Hunter}, J.~D. 2007, Computing in Science and Engineering, 9, 90, \dodoi{10.1109/MCSE.2007.55}

\bibitem[{{Iani} {et~al.}(2024){Iani}, {Caputi}, {Rinaldi}, {Annunziatella}, {Boogaard}, {{\"O}stlin}, {Costantin}, {Gillman}, {P{\'e}rez-Gonz{\'a}lez}, {Colina}, {Greve}, {Wright}, {Alonso-Herrero}, {{\'A}lvarez-M{\'a}rquez}, {Bik}, {Bosman}, {Crespo G{\'o}mez}, {Eckart}, {Hjorth}, {Jermann}, {Labiano}, {Langeroodi}, {Melinder}, {Moutard}, {Pei{\ss}ker}, {Pye}, {Tikkanen}, {van der Werf}, {Walter}, {Henning}, {Lagage}, \& {van Dishoeck}}]{Iani:2024}
{Iani}, E., {Caputi}, K.~I., {Rinaldi}, P., {et~al.} 2024, \apj, 963, 97, \dodoi{10.3847/1538-4357/ad15f6}

\bibitem[{{Illingworth}(2015)}]{HLF_DOI}
{Illingworth}, G. 2015, Hubble Legacy Fields (HLF),  STScI/MAST, \dodoi{10.17909/T91019}

\bibitem[{{Illingworth} {et~al.}(2016){Illingworth}, {Magee}, {Bouwens}, {Oesch}, {Labbe}, {van Dokkum}, {Whitaker}, {Holden}, {Franx}, \& {Gonzalez}}]{Illingworth:2016}
{Illingworth}, G., {Magee}, D., {Bouwens}, R., {et~al.} 2016, arXiv e-prints, arXiv:1606.00841, \dodoi{10.48550/arXiv.1606.00841}

\bibitem[{{Ji} {et~al.}(2024){Ji}, {Williams}, {Tacchella}, {Suess}, {Baker}, {Alberts}, {Bunker}, {Johnson}, {Robertson}, {Sun}, {Eisenstein}, {Rieke}, {Maseda}, {Hainline}, {Hausen}, {Rieke}, {Willmer}, {Egami}, {Shivaei}, {Carniani}, {Charlot}, {Chevallard}, {Curtis-Lake}, {Looser}, {Maiolino}, {Willott}, {Chen}, {Helton}, {Lyu}, {Nelson}, {Bhatawdekar}, {Boyett}, \& {Sandles}}]{Ji:2024}
{Ji}, Z., {Williams}, C.~C., {Tacchella}, S., {et~al.} 2024, \apj, 974, 135, \dodoi{10.3847/1538-4357/ad6e7f}

\bibitem[{{Johnson} {et~al.}(2021){Johnson}, {Leja}, {Conroy}, \& {Speagle}}]{Johnson:2021}
{Johnson}, B.~D., {Leja}, J., {Conroy}, C., \& {Speagle}, J.~S. 2021, \apjs, 254, 22, \dodoi{10.3847/1538-4365/abef67}

\bibitem[{{Kriek} \& {Conroy}(2013)}]{Kriek:2013}
{Kriek}, M., \& {Conroy}, C. 2013, \apjl, 775, L16, \dodoi{10.1088/2041-8205/775/1/L16}

\bibitem[{{Krist} {et~al.}(2011){Krist}, {Hook}, \& {Stoehr}}]{Krist:2011}
{Krist}, J.~E., {Hook}, R.~N., \& {Stoehr}, F. 2011, in Society of Photo-Optical Instrumentation Engineers (SPIE) Conference Series, Vol. 8127, Optical Modeling and Performance Predictions V, ed. M.~A. {Kahan}, 81270J, \dodoi{10.1117/12.892762}

\bibitem[{{Labb{\'e}} {et~al.}(2005){Labb{\'e}}, {Huang}, {Franx}, {Rudnick}, {Barmby}, {Daddi}, {van Dokkum}, {Fazio}, {F{\"o}rster Schreiber}, {Moorwood}, {Rix}, {R{\"o}ttgering}, {Trujillo}, \& {van der Werf}}]{Labbe:2005a}
{Labb{\'e}}, I., {Huang}, J., {Franx}, M., {et~al.} 2005, \apjl, 624, L81, \dodoi{10.1086/430700}

\bibitem[{{Madau}(1995)}]{Madau:1995}
{Madau}, P. 1995, \apj, 441, 18, \dodoi{10.1086/175332}

\bibitem[{{McLure} {et~al.}(2010){McLure}, {Dunlop}, {Cirasuolo}, {Koekemoer}, {Sabbi}, {Stark}, {Targett}, \& {Ellis}}]{McLure:2010}
{McLure}, R.~J., {Dunlop}, J.~S., {Cirasuolo}, M., {et~al.} 2010, \mnras, 403, 960, \dodoi{10.1111/j.1365-2966.2009.16176.x}

\bibitem[{{McLure} {et~al.}(2013){McLure}, {Dunlop}, {Bowler}, {Curtis-Lake}, {Schenker}, {Ellis}, {Robertson}, {Koekemoer}, {Rogers}, {Ono}, {Ouchi}, {Charlot}, {Wild}, {Stark}, {Furlanetto}, {Cirasuolo}, \& {Targett}}]{McLure:2013}
{McLure}, R.~J., {Dunlop}, J.~S., {Bowler}, R.~A.~A., {et~al.} 2013, \mnras, 432, 2696, \dodoi{10.1093/mnras/stt627}

\bibitem[{{Narayanan} {et~al.}(2024){Narayanan}, {Lower}, {Torrey}, {Brammer}, {Cui}, {Dav{\'e}}, {Iyer}, {Li}, {Lovell}, {Sales}, {Stark}, {Marinacci}, \& {Vogelsberger}}]{Narayanan:2024}
{Narayanan}, D., {Lower}, S., {Torrey}, P., {et~al.} 2024, \apj, 961, 73, \dodoi{10.3847/1538-4357/ad0966}

\bibitem[{{Navarro-Carrera} {et~al.}(2024){Navarro-Carrera}, {Rinaldi}, {Caputi}, {Iani}, {Kokorev}, \& {van Mierlo}}]{Navarro-Carrera:2024}
{Navarro-Carrera}, R., {Rinaldi}, P., {Caputi}, K.~I., {et~al.} 2024, \apj, 961, 207, \dodoi{10.3847/1538-4357/ad0df6}

\bibitem[{{Oesch} {et~al.}(2010){Oesch}, {Bouwens}, {Illingworth}, {Carollo}, {Franx}, {Labb{\'e}}, {Magee}, {Stiavelli}, {Trenti}, \& {van Dokkum}}]{Oesch:2010}
{Oesch}, P.~A., {Bouwens}, R.~J., {Illingworth}, G.~D., {et~al.} 2010, \apjl, 709, L16, \dodoi{10.1088/2041-8205/709/1/L16}

\bibitem[{{Oesch} {et~al.}(2012){Oesch}, {Bouwens}, {Illingworth}, {Gonzalez}, {Trenti}, {van Dokkum}, {Franx}, {Labb{\'e}}, {Carollo}, \& {Magee}}]{Oesch:2012}
---. 2012, \apj, 759, 135, \dodoi{10.1088/0004-637X/759/2/135}

\bibitem[{{Oesch} {et~al.}(2014){Oesch}, {Bouwens}, {Illingworth}, {Labb{\'e}}, {Smit}, {Franx}, {van Dokkum}, {Momcheva}, {Ashby}, {Fazio}, {Huang}, {Willner}, {Gonzalez}, {Magee}, {Trenti}, {Brammer}, {Skelton}, \& {Spitler}}]{Oesch:2014}
---. 2014, \apj, 786, 108, \dodoi{10.1088/0004-637X/786/2/108}

\bibitem[{{Oesch} {et~al.}(2023){Oesch}, {Brammer}, {Naidu}, {Bouwens}, {Chisholm}, {Illingworth}, {Matthee}, {Nelson}, {Qin}, {Reddy}, {Shapley}, {Shivaei}, {van Dokkum}, {Weibel}, {Whitaker}, {Wuyts}, {Covelo-Paz}, {Endsley}, {Fudamoto}, {Giovinazzo}, {Herard-Demanche}, {Kerutt}, {Kramarenko}, {Labbe}, {Leonova}, {Lin}, {Magee}, {Marchesini}, {Maseda}, {Mason}, {Matharu}, {Meyer}, {Neufeld}, {Prieto Lyon}, {Schaerer}, {Sharma}, {Shuntov}, {Smit}, {Stefanon}, {Wyithe}, \& {Xiao}}]{Oesch:2023}
{Oesch}, P.~A., {Brammer}, G., {Naidu}, R.~P., {et~al.} 2023, \mnras, 525, 2864, \dodoi{10.1093/mnras/stad2411}

\bibitem[{{Oke} \& {Gunn}(1983)}]{Oke:1983}
{Oke}, J.~B., \& {Gunn}, J.~E. 1983, \apj, 266, 713, \dodoi{10.1086/160817}

\bibitem[{{Papovich} {et~al.}(2023){Papovich}, {Cole}, {Yang}, {Finkelstein}, {Barro}, {Buat}, {Burgarella}, {P{\'e}rez-Gonz{\'a}lez}, {Santini}, {Seill{\'e}}, {Shen}, {Arrabal Haro}, {Bagley}, {Bell}, {Bisigello}, {Calabr{\`o}}, {Casey}, {Castellano}, {Chworowsky}, {Cleri}, {Costantin}, {Cooper}, {Dickinson}, {Ferguson}, {Fontana}, {Giavalisco}, {Grazian}, {Grogin}, {Hathi}, {Holwerda}, {Hutchison}, {Kartaltepe}, {Kewley}, {Kirkpatrick}, {Kocevski}, {Koekemoer}, {Larson}, {Long}, {Lucas}, {Pentericci}, {Pirzkal}, {Ravindranath}, {Somerville}, {Trump}, {Urbano Stawinski}, {Weiner}, {Wilkins}, {Yung}, \& {Zavala}}]{Papovich:2023}
{Papovich}, C., {Cole}, J.~W., {Yang}, G., {et~al.} 2023, \apjl, 949, L18, \dodoi{10.3847/2041-8213/acc948}

\bibitem[{{Paxton} {et~al.}(2011){Paxton}, {Bildsten}, {Dotter}, {Herwig}, {Lesaffre}, \& {Timmes}}]{Paxton:2011}
{Paxton}, B., {Bildsten}, L., {Dotter}, A., {et~al.} 2011, \apjs, 192, 3, \dodoi{10.1088/0067-0049/192/1/3}

\bibitem[{{Paxton} {et~al.}(2013){Paxton}, {Cantiello}, {Arras}, {Bildsten}, {Brown}, {Dotter}, {Mankovich}, {Montgomery}, {Stello}, {Timmes}, \& {Townsend}}]{Paxton:2013}
{Paxton}, B., {Cantiello}, M., {Arras}, P., {et~al.} 2013, \apjs, 208, 4, \dodoi{10.1088/0067-0049/208/1/4}

\bibitem[{{Paxton} {et~al.}(2015){Paxton}, {Marchant}, {Schwab}, {Bauer}, {Bildsten}, {Cantiello}, {Dessart}, {Farmer}, {Hu}, {Langer}, {Townsend}, {Townsley}, \& {Timmes}}]{Paxton:2015}
{Paxton}, B., {Marchant}, P., {Schwab}, J., {et~al.} 2015, \apjs, 220, 15, \dodoi{10.1088/0067-0049/220/1/15}

\bibitem[{{Paxton} {et~al.}(2018){Paxton}, {Schwab}, {Bauer}, {Bildsten}, {Blinnikov}, {Duffell}, {Farmer}, {Goldberg}, {Marchant}, {Sorokina}, {Thoul}, {Townsend}, \& {Timmes}}]{Paxton:2018}
{Paxton}, B., {Schwab}, J., {Bauer}, E.~B., {et~al.} 2018, \apjs, 234, 34, \dodoi{10.3847/1538-4365/aaa5a8}

\bibitem[{{Perrin} {et~al.}(2014){Perrin}, {Sivaramakrishnan}, {Lajoie}, {Elliott}, {Pueyo}, {Ravindranath}, \& {Albert}}]{Perrin:2014}
{Perrin}, M.~D., {Sivaramakrishnan}, A., {Lajoie}, C.-P., {et~al.} 2014, in Society of Photo-Optical Instrumentation Engineers (SPIE) Conference Series, Vol. 9143, Space Telescopes and Instrumentation 2014: Optical, Infrared, and Millimeter Wave, ed. J.~{Oschmann}, Jacobus~M., M.~{Clampin}, G.~G. {Fazio}, \& H.~A. {MacEwen}, 91433X, \dodoi{10.1117/12.2056689}

\bibitem[{{Planck Collaboration} {et~al.}(2020){Planck Collaboration}, {Aghanim}, {Akrami}, {Ashdown}, {Aumont}, {Baccigalupi}, {Ballardini}, {Banday}, {Barreiro}, {Bartolo}, {Basak}, {Battye}, {Benabed}, {Bernard}, {Bersanelli}, {Bielewicz}, {Bock}, {Bond}, {Borrill}, {Bouchet}, {Boulanger}, {Bucher}, {Burigana}, {Butler}, {Calabrese}, {Cardoso}, {Carron}, {Challinor}, {Chiang}, {Chluba}, {Colombo}, {Combet}, {Contreras}, {Crill}, {Cuttaia}, {de Bernardis}, {de Zotti}, {Delabrouille}, {Delouis}, {Di Valentino}, {Diego}, {Dor{\'e}}, {Douspis}, {Ducout}, {Dupac}, {Dusini}, {Efstathiou}, {Elsner}, {En{\ss}lin}, {Eriksen}, {Fantaye}, {Farhang}, {Fergusson}, {Fernandez-Cobos}, {Finelli}, {Forastieri}, {Frailis}, {Fraisse}, {Franceschi}, {Frolov}, {Galeotta}, {Galli}, {Ganga}, {G{\'e}nova-Santos}, {Gerbino}, {Ghosh}, {Gonz{\'a}lez-Nuevo}, {G{\'o}rski}, {Gratton}, {Gruppuso}, {Gudmundsson}, {Hamann}, {Handley}, {Hansen}, {Herranz}, {Hildebrandt}, {Hivon}, {Huang}, {Jaffe}, {Jones}, {Karakci}, {Keih{\"a}nen},
  {Keskitalo}, {Kiiveri}, {Kim}, {Kisner}, {Knox}, {Krachmalnicoff}, {Kunz}, {Kurki-Suonio}, {Lagache}, {Lamarre}, {Lasenby}, {Lattanzi}, {Lawrence}, {Le Jeune}, {Lemos}, {Lesgourgues}, {Levrier}, {Lewis}, {Liguori}, {Lilje}, {Lilley}, {Lindholm}, {L{\'o}pez-Caniego}, {Lubin}, {Ma}, {Mac{\'\i}as-P{\'e}rez}, {Maggio}, {Maino}, {Mandolesi}, {Mangilli}, {Marcos-Caballero}, {Maris}, {Martin}, {Martinelli}, {Mart{\'\i}nez-Gonz{\'a}lez}, {Matarrese}, {Mauri}, {McEwen}, {Meinhold}, {Melchiorri}, {Mennella}, {Migliaccio}, {Millea}, {Mitra}, {Miville-Desch{\^e}nes}, {Molinari}, {Montier}, {Morgante}, {Moss}, {Natoli}, {N{\o}rgaard-Nielsen}, {Pagano}, {Paoletti}, {Partridge}, {Patanchon}, {Peiris}, {Perrotta}, {Pettorino}, {Piacentini}, {Polastri}, {Polenta}, {Puget}, {Rachen}, {Reinecke}, {Remazeilles}, {Renzi}, {Rocha}, {Rosset}, {Roudier}, {Rubi{\~n}o-Mart{\'\i}n}, {Ruiz-Granados}, {Salvati}, {Sandri}, {Savelainen}, {Scott}, {Shellard}, {Sirignano}, {Sirri}, {Spencer}, {Sunyaev}, {Suur-Uski}, {Tauber}, {Tavagnacco},
  {Tenti}, {Toffolatti}, {Tomasi}, {Trombetti}, {Valenziano}, {Valiviita}, {Van Tent}, {Vibert}, {Vielva}, {Villa}, {Vittorio}, {Wandelt}, {Wehus}, {White}, {White}, {Zacchei}, \& {Zonca}}]{Planck:2020}
{Planck Collaboration}, {Aghanim}, N., {Akrami}, Y., {et~al.} 2020, \aap, 641, A6, \dodoi{10.1051/0004-6361/201833910}

\bibitem[{{Popesso} {et~al.}(2023){Popesso}, {Concas}, {Cresci}, {Belli}, {Rodighiero}, {Inami}, {Dickinson}, {Ilbert}, {Pannella}, \& {Elbaz}}]{Popesso:2023}
{Popesso}, P., {Concas}, A., {Cresci}, G., {et~al.} 2023, \mnras, 519, 1526, \dodoi{10.1093/mnras/stac3214}

\bibitem[{{Quadri} {et~al.}(2007){Quadri}, {Marchesini}, {van Dokkum}, {Gawiser}, {Franx}, {Lira}, {Rudnick}, {Urry}, {Maza}, {Kriek}, {Barrientos}, {Blanc}, {Castander}, {Christlein}, {Coppi}, {Hall}, {Herrera}, {Infante}, {Taylor}, {Treister}, \& {Willis}}]{Quadri:2007}
{Quadri}, R., {Marchesini}, D., {van Dokkum}, P., {et~al.} 2007, \aj, 134, 1103, \dodoi{10.1086/520330}

\bibitem[{{Rieke} {et~al.}(2023{\natexlab{a}}){Rieke}, {Robertson}, {Tacchella}, {Willmer}, {Johnson}, {Carniani}, \& {Bunker}}]{JADES_DR2_DOI}
{Rieke}, M.~J., {Robertson}, B., {Tacchella}, S., {et~al.} 2023{\natexlab{a}}, Data from the JWST Advanced Deep Extragalactic Survey (JADES) DR2,  STScI/MAST, \dodoi{10.17909/Z2GW-MK31}

\bibitem[{{Rieke} {et~al.}(2023{\natexlab{b}}){Rieke}, {Robertson}, {Tacchella}, {Hainline}, {Johnson}, {Hausen}, {Ji}, {Willmer}, {Eisenstein}, {Pusk{\'a}s}, {Alberts}, {Arribas}, {Baker}, {Baum}, {Bhatawdekar}, {Bonaventura}, {Boyett}, {Bunker}, {Cameron}, {Carniani}, {Charlot}, {Chevallard}, {Chen}, {Curti}, {Curtis-Lake}, {Danhaive}, {DeCoursey}, {Dressler}, {Egami}, {Endsley}, {Helton}, {Hviding}, {Kumari}, {Looser}, {Lyu}, {Maiolino}, {Maseda}, {Nelson}, {Rieke}, {Rix}, {Sandles}, {Saxena}, {Sharpe}, {Shivaei}, {Skarbinski}, {Smit}, {Stark}, {Stone}, {Suess}, {Sun}, {Topping}, {{\"U}bler}, {Villanueva}, {Wallace}, {Williams}, {Willott}, {Whitler}, {Witstok}, \& {Woodrum}}]{Rieke:2023}
---. 2023{\natexlab{b}}, \apjs, 269, 16, \dodoi{10.3847/1538-4365/acf44d}

\bibitem[{{Rinaldi} {et~al.}(2025){Rinaldi}, {P{\'e}rez-Gonz{\'a}lez}, {Rieke}, {Lyu}, {D'Eugenio}, {Wu}, {Carniani}, {Looser}, {Shivaei}, {Boogaard}, {Diaz-Santos}, {Colina}, {{\"O}stlin}, {Alberts}, {{\'A}lvarez-M{\'a}rquez}, {Annuziatella}, {Aravena}, {Bhatawdekar}, {Bunker}, {Caputi}, {Charlot}, {Crespo G{\'o}mez}, {Curti}, {Eckart}, {Gillman}, {Hainline}, {Kumari}, {Hjorth}, {Iani}, {Inami}, {Ji}, {Johnson}, {Jones}, {Labiano}, {Maiolino}, {Melinder}, {Moutard}, {Peissker}, {Rieke}, {Robertson}, {Scholtz}, {Tacchella}, {van der Werf}, {Walter}, {Williams}, {Willott}, {Witstok}, {{\"U}bler}, \& {Zhu}}]{Rinaldi:2025}
{Rinaldi}, P., {P{\'e}rez-Gonz{\'a}lez}, P.~G., {Rieke}, G.~H., {et~al.} 2025, \apj, 994, 86, \dodoi{10.3847/1538-4357/ae089c}

\bibitem[{{Robertson} {et~al.}(2024){Robertson}, {Johnson}, {Tacchella}, {Eisenstein}, {Hainline}, {Arribas}, {Baker}, {Bunker}, {Carniani}, {Cargile}, {Carreira}, {Charlot}, {Chevallard}, {Curti}, {Curtis-Lake}, {D'Eugenio}, {Egami}, {Hausen}, {Helton}, {Jakobsen}, {Ji}, {Jones}, {Maiolino}, {Maseda}, {Nelson}, {P{\'e}rez-Gonz{\'a}lez}, {Pusk{\'a}s}, {Rieke}, {Smit}, {Sun}, {{\"U}bler}, {Whitler}, {Williams}, {Willmer}, {Willott}, \& {Witstok}}]{Robertson:2024}
{Robertson}, B., {Johnson}, B.~D., {Tacchella}, S., {et~al.} 2024, \apj, 970, 31, \dodoi{10.3847/1538-4357/ad463d}

\bibitem[{{Schaerer} \& {de Barros}(2009)}]{Schaerer:2009}
{Schaerer}, D., \& {de Barros}, S. 2009, \aap, 502, 423, \dodoi{10.1051/0004-6361/200911781}

\bibitem[{{Simmonds} {et~al.}(2024){Simmonds}, {Tacchella}, {Hainline}, {Johnson}, {McClymont}, {Robertson}, {Saxena}, {Sun}, {Witten}, {Baker}, {Bhatawdekar}, {Boyett}, {Bunker}, {Charlot}, {Curtis-Lake}, {Egami}, {Eisenstein}, {Hausen}, {Maiolino}, {Maseda}, {Scholtz}, {Williams}, {Willott}, \& {Witstok}}]{Simmonds:2024}
{Simmonds}, C., {Tacchella}, S., {Hainline}, K., {et~al.} 2024, \mnras, 527, 6139, \dodoi{10.1093/mnras/stad3605}

\bibitem[{{Smit} {et~al.}(2014){Smit}, {Bouwens}, {Labb{\'e}}, {Zheng}, {Bradley}, {Donahue}, {Lemze}, {Moustakas}, {Umetsu}, {Zitrin}, {Coe}, {Postman}, {Gonzalez}, {Bartelmann}, {Ben{\'\i}tez}, {Broadhurst}, {Ford}, {Grillo}, {Infante}, {Jimenez-Teja}, {Jouvel}, {Kelson}, {Lahav}, {Maoz}, {Medezinski}, {Melchior}, {Meneghetti}, {Merten}, {Molino}, {Moustakas}, {Nonino}, {Rosati}, \& {Seitz}}]{Smit:2014}
{Smit}, R., {Bouwens}, R.~J., {Labb{\'e}}, I., {et~al.} 2014, \apj, 784, 58, \dodoi{10.1088/0004-637X/784/1/58}

\bibitem[{{Speagle}(2020)}]{Speagle:2020}
{Speagle}, J.~S. 2020, \mnras, 493, 3132, \dodoi{10.1093/mnras/staa278}

\bibitem[{{Stark} {et~al.}(2013){Stark}, {Schenker}, {Ellis}, {Robertson}, {McLure}, \& {Dunlop}}]{Stark:2013}
{Stark}, D.~P., {Schenker}, M.~A., {Ellis}, R., {et~al.} 2013, \apj, 763, 129, \dodoi{10.1088/0004-637X/763/2/129}

\bibitem[{{Tacchella} {et~al.}(2018){Tacchella}, {Bose}, {Conroy}, {Eisenstein}, \& {Johnson}}]{Tacchella:2018}
{Tacchella}, S., {Bose}, S., {Conroy}, C., {Eisenstein}, D.~J., \& {Johnson}, B.~D. 2018, \apj, 868, 92, \dodoi{10.3847/1538-4357/aae8e0}

\bibitem[{{Tacchella} {et~al.}(2022){Tacchella}, {Finkelstein}, {Bagley}, {Dickinson}, {Ferguson}, {Giavalisco}, {Graziani}, {Grogin}, {Hathi}, {Hutchison}, {Jung}, {Koekemoer}, {Larson}, {Papovich}, {Pirzkal}, {Rojas-Ruiz}, {Song}, {Schneider}, {Somerville}, {Wilkins}, \& {Yung}}]{Tacchella:2022}
{Tacchella}, S., {Finkelstein}, S.~L., {Bagley}, M., {et~al.} 2022, \apj, 927, 170, \dodoi{10.3847/1538-4357/ac4cad}

\bibitem[{{Tacchella} {et~al.}(2023){Tacchella}, {Johnson}, {Robertson}, {Carniani}, {D'Eugenio}, {Kumari}, {Maiolino}, {Nelson}, {Suess}, {{\"U}bler}, {Williams}, {Adebusola}, {Alberts}, {Arribas}, {Bhatawdekar}, {Bonaventura}, {Bowler}, {Bunker}, {Cameron}, {Curti}, {Egami}, {Eisenstein}, {Frye}, {Hainline}, {Helton}, {Ji}, {Looser}, {Lyu}, {Perna}, {Rawle}, {Rieke}, {Rieke}, {Saxena}, {Sandles}, {Shivaei}, {Simmonds}, {Sun}, {Willmer}, {Willott}, \& {Witstok}}]{Tacchella:2023}
{Tacchella}, S., {Johnson}, B.~D., {Robertson}, B.~E., {et~al.} 2023, \mnras, 522, 6236, \dodoi{10.1093/mnras/stad1408}

\bibitem[{{The Pandas Development Team}(2022)}]{Pandas:2022}
{The Pandas Development Team}. 2022, {pandas-dev/pandas: Pandas}, v1.5.0, Zenodo,  Zenodo, \dodoi{10.5281/zenodo.7093122}

\bibitem[{{Topping} {et~al.}(2024){Topping}, {Stark}, {Endsley}, {Whitler}, {Hainline}, {Johnson}, {Robertson}, {Tacchella}, {Chen}, {Alberts}, {Baker}, {Bunker}, {Carniani}, {Charlot}, {Chevallard}, {Curtis-Lake}, {DeCoursey}, {Egami}, {Eisenstein}, {Ji}, {Maiolino}, {Williams}, {Willmer}, {Willott}, \& {Witstok}}]{Topping:2024}
{Topping}, M.~W., {Stark}, D.~P., {Endsley}, R., {et~al.} 2024, \mnras, 529, 4087, \dodoi{10.1093/mnras/stae800}

\bibitem[{{van der Walt} {et~al.}(2011){van der Walt}, {Colbert}, \& {Varoquaux}}]{NumPy:2011}
{van der Walt}, S., {Colbert}, S.~C., \& {Varoquaux}, G. 2011, Computing in Science and Engineering, 13, 22, \dodoi{10.1109/MCSE.2011.37}

\bibitem[{{Virtanen} {et~al.}(2020){Virtanen}, {Gommers}, {Oliphant}, {Haberland}, {Reddy}, {Cournapeau}, {Burovski}, {Peterson}, {Weckesser}, {Bright}, {van der Walt}, {Brett}, {Wilson}, {Millman}, {Mayorov}, {Nelson}, {Jones}, {Kern}, {Larson}, {Carey}, {Polat}, {Feng}, {Moore}, {VanderPlas}, {Laxalde}, {Perktold}, {Cimrman}, {Henriksen}, {Quintero}, {Harris}, {Archibald}, {Ribeiro}, {Pedregosa}, {van Mulbregt}, \& {SciPy 1. 0 Contributors}}]{SciPy:2020}
{Virtanen}, P., {Gommers}, R., {Oliphant}, T.~E., {et~al.} 2020, Nature Methods, 17, 261, \dodoi{10.1038/s41592-019-0686-2}

\bibitem[{{Wang} {et~al.}(2024){Wang}, {Leja}, {Atek}, {Labb{\'e}}, {Li}, {Bezanson}, {Brammer}, {Cutler}, {Dayal}, {Furtak}, {Greene}, {Kokorev}, {Pan}, {Price}, {Suess}, {Weaver}, {Whitaker}, \& {Williams}}]{Wang:2024}
{Wang}, B., {Leja}, J., {Atek}, H., {et~al.} 2024, \apj, 963, 74, \dodoi{10.3847/1538-4357/ad187c}

\bibitem[{{Wang} {et~al.}(2025){Wang}, {Sun}, {Zhou}, {Xu}, {Cheng}, {Li}, {Chen}, {Mo}, {Dekel}, {Yang}, {Wang}, {Chen}, {Zheng}, {Cai}, {Elbaz}, {Dai}, \& {Huang}}]{Wang_2025_Massive}
{Wang}, T., {Sun}, H., {Zhou}, L., {et~al.} 2025, \apjl, 988, L35, \dodoi{10.3847/2041-8213/adebe7}

\bibitem[{{Waskom}(2021)}]{Waskom:2021}
{Waskom}, M. 2021, The Journal of Open Source Software, 6, 3021, \dodoi{10.21105/joss.03021}

\bibitem[{{Whitaker} {et~al.}(2011){Whitaker}, {Labb{\'e}}, {van Dokkum}, {Brammer}, {Kriek}, {Marchesini}, {Quadri}, {Franx}, {Muzzin}, {Williams}, {Bezanson}, {Illingworth}, {Lee}, {Lundgren}, {Nelson}, {Rudnick}, {Tal}, \& {Wake}}]{Whitaker:2011}
{Whitaker}, K.~E., {Labb{\'e}}, I., {van Dokkum}, P.~G., {et~al.} 2011, \apj, 735, 86, \dodoi{10.1088/0004-637X/735/2/86}

\bibitem[{{Whitaker} {et~al.}(2019){Whitaker}, {Ashas}, {Illingworth}, {Magee}, {Leja}, {Oesch}, {van Dokkum}, {Mowla}, {Bouwens}, {Franx}, {Holden}, {Labb{\'e}}, {Rafelski}, {Teplitz}, \& {Gonzalez}}]{Whitaker:2019}
{Whitaker}, K.~E., {Ashas}, M., {Illingworth}, G., {et~al.} 2019, \apjs, 244, 16, \dodoi{10.3847/1538-4365/ab3853}

\bibitem[{{Williams} {et~al.}(2024){Williams}, {Alberts}, {Ji}, {Hainline}, {Lyu}, {Rieke}, {Endsley}, {Suess}, {Sun}, {Johnson}, {Florian}, {Shivaei}, {Rujopakarn}, {Baker}, {Bhatawdekar}, {Boyett}, {Bunker}, {Cameron}, {Carniani}, {Charlot}, {Curtis-Lake}, {DeCoursey}, {de Graaff}, {Egami}, {Eisenstein}, {Gibson}, {Hausen}, {Helton}, {Maiolino}, {Maseda}, {Nelson}, {P{\'e}rez-Gonz{\'a}lez}, {Rieke}, {Robertson}, {Saxena}, {Tacchella}, {Willmer}, \& {Willott}}]{Williams:2024}
{Williams}, C.~C., {Alberts}, S., {Ji}, Z., {et~al.} 2024, \apj, 968, 34, \dodoi{10.3847/1538-4357/ad3f17}

\bibitem[{{Woodrum} {et~al.}(2024){Woodrum}, {Rieke}, {Ji}, {Baker}, {Bhatawdekar}, {Bunker}, {Charlot}, {Curtis-Lake}, {Eisenstein}, {Hainline}, {Hausen}, {Helton}, {Hviding}, {Johnson}, {Robertson}, {Sun}, {Tacchella}, {Whitler}, {Williams}, \& {Willmer}}]{Woodrum:2024}
{Woodrum}, C., {Rieke}, M., {Ji}, Z., {et~al.} 2024, Proceedings of the National Academy of Science, 121, e2317375121, \dodoi{10.1073/pnas.2317375121}

\bibitem[{{Yan} {et~al.}(2012){Yan}, {Finkelstein}, {Huang}, {Ryan}, {Ferguson}, {Koekemoer}, {Grogin}, {Dickinson}, {Newman}, {Somerville}, {Dav{\'e}}, {Faber}, {Papovich}, {Guo}, {Giavalisco}, {Lee}, {Reddy}, {Cooray}, {Siana}, {Hathi}, {Fazio}, {Ashby}, {Weiner}, {Lucas}, {Dekel}, {Pentericci}, {Conselice}, {Kocevski}, \& {Lai}}]{Yan:2012}
{Yan}, H., {Finkelstein}, S.~L., {Huang}, K.-H., {et~al.} 2012, \apj, 761, 177, \dodoi{10.1088/0004-637X/761/2/177}

\end{thebibliography}
\appendix
\vspace{-4mm}
\section{Physical Properties for the Final Photometric Sample of MIRI/F770W Non-Detections}
\label{AppendixA}

\setcounter{table}{0}
\renewcommand{\thetable}{\thesection\arabic{table}}

Table~\ref{tab:FinalSampleTable_Appendix} provides a summary of physical quantities for the final photometric sample of high-redshift star-forming galaxies with MIRI/F770W non-detections ($\mathrm{S/N} \leq 1.5$), as described in Section~\ref{SectionThreeOne}. These quantities include: (1) identification number; (2) right ascension in degrees; (3) declination in degrees; (4) spectroscopic redshift, when available; (5) best-fit photometric redshift from \texttt{EAZY}, corresponding to the fit where the $\chi^{2}$ was minimized, along with the associated $1\sigma$ uncertainty; (6) summed probability of being at $z > 7$ from \texttt{EAZY}, assuming the uniform redshift prior $P(z) = \mathrm{exp}[-\chi^{2}(z)/2]$; (7) apparent magnitude in F277W, assuming the fiducial photometry described in Section~\ref{SectionTwo}, along with the associated $1\sigma$ uncertainty; (8) rest-UV absolute magnitude, as measured in Section~\ref{SectionFourOne}, along with the associated $1\sigma$ uncertainty; (9) rest-UV continuum slope, as measured in Section~\ref{SectionFourOne}, along with the associated $1\sigma$ uncertainty; and (10) original reference, when available. The identification numbers and coordinates are identical to the quantities reported in the JADES internal catalog and public data releases \citep[][]{Rieke:2023, D'Eugenio:2024, Eisenstein:2025}. The original reference was determined by searching SIMBAD with \texttt{astroquery} \citep[][]{Ginsburg:2019} and identifying sources from the literature that are within a radius of $1.0^{\prime\prime}$ around each of the galaxies that are part of the final photometric sample.

\begin{table*}[hbp]
	\caption{A summary of physical quantities for the final photometric sample of flux-limited high-redshift star-forming galaxies with MIRI/F770W non-detections ($\mathrm{S/N} \leq 1.5$), as described in Section~\ref{SectionThreeOne}.}
	\label{tab:FinalSampleTable_Appendix}
	\begin{threeparttable}
	\makebox[\textwidth]{
        \begin{tabular}{c|cc|ccc|ccc|c}
		\hline
		\hline
		$\mathrm{ID}$\tnote{a} & $\mathrm{R.A.}$\tnote{b} & $\mathrm{Decl.}$\tnote{c} & $z_{\mathrm{spec}}$\tnote{d} & $z_{\mathrm{phot}}$\tnote{e} & $P\left( z > 7 \right)$\tnote{f} & $m_{\mathrm{F277W}}$\tnote{g} & $M_{\mathrm{UV}}$\tnote{h} & $\beta_{\mathrm{UV}}$\tnote{i} & $\mathrm{Ref.}$\tnote{j} \\
		\hline
            $5141$ & $53.08446$ & $-27.90304$ & n/a & $7.93 \pm 0.21$ & $0.998$ & $28.36 \pm 0.10$ & $-19.54 \pm 0.08$ & $-2.90 \pm 0.17$ & n/a \\
            $9729$ & $53.06237$ & $-27.89550$ & n/a & $7.83 \pm 0.28$ & $0.993$ & $28.64 \pm 0.26$ & $-18.32 \pm 0.37$ & $-1.75 \pm 0.61$ & n/a \\
            $11788$ & $53.09092$ & $-27.89257$ & n/a & $8.84 \pm 0.17$ & $1.000$ & $28.47 \pm 0.13$ & $-18.90 \pm 0.13$ & $-2.04 \pm 0.26$ & n/a \\
            $12678$ & $53.07630$ & $-27.89145$ & n/a & $8.12 \pm 0.29$ & $1.000$ & $28.79 \pm 0.14$ & $-18.57 \pm 0.14$ & $-2.28 \pm 0.27$ & n/a \\
            $15116$ & $53.10001$ & $-27.88876$ & n/a & $7.51 \pm 0.05$ & $1.000$ & $28.40 \pm 0.14$ & $-19.01 \pm 0.15$ & $-2.34 \pm 0.26$ & n/a \\
            $17264$ & $53.06669$ & $-27.88659$ & $7.961$ & $8.04 \pm 0.13$ & $1.000$ & $28.30 \pm 0.12$ & $-19.25 \pm 0.13$ & $-2.48 \pm 0.23$ & n/a \\
            $22015$ & $53.09753$ & $-27.88269$ & n/a & $8.04 \pm 0.31$ & $1.000$ & $28.95 \pm 0.21$ & $-19.03 \pm 0.19$ & $-2.53 \pm 0.39$ & n/a \\
            $23668$ & $53.05305$ & $-27.88155$ & n/a & $7.54 \pm 0.12$ & $1.000$ & $28.94 \pm 0.11$ & $-18.39 \pm 0.11$ & $-2.34 \pm 0.19$ & n/a \\
            $28799$ & $53.04792$ & $-27.87873$ & $7.874$ & $8.02 \pm 0.08$ & $1.000$ & $28.06 \pm 0.04$ & $-19.52 \pm 0.05$ & $-2.52 \pm 0.10$ & n/a \\
            $39639$ & $53.05718$ & $-27.87075$ & n/a & $8.03 \pm 0.22$ & $1.000$ & $28.90 \pm 0.22$ & $-18.87 \pm 0.15$ & $-2.82 \pm 0.34$ & n/a \\
            $41769$ & $53.05580$ & $-27.86901$ & n/a & $7.63 \pm 0.40$ & $0.998$ & $28.46 \pm 0.16$ & $-18.12 \pm 0.22$ & $-1.37 \pm 0.36$ & n/a \\
            $53906$ & $53.03719$ & $-27.86107$ & n/a & $7.43 \pm 0.07$ & $1.000$ & $28.52 \pm 0.18$ & $-18.93 \pm 0.20$ & $-2.32 \pm 0.34$ & n/a \\
            $65024$ & $53.05283$ & $-27.85490$ & n/a & $7.44 \pm 0.14$ & $0.997$ & $28.34 \pm 0.21$ & $-18.51 \pm 0.23$ & $-1.73 \pm 0.40$ & n/a \\
            $160426$ & $53.08796$ & $-27.89883$ & n/a & $8.09 \pm 0.28$ & $0.995$ & $28.96 \pm 0.18$ & $-18.48 \pm 0.19$ & $-2.32 \pm 0.35$ & n/a \\
            $165285$ & $53.05802$ & $-27.88558$ & n/a & $7.47 \pm 0.03$ & $1.000$ & $28.34 \pm 0.06$ & $-18.71 \pm 0.07$ & $-1.98 \pm 0.12$ & n/a \\
            $168445$ & $53.09916$ & $-27.87942$ & n/a & $8.80 \pm 0.33$ & $1.000$ & $28.67 \pm 0.18$ & $-18.85 \pm 0.21$ & $-2.29 \pm 0.41$ & n/a \\
            $169788$ & $53.03269$ & $-27.87735$ & n/a & $7.46 \pm 0.17$ & $0.997$ & $28.75 \pm 0.08$ & $-18.55 \pm 0.10$ & $-2.21 \pm 0.17$ & n/a \\
            $173080$ & $53.05049$ & $-27.87104$ & n/a & $8.84 \pm 0.28$ & $1.000$ & $28.93 \pm 0.16$ & $-18.91 \pm 0.07$ & $-2.50 \pm 0.23$ & n/a \\
            $174601$ & $53.08223$ & $-27.86811$ & n/a & $8.52 \pm 0.05$ & $1.000$ & $28.80 \pm 0.11$ & $-18.62 \pm 0.08$ & $-2.26 \pm 0.18$ & n/a \\
            $174653$ & $53.03825$ & $-27.86798$ & n/a & $7.08 \pm 0.13$ & $0.994$ & $28.86 \pm 0.10$ & $-18.21 \pm 0.11$ & $-1.67 \pm 0.29$ & n/a \\
            $180445$ & $53.08633$ & $-27.85939$ & $7.959$ & $8.94 \pm 0.28$ & $1.000$ & $28.98 \pm 0.13$ & $-17.30 \pm 0.22$ & $-0.85 \pm 0.33$ & n/a \\
            $180642$ & $53.08679$ & $-27.85916$ & $7.953$ & $7.91 \pm 0.10$ & $1.000$ & $28.37 \pm 0.08$ & $-19.11 \pm 0.06$ & $-2.47 \pm 0.13$ & n/a \\
            $283024$ & $53.05834$ & $-27.88481$ & n/a & $8.73 \pm 0.12$ & $1.000$ & $28.52 \pm 0.07$ & $-18.95 \pm 0.05$ & $-2.16 \pm 0.12$ & n/a \\
            $283487$ & $53.07772$ & $-27.87116$ & n/a & $8.38 \pm 0.29$ & $1.000$ & $28.80 \pm 0.09$ & $-18.24 \pm 0.10$ & $-1.70 \pm 0.19$ & n/a \\
            $300298$ & $53.07897$ & $-27.90803$ & n/a & $7.48 \pm 0.07$ & $1.000$ & $28.33 \pm 0.14$ & $-19.03 \pm 0.15$ & $-2.42 \pm 0.26$ & n/a \\
            \hline
	\end{tabular}
        }
	\begin{tablenotes}
	    \footnotesize
            \item \textbf{Notes.}
            \item[a] Identification number, from the JADES internal catalog and public data releases.
            \item[b] Right ascension, in degrees from the epoch J2000.
            \item[c] Declination, in degrees from the epoch J2000.
            \item[d] Spectroscopic redshift, when available.
            \item[e] Best-fit photometric redshift from \texttt{EAZY}, where the $\chi^{2}$ is minimized, and the associated $1\sigma$ uncertainty.
            \item[f] Summed probability of being at $z > 7$ from \texttt{EAZY}, assuming the uniform redshift prior $P(z) = \mathrm{exp}[-\chi^{2}(z)/2]$.
            \item[g] Apparent magnitude in F277W, assuming the fiducial photometry described in Section~\ref{SectionTwo}, and the associated $1\sigma$ uncertainty.
            \item[h] Rest-UV absolute magnitude, as measured in Section~\ref{SectionFourOne}, and the associated $1\sigma$ uncertainty.
            \item[i] Rest-UV continuum slope, as measured in Section~\ref{SectionFourOne}, and the associated $1\sigma$ uncertainty.
            \item[j] Original reference, when available, as determined by \texttt{astroquery}.
        \end{tablenotes}
	\end{threeparttable}
\end{table*}

\addtocounter{table}{-1}

\section{Derived Stellar Population Properties for the Final Photometric Sample of MIRI/F770W Detections}
\label{AppendixB}

Table~\ref{tab:StellarPopulations_Appendix} provides a summary of the most relevant stellar population properties for the final photometric sample of high-redshift star-forming galaxies with MIRI/F770W detections, as described in Section~\ref{SectionFourThree}. These properties are derived assuming the JOF filter set (including MIRI/F770W) and using the fiducial \texttt{Prospector} model as described in Section~\ref{SectionFourTwo}. The medians and $68\%$ confidence intervals of the derived quantities are provided. From left to right: stellar mass ($M_{\ast}$), mass-weighted stellar age ($t_{\mathrm{MW}}$), specific star formation rate ($\mathrm{sSFR}_{10}$) averaged over the last $10\ \mathrm{Myr}$ of lookback time, star formation rate ($\mathrm{SFR}_{10}$) averaged over the same timescale, and rest-frame equivalent width ($\mathrm{EW}_{\mathrm{[OIII]}+\mathrm{H}\beta}$) of the rest-optical emission lines $\mathrm{[OIII]}+\mathrm{H}\beta$. The results presented in these tables assuming the JOF filter set are consistent with results using the JADES and CEERS filter sets.

\begin{table*}[htp!]
	\caption{A summary of the derived stellar population properties for the final photometric sample of high-redshift star-forming galaxies with MIRI/F770W detections ($\mathrm{S/N} > 1.5$), as described in Section~\ref{SectionFourThree}. These properties are derived assuming the JOF filter set (including MIRI/F770W), while using the fiducial \texttt{Prospector} model and the four different models for the SFH, as described in Section~\ref{SectionFourTwo}. Medians and $68\%$ confidence intervals of the derived quantities are reported.}
	\label{tab:StellarPopulations_Appendix}
	\makebox[\textwidth]{
        \begin{tabular}{c|c||c|c|c|c|c}
		\hline
		\hline
            & & \multicolumn{5}{c}{Derived Stellar Population Property} \\
		\hline
		& ID & $\mathrm{log}_{10}\left( M_{\ast}/M_{\odot} \right)$ & $\mathrm{log}_{10}\left( t_\mathrm{MW}/\mathrm{yr} \right)$ & $\mathrm{log}_{10}\left( \mathrm{SFR}_{10}/[M_{\ast}/\mathrm{yr}] \right)$ & $\mathrm{log}_{10}\left( \mathrm{sSFR}_{10}/\mathrm{yr}^{-1} \right)$ & $\mathrm{log}_{10}\left( \mathrm{EW}_{\mathrm{[OIII]}+\mathrm{H}\beta}/\mathrm{\AA} \right)$ \\
		\hline
            \multirow{22}{*}{\rotatebox{90}{CSFH }} & $14647$ & $7.451^{+0.110}_{-0.089}$ & $6.576^{+0.250}_{-0.302}$ & $0.407^{+0.069}_{-0.061}$ & $-7.000^{+0.000}_{-0.127}$ & $3.294^{+0.144}_{-0.047}$ \\
            & $21468$ & $7.940^{+0.219}_{-0.204}$ & $7.131^{+0.283}_{-0.264}$ & $0.501^{+0.079}_{-0.078}$ & $-7.432^{+0.264}_{-0.283}$ & $3.073^{+0.056}_{-0.056}$ \\
            & $25526$ & $7.806^{+0.088}_{-0.075}$ & $6.698^{+0.110}_{-0.177}$ & $0.765^{+0.054}_{-0.048}$ & $-7.000^{+0.000}_{-0.110}$ & $3.309^{+0.077}_{-0.040}$ \\
            & $27503$ & $8.111^{+0.101}_{-0.106}$ & $6.840^{+0.123}_{-0.226}$ & $0.964^{+0.055}_{-0.056}$ & $-7.141^{+0.141}_{-0.123}$ & $3.224^{+0.049}_{-0.047}$ \\
            & $30333$ & $7.860^{+0.060}_{-0.089}$ & $6.834^{+0.056}_{-0.063}$ & $0.726^{+0.046}_{-0.074}$ & $-7.135^{+0.063}_{-0.056}$ & $3.256^{+0.026}_{-0.028}$ \\
            & $37458$ & $7.629^{+0.135}_{-0.074}$ & $6.883^{+0.189}_{-0.104}$ & $0.440^{+0.041}_{-0.051}$ & $-7.184^{+0.104}_{-0.189}$ & $3.223^{+0.031}_{-0.041}$ \\
            & $57378$ & $8.346^{+0.112}_{-0.123}$ & $7.124^{+0.153}_{-0.168}$ & $0.923^{+0.057}_{-0.063}$ & $-7.425^{+0.168}_{-0.153}$ & $3.083^{+0.052}_{-0.043}$ \\
            & $66293$ & $9.001^{+0.081}_{-0.081}$ & $7.056^{+0.106}_{-0.099}$ & $1.641^{+0.077}_{-0.085}$ & $-7.357^{+0.099}_{-0.105}$ & $2.820^{+0.056}_{-0.054}$ \\
            & $164055$ & $7.515^{+0.199}_{-0.154}$ & $6.883^{+0.302}_{-0.261}$ & $0.306^{+0.084}_{-0.102}$ & $-7.184^{+0.184}_{-0.302}$ & $3.221^{+0.090}_{-0.102}$ \\
            & $165595$ & $8.371^{+0.087}_{-0.084}$ & $7.189^{+0.111}_{-0.098}$ & $0.879^{+0.051}_{-0.054}$ & $-7.490^{+0.098}_{-0.111}$ & $3.043^{+0.032}_{-0.035}$ \\
            & $173624$ & $7.908^{+0.082}_{-0.082}$ & $6.860^{+0.098}_{-0.074}$ & $0.733^{+0.065}_{-0.066}$ & $-7.161^{+0.074}_{-0.098}$ & $3.251^{+0.029}_{-0.034}$ \\
            & $173679$ & $9.102^{+0.182}_{-0.189}$ & $7.807^{+0.248}_{-0.232}$ & $0.981^{+0.110}_{-0.099}$ & $-8.108^{+0.232}_{-0.248}$ & $2.670^{+0.088}_{-0.067}$ \\
            & $174121$ & $7.860^{+0.073}_{-0.054}$ & $6.409^{+0.125}_{-0.264}$ & $0.860^{+0.072}_{-0.054}$ & $-7.000^{+0.000}_{+0.000}$ & $3.548^{+0.073}_{-0.074}$ \\
            & $174693$ & $8.313^{+0.235}_{-0.238}$ & $7.660^{+0.337}_{-0.342}$ & $0.346^{+0.122}_{-0.109}$ & $-7.961^{+0.342}_{-0.337}$ & $2.896^{+0.070}_{-0.060}$ \\
            & $175729$ & $8.106^{+0.166}_{-0.172}$ & $7.119^{+0.235}_{-0.249}$ & $0.685^{+0.099}_{-0.091}$ & $-7.420^{+0.249}_{-0.235}$ & $3.127^{+0.080}_{-0.066}$ \\
            & $175837$ & $8.538^{+0.217}_{-0.231}$ & $7.503^{+0.370}_{-0.385}$ & $0.705^{+0.237}_{-0.176}$ & $-7.804^{+0.385}_{-0.370}$ & $2.880^{+0.082}_{-0.080}$ \\
            & $177322$ & $7.860^{+0.252}_{-0.215}$ & $7.019^{+0.334}_{-0.319}$ & $0.530^{+0.086}_{-0.090}$ & $-7.320^{+0.319}_{-0.333}$ & $3.121^{+0.078}_{-0.067}$ \\
            & $179485$ & $8.806^{+0.133}_{-0.120}$ & $7.392^{+0.205}_{-0.182}$ & $1.113^{+0.075}_{-0.087}$ & $-7.693^{+0.182}_{-0.205}$ & $2.860^{+0.041}_{-0.050}$ \\
            & $180446$ & $8.466^{+0.107}_{-0.177}$ & $8.140^{+0.154}_{-0.267}$ & $0.031^{+0.108}_{-0.075}$ & $-8.441^{+0.267}_{-0.154}$ & $2.719^{+0.050}_{-0.050}$ \\
            & $300287$ & $8.067^{+0.077}_{-0.074}$ & $6.404^{+0.088}_{-0.128}$ & $1.067^{+0.077}_{-0.074}$ & $-7.000^{+0.000}_{+0.000}$ & $2.730^{+0.103}_{-0.101}$ \\
            & $300391$ & $7.882^{+0.083}_{-0.052}$ & $6.543^{+0.185}_{-0.130}$ & $0.878^{+0.051}_{-0.048}$ & $-7.000^{+0.000}_{-0.029}$ & $3.394^{+0.073}_{-0.068}$ \\
            & $380203$ & $7.545^{+0.148}_{-0.077}$ & $6.297^{+0.464}_{-0.330}$ & $0.541^{+0.089}_{-0.075}$ & $-7.000^{+0.000}_{-0.062}$ & $3.088^{+0.262}_{-0.082}$ \\
		      \hline
            \multirow{22}{*}{\rotatebox{90}{DtSFH }} & $14647$ & $7.525^{+0.084}_{-0.104}$ & $6.516^{+0.430}_{-0.457}$ & $0.429^{+0.129}_{-0.080}$ & $-7.000^{+0.000}_{-0.214}$ & $3.341^{+0.147}_{-0.067}$ \\
            & $21468$ & $7.970^{+0.200}_{-0.187}$ & $7.242^{+0.270}_{-0.271}$ & $0.498^{+0.072}_{-0.070}$ & $-7.465^{+0.232}_{-0.252}$ & $3.097^{+0.058}_{-0.058}$ \\
            & $25526$ & $7.802^{+0.077}_{-0.068}$ & $6.711^{+0.143}_{-0.159}$ & $0.730^{+0.045}_{-0.035}$ & $-7.057^{+0.055}_{-0.089}$ & $3.367^{+0.055}_{-0.048}$ \\
            & $27503$ & $8.173^{+0.094}_{-0.095}$ & $6.928^{+0.162}_{-0.172}$ & $0.966^{+0.071}_{-0.068}$ & $-7.200^{+0.118}_{-0.131}$ & $3.256^{+0.062}_{-0.052}$ \\
            & $30333$ & $7.902^{+0.090}_{-0.091}$ & $6.881^{+0.176}_{-0.116}$ & $0.727^{+0.057}_{-0.093}$ & $-7.165^{+0.077}_{-0.139}$ & $3.295^{+0.042}_{-0.053}$ \\
            & $37458$ & $7.710^{+0.117}_{-0.102}$ & $7.052^{+0.169}_{-0.161}$ & $0.409^{+0.032}_{-0.033}$ & $-7.299^{+0.127}_{-0.148}$ & $3.224^{+0.043}_{-0.039}$ \\
            & $57378$ & $8.355^{+0.101}_{-0.095}$ & $7.232^{+0.146}_{-0.136}$ & $0.903^{+0.052}_{-0.068}$ & $-7.456^{+0.120}_{-0.136}$ & $3.117^{+0.043}_{-0.042}$ \\
            & $66293$ & $9.062^{+0.076}_{-0.073}$ & $7.086^{+0.094}_{-0.077}$ & $1.726^{+0.087}_{-0.089}$ & $-7.327^{+0.064}_{-0.081}$ & $2.874^{+0.048}_{-0.055}$ \\
            & $164055$ & $7.571^{+0.152}_{-0.128}$ & $7.005^{+0.235}_{-0.276}$ & $0.304^{+0.083}_{-0.080}$ & $-7.260^{+0.194}_{-0.204}$ & $3.247^{+0.071}_{-0.078}$ \\
            & $165595$ & $8.378^{+0.087}_{-0.086}$ & $7.274^{+0.122}_{-0.116}$ & $0.882^{+0.048}_{-0.051}$ & $-7.494^{+0.105}_{-0.112}$ & $3.085^{+0.034}_{-0.035}$ \\
            & $173624$ & $7.924^{+0.100}_{-0.099}$ & $6.932^{+0.162}_{-0.155}$ & $0.709^{+0.060}_{-0.049}$ & $-7.203^{+0.108}_{-0.132}$ & $3.279^{+0.048}_{-0.045}$ \\
            & $173679$ & $9.102^{+0.131}_{-0.138}$ & $7.884^{+0.177}_{-0.183}$ & $1.005^{+0.085}_{-0.088}$ & $-8.099^{+0.182}_{-0.172}$ & $2.692^{+0.066}_{-0.050}$ \\
            & $174121$ & $7.933^{+0.093}_{-0.074}$ & $6.329^{+0.288}_{-0.486}$ & $0.923^{+0.097}_{-0.074}$ & $-7.000^{+0.000}_{-0.017}$ & $3.566^{+0.056}_{-0.091}$ \\
            & $174693$ & $8.287^{+0.189}_{-0.207}$ & $7.706^{+0.264}_{-0.292}$ & $0.376^{+0.093}_{-0.083}$ & $-7.907^{+0.283}_{-0.261}$ & $2.936^{+0.065}_{-0.057}$ \\
            & $175729$ & $8.101^{+0.159}_{-0.159}$ & $7.187^{+0.243}_{-0.263}$ & $0.681^{+0.083}_{-0.083}$ & $-7.415^{+0.219}_{-0.223}$ & $3.166^{+0.078}_{-0.067}$ \\
            & $175837$ & $8.525^{+0.238}_{-0.231}$ & $7.529^{+0.352}_{-0.355}$ & $0.762^{+0.202}_{-0.170}$ & $-7.740^{+0.335}_{-0.352}$ & $2.934^{+0.118}_{-0.080}$ \\
            & $177322$ & $7.903^{+0.236}_{-0.202}$ & $7.113^{+0.339}_{-0.314}$ & $0.530^{+0.085}_{-0.081}$ & $-7.352^{+0.243}_{-0.311}$ & $3.130^{+0.072}_{-0.065}$ \\
            & $179485$ & $8.836^{+0.127}_{-0.119}$ & $7.521^{+0.192}_{-0.186}$ & $1.107^{+0.072}_{-0.077}$ & $-7.729^{+0.178}_{-0.192}$ & $2.866^{+0.044}_{-0.046}$ \\
            & $180446$ & $8.316^{+0.173}_{-0.211}$ & $7.886^{+0.229}_{-0.312}$ & $0.176^{+0.186}_{-0.140}$ & $-8.113^{+0.327}_{-0.274}$ & $2.756^{+0.092}_{-0.106}$ \\
            & $300287$ & $8.066^{+0.055}_{-0.059}$ & $6.464^{+0.161}_{-0.133}$ & $1.054^{+0.054}_{-0.059}$ & $-7.000^{+0.000}_{-0.019}$ & $2.674^{+0.112}_{-0.093}$ \\
            & $300391$ & $7.940^{+0.086}_{-0.067}$ & $6.548^{+0.204}_{-0.597}$ & $0.906^{+0.060}_{-0.048}$ & $-7.001^{+0.001}_{-0.079}$ & $3.436^{+0.108}_{-0.058}$ \\
            & $380203$ & $7.577^{+0.094}_{-0.083}$ & $6.166^{+0.632}_{-0.350}$ & $0.537^{+0.083}_{-0.090}$ & $-7.000^{+0.000}_{-0.108}$ & $3.247^{+0.136}_{-0.207}$ \\
            \hline
            \rotatebox{90}{... } & \rotatebox{90}{... } & \rotatebox{90}{... } & \rotatebox{90}{... } & \rotatebox{90}{... } & \rotatebox{90}{... } & \rotatebox{90}{... } \\
            \hline
	\end{tabular}
        }
\end{table*}

\section{Constraints on the Joint Posterior Distributions and Star Formation Histories for the Final Photometric Sample of MIRI/F770W Detections}
\label{AppendixC}

\setcounter{figure}{0}
\renewcommand{\thefigure}{\thesection\arabic{figure}}

\begin{figure*}
    \centering
    \figurenum{C1}
    \includegraphics[width=1.0\linewidth]{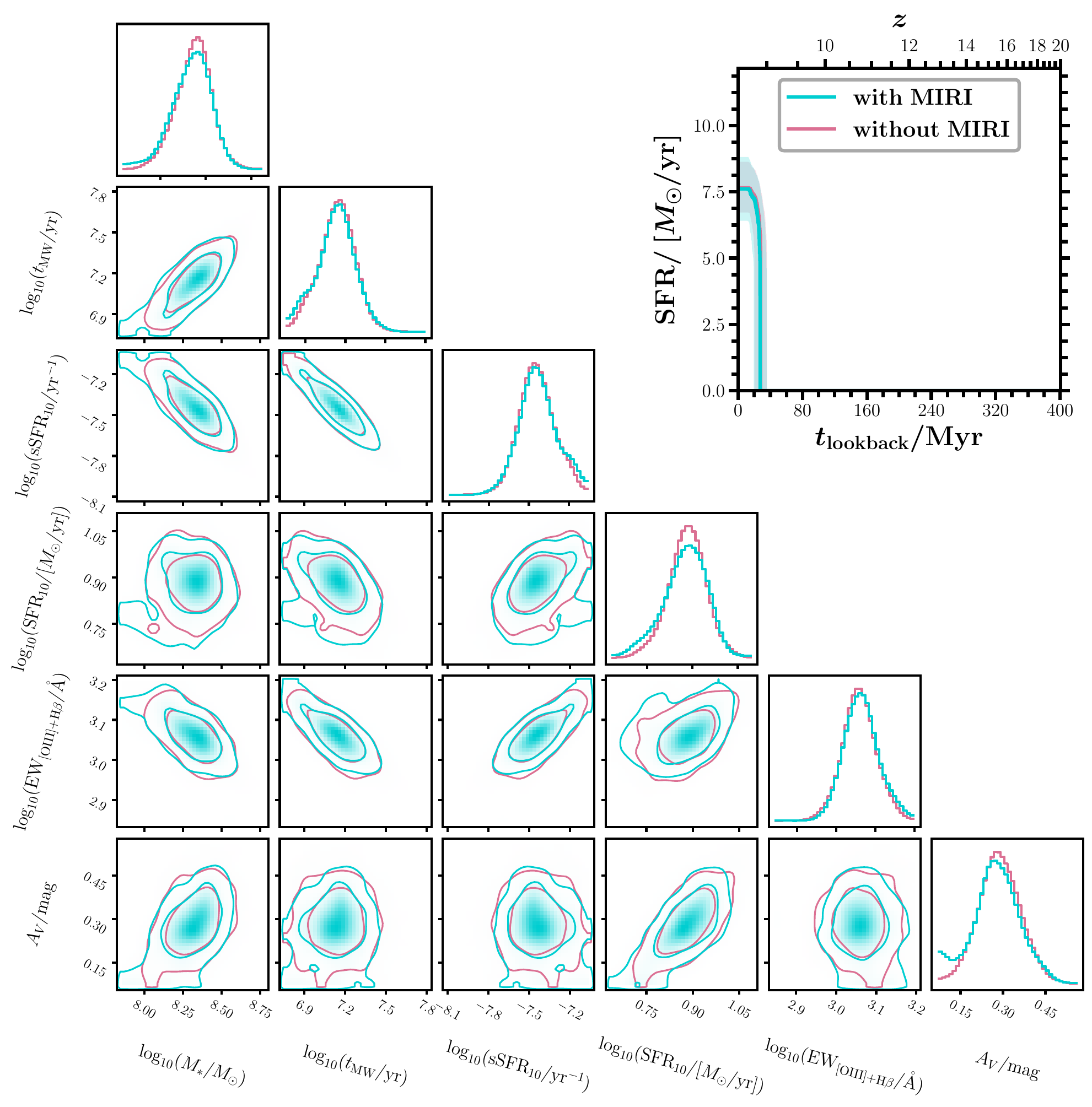}
    \caption{\textit{Lower left}: Example of joint posterior distributions for the most relevant stellar populations properties for a typical galaxy (JADES$-$GS$-$ID$-$$165595$, or JADES$-$GS$+53.05830$$-$$27.88486$, at $z_{\mathrm{spec}} = 8.585$) from the final photometric sample of MIRI/F770W detections described in Section~\ref{SectionThreeOne}. From left to right: stellar mass, mass-weighted stellar age, sSFR averaged over the last $10$ Myr, SFR averaged over the same timescale, rest-frame EW of $\mathrm{[OIII]}+\mathrm{H}\beta$, and diffuse dust attenuation as measured in the $V$-band. \textit{Upper right}: Example of the derived SFHs. The solid lines (shaded regions) represent the median ($68\%$ confidence interval) of the \texttt{Prospector} models. \textit{Lower left and upper right}: The fiducial \texttt{Prospector} model is described in Section~\ref{SectionFourTwo} and assumes the constant SFH model. The blue (pink) lines and shaded regions represent results from fitting to the full JADES filter set, including (excluding) MIRI/F770W. Similar to Figure~\ref{figset:ExampleSEDs_1}, it is evident that for this typical galaxy, MIRI/F770W has little to no effect on the inferred stellar population properties and SFH. The complete figure set of these typical galaxies ($18$ figures) is available in the online journal. \label{figset:ExampleCorners_1}}
\end{figure*}

Figure~\ref{figset:ExampleCorners_1} provides an example of the joint posterior distributions in the lower left for one of the typical galaxies (JADES$-$GS$-$ID$-$$165595$, or JADES$-$GS$+53.05830$$-$$27.88486$, at $z_{\mathrm{spec}} = 8.585$) from the final photometric sample of MIRI/F770W detections described in Section~\ref{SectionThreeOne}. The joint posterior distributions are presented for the stellar population properties in the off-diagonal panels, with the shaded regions representing the $68\%$ and $95\%$ confidence intervals. From left to right, these properties are: stellar mass ($M_{\ast}$), mass-weighted stellar age ($t_{\mathrm{MW}}$), specific star formation rate ($\mathrm{sSFR}_{10}$) averaged over the last $10$ Myr of lookback time, rest-frame equivalent width ($\mathrm{EW}_{\mathrm{[OIII]}+\mathrm{H}\beta}$) of the rest-optical emission lines $\mathrm{[OIII]}+\mathrm{H}\beta$, and diffuse dust attenuation ($A_{V}$) as measured in the $V$-band. The marginal posterior distributions are provided in the diagonal panels. For galaxies without an available spectroscopic redshift, we additionally include the joint and marginal posterior distributions for the derived photometric redshifts. These distributions are presented for the fiducial \texttt{Prospector} model described in Section~\ref{SectionFourTwo} and assume the constant SFH model. This figure also provides an example of the derived SFHs in the upper right for the same typical galaxy. The lines and shaded regions represent the median and $68\%$ confidence interval of the \texttt{Prospector} models.

The results from \texttt{Prospector} assuming the full JADES filter set, which includes MIRI/F770W, are represented by the blue lines and shaded regions. Those results assuming the partial JADES filter set, which excludes MIRI/F770W, are represented by the pink lines and shaded regions. The joint posterior distributions in the lower left demonstrate degeneracies between some of the inferred physical parameters. For example, stellar mass is degenerate with the following quantities: mass-weighted stellar age, specific star formation rate, rest-optical equivalent width, and diffuse dust attenuation. The low-mass solutions correspond to younger stellar ages, larger specific star formation rates, larger equivalent widths, and less dust attenuation. Similar to Figure~\ref{figset:ExampleSEDs_1}, it is evident from Figure~\ref{figset:ExampleCorners_1} that for this typical galaxy, MIRI/F770W has little to no effect on the inferred stellar population properties and SFH. These results are true for the vast majority ($\approx 80\%$) of galaxies in the final photometric sample.

\begin{figure*}
    \centering
    \figurenum{C2}
    \includegraphics[width=1.0\linewidth]{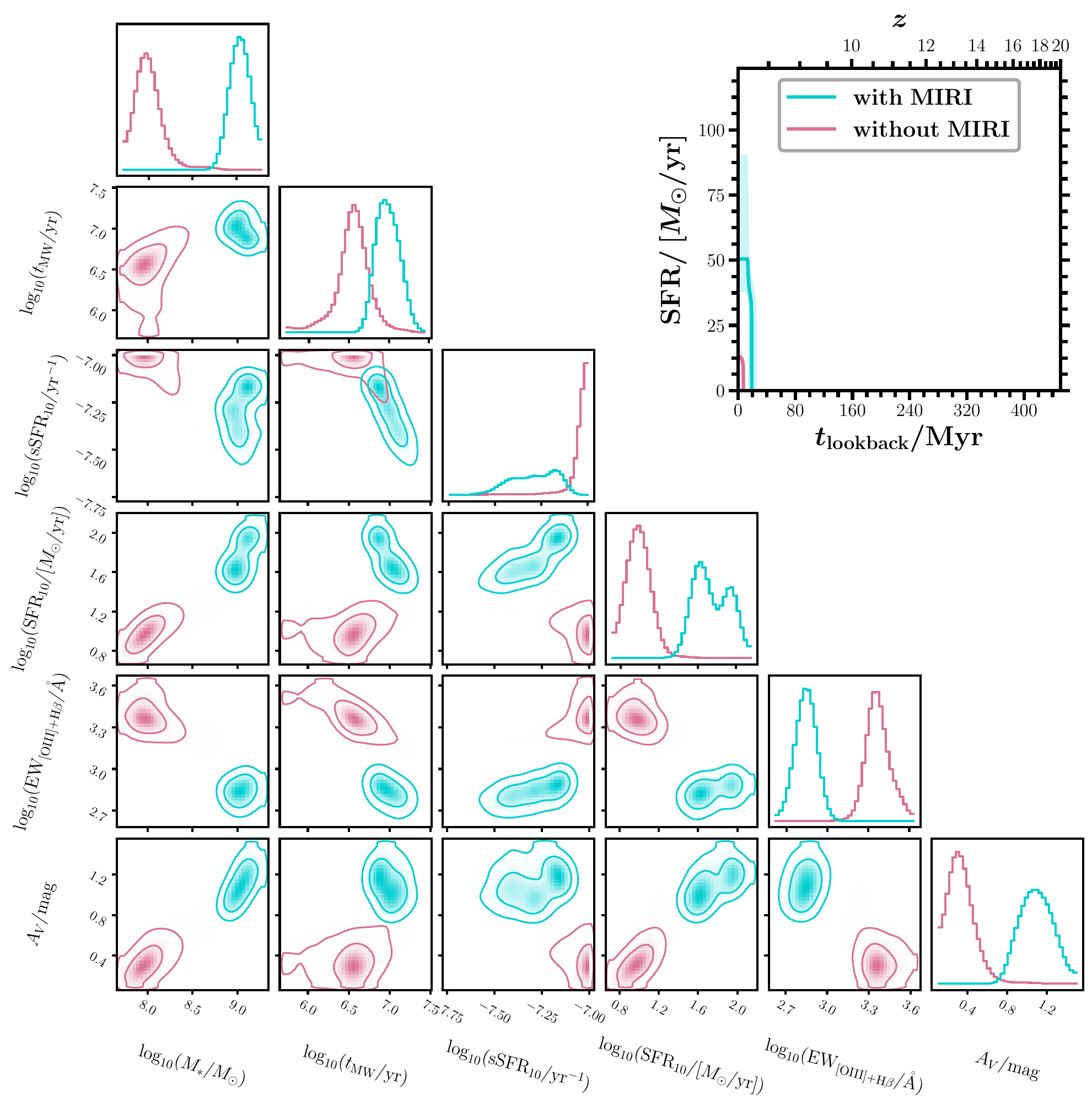}
    \caption{Similar to Figure~\ref{figset:ExampleCorners_1}, but for an atypical galaxy (JADES$-$GS$-$ID$-$$66293$, or JADES$-$GS$+53.04601$$-$$27.85399$, at $z_{\mathrm{spec}} = 8.065$) from the final photometric sample of MIRI/F770W detections described in Section~\ref{SectionThreeOne}. It is evident that for this atypical galaxy, MIRI/F770W has a significant effect on the inferred stellar population properties and SFH. The complete figure set of these atypical galaxies (four figures) is available in the online journal. \label{figset:ExampleCorners_2}}
\end{figure*}

Similar to Figure~\ref{figset:ExampleCorners_1}, Figure~\ref{figset:ExampleCorners_2} provides an example of the joint posterior distributions and derived SFHs for one of the atypical galaxies (JADES$-$GS$-$ID$-$$66293$, or JADES$-$GS$+53.04601$$-$$27.85399$, at $z_{\mathrm{spec}} = 8.065$) from the final photometric sample of MIRI/F770W detections described in Section~\ref{SectionThreeOne}. As before, the joint posterior distributions in the lower left demonstrate the same degeneracies between some of the inferred physical parameters. Unlike the example of the typical galaxy in Figure~\ref{figset:ExampleCorners_1}, the inclusion of MIRI/F770W has a significant effect on the inferred stellar population properties and SFH, similar to what is presented in Figure~\ref{figset:ExampleSEDs_2}. These results are true for the minority ($\approx 20\%$) of galaxies in the final photometric sample. As described in \ref{SectionFourThree}, the four atypical galaxies include two that are anomalously red and two that are anomalously blue. These atypical galaxies are some of the reddest and bluest galaxies in our sample, based on their rest-UV continuum slopes.

\clearpage

\figsetstart
\figsetnum{C}
\graphicspath{{./}{files/ExampleCorners/}}

\figsetgrpstart
\figsetgrpnum{C1.1}
\figsetplot{Corner_014647_ConstantPrior_JADES_freeRedshift.pdf}
\figsetgrpnote{\textit{Lower left}: Example of joint posterior distributions for the most relevant stellar populations properties for a typical galaxy (JADES$-$GS$-$ID$-$$14647$, or JADES$-$GS$+53.08646$$-$$27.88925$, at $z_{\mathrm{phot}} = 7.38 \pm 0.05$) from the final photometric sample of MIRI/F770W detections described in Section~\ref{SectionThreeOne}. From left to right: stellar mass, mass-weighted stellar age, sSFR averaged over the last $10$ Myr, SFR averaged over the same timescale, rest-frame EW of $\mathrm{[OIII]}+\mathrm{H}\beta$, and diffuse dust attenuation as measured in the $V$-band. \textit{Upper right}: Example of the derived SFHs. The solid lines (shaded regions) represent the median ($68\%$ confidence interval) of the \texttt{Prospector} models. \textit{Lower left and upper right}: The fiducial \texttt{Prospector} model is described in Section~\ref{SectionFourTwo} and assumes the constant SFH model. The blue (pink) lines and shaded regions represent results from fitting to the full JADES filter set, including (excluding) MIRI/F770W. Similar to Figure~\ref{figset:ExampleSEDs_1}, it is evident that for this typical galaxy, MIRI/F770W has little to no effect on the inferred stellar population properties and SFH.}
\figsetgrpend

\figsetgrpstart
\figsetgrpnum{C1.2}
\figsetplot{Corner_025526_ConstantPrior_JADES_fixedRedshift.pdf}
\figsetgrpnote{\textit{Lower left}: Example of joint posterior distributions for the most relevant stellar populations properties for a typical galaxy (JADES$-$GS$-$ID$-$$25526$, or JADES$-$GS$+53.09942$$-$$27.88038$, at $z_{\mathrm{spec}} = 7.957$) from the final photometric sample of MIRI/F770W detections described in Section~\ref{SectionThreeOne}. From left to right: stellar mass, mass-weighted stellar age, sSFR averaged over the last $10$ Myr, SFR averaged over the same timescale, rest-frame EW of $\mathrm{[OIII]}+\mathrm{H}\beta$, and diffuse dust attenuation as measured in the $V$-band. \textit{Upper right}: Example of the derived SFHs. The solid lines (shaded regions) represent the median ($68\%$ confidence interval) of the \texttt{Prospector} models. \textit{Lower left and upper right}: The fiducial \texttt{Prospector} model is described in Section~\ref{SectionFourTwo} and assumes the constant SFH model. The blue (pink) lines and shaded regions represent results from fitting to the full JADES filter set, including (excluding) MIRI/F770W. Similar to Figure~\ref{figset:ExampleSEDs_1}, it is evident that for this typical galaxy, MIRI/F770W has little to no effect on the inferred stellar population properties and SFH.}
\figsetgrpend

\figsetgrpstart
\figsetgrpnum{C1.3}
\figsetplot{Corner_027503_ConstantPrior_JADES_fixedRedshift.pdf}
\figsetgrpnote{\textit{Lower left}: Example of joint posterior distributions for the most relevant stellar populations properties for a typical galaxy (JADES$-$GS$-$ID$-$$27503$, or JADES$-$GS$+53.07581$$-$$27.87938$, at $z_{\mathrm{spec}} = 8.196$) from the final photometric sample of MIRI/F770W detections described in Section~\ref{SectionThreeOne}. From left to right: stellar mass, mass-weighted stellar age, sSFR averaged over the last $10$ Myr, SFR averaged over the same timescale, rest-frame EW of $\mathrm{[OIII]}+\mathrm{H}\beta$, and diffuse dust attenuation as measured in the $V$-band. \textit{Upper right}: Example of the derived SFHs. The solid lines (shaded regions) represent the median ($68\%$ confidence interval) of the \texttt{Prospector} models. \textit{Lower left and upper right}: The fiducial \texttt{Prospector} model is described in Section~\ref{SectionFourTwo} and assumes the constant SFH model. The blue (pink) lines and shaded regions represent results from fitting to the full JADES filter set, including (excluding) MIRI/F770W. Similar to Figure~\ref{figset:ExampleSEDs_1}, it is evident that for this typical galaxy, MIRI/F770W has little to no effect on the inferred stellar population properties and SFH.}
\figsetgrpend

\figsetgrpstart
\figsetgrpnum{C1.4}
\figsetplot{Corner_030333_ConstantPrior_JADES_fixedRedshift.pdf}
\figsetgrpnote{\textit{Lower left}: Example of joint posterior distributions for the most relevant stellar populations properties for a typical galaxy (JADES$-$GS$-$ID$-$$30333$, or JADES$-$GS$+53.05373$$-$$27.87789$, at $z_{\mathrm{spec}} = 7.891$) from the final photometric sample of MIRI/F770W detections described in Section~\ref{SectionThreeOne}. From left to right: stellar mass, mass-weighted stellar age, sSFR averaged over the last $10$ Myr, SFR averaged over the same timescale, rest-frame EW of $\mathrm{[OIII]}+\mathrm{H}\beta$, and diffuse dust attenuation as measured in the $V$-band. \textit{Upper right}: Example of the derived SFHs. The solid lines (shaded regions) represent the median ($68\%$ confidence interval) of the \texttt{Prospector} models. \textit{Lower left and upper right}: The fiducial \texttt{Prospector} model is described in Section~\ref{SectionFourTwo} and assumes the constant SFH model. The blue (pink) lines and shaded regions represent results from fitting to the full JADES filter set, including (excluding) MIRI/F770W. Similar to Figure~\ref{figset:ExampleSEDs_1}, it is evident that for this typical galaxy, MIRI/F770W has little to no effect on the inferred stellar population properties and SFH.}
\figsetgrpend

\figsetgrpstart
\figsetgrpnum{C1.5}
\figsetplot{Corner_037458_ConstantPrior_JADES_fixedRedshift.pdf}
\figsetgrpnote{\textit{Lower left}: Example of joint posterior distributions for the most relevant stellar populations properties for a typical galaxy (JADES$-$GS$-$ID$-$$37458$, or JADES$-$GS$+53.08932$$-$$27.87269$, at $z_{\mathrm{spec}} = 8.225$) from the final photometric sample of MIRI/F770W detections described in Section~\ref{SectionThreeOne}. From left to right: stellar mass, mass-weighted stellar age, sSFR averaged over the last $10$ Myr, SFR averaged over the same timescale, rest-frame EW of $\mathrm{[OIII]}+\mathrm{H}\beta$, and diffuse dust attenuation as measured in the $V$-band. \textit{Upper right}: Example of the derived SFHs. The solid lines (shaded regions) represent the median ($68\%$ confidence interval) of the \texttt{Prospector} models. \textit{Lower left and upper right}: The fiducial \texttt{Prospector} model is described in Section~\ref{SectionFourTwo} and assumes the constant SFH model. The blue (pink) lines and shaded regions represent results from fitting to the full JADES filter set, including (excluding) MIRI/F770W. Similar to Figure~\ref{figset:ExampleSEDs_1}, it is evident that for this typical galaxy, MIRI/F770W has little to no effect on the inferred stellar population properties and SFH.}
\figsetgrpend

\figsetgrpstart
\figsetgrpnum{C1.6}
\figsetplot{Corner_057378_ConstantPrior_JADES_fixedRedshift.pdf}
\figsetgrpnote{\textit{Lower left}: Example of joint posterior distributions for the most relevant stellar populations properties for a typical galaxy (JADES$-$GS$-$ID$-$$57378$, or JADES$-$GS$+53.08650$$-$$27.85920$, at $z_{\mathrm{spec}} = 7.950$) from the final photometric sample of MIRI/F770W detections described in Section~\ref{SectionThreeOne}. From left to right: stellar mass, mass-weighted stellar age, sSFR averaged over the last $10$ Myr, SFR averaged over the same timescale, rest-frame EW of $\mathrm{[OIII]}+\mathrm{H}\beta$, and diffuse dust attenuation as measured in the $V$-band. \textit{Upper right}: Example of the derived SFHs. The solid lines (shaded regions) represent the median ($68\%$ confidence interval) of the \texttt{Prospector} models. \textit{Lower left and upper right}: The fiducial \texttt{Prospector} model is described in Section~\ref{SectionFourTwo} and assumes the constant SFH model. The blue (pink) lines and shaded regions represent results from fitting to the full JADES filter set, including (excluding) MIRI/F770W. Similar to Figure~\ref{figset:ExampleSEDs_1}, it is evident that for this typical galaxy, MIRI/F770W has little to no effect on the inferred stellar population properties and SFH.}
\figsetgrpend

\figsetgrpstart
\figsetgrpnum{C1.7}
\figsetplot{Corner_164055_ConstantPrior_JADES_fixedRedshift.pdf}
\figsetgrpnote{\textit{Lower left}: Example of joint posterior distributions for the most relevant stellar populations properties for a typical galaxy (JADES$-$GS$-$ID$-$$164055$, or JADES$-$GS$+53.08168$$-$$27.88858$, at $z_{\mathrm{spec}} = 7.397$) from the final photometric sample of MIRI/F770W detections described in Section~\ref{SectionThreeOne}. From left to right: stellar mass, mass-weighted stellar age, sSFR averaged over the last $10$ Myr, SFR averaged over the same timescale, rest-frame EW of $\mathrm{[OIII]}+\mathrm{H}\beta$, and diffuse dust attenuation as measured in the $V$-band. \textit{Upper right}: Example of the derived SFHs. The solid lines (shaded regions) represent the median ($68\%$ confidence interval) of the \texttt{Prospector} models. \textit{Lower left and upper right}: The fiducial \texttt{Prospector} model is described in Section~\ref{SectionFourTwo} and assumes the constant SFH model. The blue (pink) lines and shaded regions represent results from fitting to the full JADES filter set, including (excluding) MIRI/F770W. Similar to Figure~\ref{figset:ExampleSEDs_1}, it is evident that for this typical galaxy, MIRI/F770W has little to no effect on the inferred stellar population properties and SFH.}
\figsetgrpend

\figsetgrpstart
\figsetgrpnum{C1.8}
\figsetplot{Corner_165595_ConstantPrior_JADES_fixedRedshift.pdf}
\figsetgrpnote{\textit{Lower left}: Example of joint posterior distributions for the most relevant stellar populations properties for a typical galaxy (JADES$-$GS$-$ID$-$$165595$, or JADES$-$GS$+53.05830$$-$$27.88486$, at $z_{\mathrm{spec}} = 8.585$) from the final photometric sample of MIRI/F770W detections described in Section~\ref{SectionThreeOne}. From left to right: stellar mass, mass-weighted stellar age, sSFR averaged over the last $10$ Myr, SFR averaged over the same timescale, rest-frame EW of $\mathrm{[OIII]}+\mathrm{H}\beta$, and diffuse dust attenuation as measured in the $V$-band. \textit{Upper right}: Example of the derived SFHs. The solid lines (shaded regions) represent the median ($68\%$ confidence interval) of the \texttt{Prospector} models. \textit{Lower left and upper right}: The fiducial \texttt{Prospector} model is described in Section~\ref{SectionFourTwo} and assumes the constant SFH model. The blue (pink) lines and shaded regions represent results from fitting to the full JADES filter set, including (excluding) MIRI/F770W. Similar to Figure~\ref{figset:ExampleSEDs_1}, it is evident that for this typical galaxy, MIRI/F770W has little to no effect on the inferred stellar population properties and SFH.}
\figsetgrpend

\figsetgrpstart
\figsetgrpnum{C1.9}
\figsetplot{Corner_173624_ConstantPrior_JADES_fixedRedshift.pdf}
\figsetgrpnote{\textit{Lower left}: Example of joint posterior distributions for the most relevant stellar populations properties for a typical galaxy (JADES$-$GS$-$ID$-$$173624$, or JADES$-$GS$+53.07688$$-$$27.86967$, at $z_{\mathrm{spec}} = 8.270$) from the final photometric sample of MIRI/F770W detections described in Section~\ref{SectionThreeOne}. From left to right: stellar mass, mass-weighted stellar age, sSFR averaged over the last $10$ Myr, SFR averaged over the same timescale, rest-frame EW of $\mathrm{[OIII]}+\mathrm{H}\beta$, and diffuse dust attenuation as measured in the $V$-band. \textit{Upper right}: Example of the derived SFHs. The solid lines (shaded regions) represent the median ($68\%$ confidence interval) of the \texttt{Prospector} models. \textit{Lower left and upper right}: The fiducial \texttt{Prospector} model is described in Section~\ref{SectionFourTwo} and assumes the constant SFH model. The blue (pink) lines and shaded regions represent results from fitting to the full JADES filter set, including (excluding) MIRI/F770W. Similar to Figure~\ref{figset:ExampleSEDs_1}, it is evident that for this typical galaxy, MIRI/F770W has little to no effect on the inferred stellar population properties and SFH.}
\figsetgrpend

\figsetgrpstart
\figsetgrpnum{C1.10}
\figsetplot{Corner_173679_ConstantPrior_JADES_freeRedshift.pdf}
\figsetgrpnote{\textit{Lower left}: Example of joint posterior distributions for the most relevant stellar populations properties for a typical galaxy (JADES$-$GS$-$ID$-$$173679$, or JADES$-$GS$+53.07277$$-$$27.86929$, at $z_{\mathrm{phot}} = 8.08 \pm 0.19$) from the final photometric sample of MIRI/F770W detections described in Section~\ref{SectionThreeOne}. From left to right: stellar mass, mass-weighted stellar age, sSFR averaged over the last $10$ Myr, SFR averaged over the same timescale, rest-frame EW of $\mathrm{[OIII]}+\mathrm{H}\beta$, and diffuse dust attenuation as measured in the $V$-band. \textit{Upper right}: Example of the derived SFHs. The solid lines (shaded regions) represent the median ($68\%$ confidence interval) of the \texttt{Prospector} models. \textit{Lower left and upper right}: The fiducial \texttt{Prospector} model is described in Section~\ref{SectionFourTwo} and assumes the constant SFH model. The blue (pink) lines and shaded regions represent results from fitting to the full JADES filter set, including (excluding) MIRI/F770W. Similar to Figure~\ref{figset:ExampleSEDs_1}, it is evident that for this typical galaxy, MIRI/F770W has little to no effect on the inferred stellar population properties and SFH.}
\figsetgrpend

\figsetgrpstart
\figsetgrpnum{C1.11}
\figsetplot{Corner_174121_ConstantPrior_JADES_fixedRedshift.pdf}
\figsetgrpnote{\textit{Lower left}: Example of joint posterior distributions for the most relevant stellar populations properties for a typical galaxy (JADES$-$GS$-$ID$-$$174121$, or JADES$-$GS$+53.05567$$-$$27.86882$, at $z_{\mathrm{spec}} = 7.623$) from the final photometric sample of MIRI/F770W detections described in Section~\ref{SectionThreeOne}. From left to right: stellar mass, mass-weighted stellar age, sSFR averaged over the last $10$ Myr, SFR averaged over the same timescale, rest-frame EW of $\mathrm{[OIII]}+\mathrm{H}\beta$, and diffuse dust attenuation as measured in the $V$-band. \textit{Upper right}: Example of the derived SFHs. The solid lines (shaded regions) represent the median ($68\%$ confidence interval) of the \texttt{Prospector} models. \textit{Lower left and upper right}: The fiducial \texttt{Prospector} model is described in Section~\ref{SectionFourTwo} and assumes the constant SFH model. The blue (pink) lines and shaded regions represent results from fitting to the full JADES filter set, including (excluding) MIRI/F770W. Similar to Figure~\ref{figset:ExampleSEDs_1}, it is evident that for this typical galaxy, MIRI/F770W has little to no effect on the inferred stellar population properties and SFH.}
\figsetgrpend

\figsetgrpstart
\figsetgrpnum{C1.12}
\figsetplot{Corner_175729_ConstantPrior_JADES_fixedRedshift.pdf}
\figsetgrpnote{\textit{Lower left}: Example of joint posterior distributions for the most relevant stellar populations properties for a typical galaxy (JADES$-$GS$-$ID$-$$175729$, or JADES$-$GS$+53.06058$$-$$27.86603$, at $z_{\mathrm{spec}} = 7.883$) from the final photometric sample of MIRI/F770W detections described in Section~\ref{SectionThreeOne}. From left to right: stellar mass, mass-weighted stellar age, sSFR averaged over the last $10$ Myr, SFR averaged over the same timescale, rest-frame EW of $\mathrm{[OIII]}+\mathrm{H}\beta$, and diffuse dust attenuation as measured in the $V$-band. \textit{Upper right}: Example of the derived SFHs. The solid lines (shaded regions) represent the median ($68\%$ confidence interval) of the \texttt{Prospector} models. \textit{Lower left and upper right}: The fiducial \texttt{Prospector} model is described in Section~\ref{SectionFourTwo} and assumes the constant SFH model. The blue (pink) lines and shaded regions represent results from fitting to the full JADES filter set, including (excluding) MIRI/F770W. Similar to Figure~\ref{figset:ExampleSEDs_1}, it is evident that for this typical galaxy, MIRI/F770W has little to no effect on the inferred stellar population properties and SFH.}
\figsetgrpend

\figsetgrpstart
\figsetgrpnum{C1.13}
\figsetplot{Corner_175837_ConstantPrior_JADES_fixedRedshift.pdf}
\figsetgrpnote{\textit{Lower left}: Example of joint posterior distributions for the most relevant stellar populations properties for a typical galaxy (JADES$-$GS$-$ID$-$$175837$, or JADES$-$GS$+53.06021$$-$$27.86572$, at $z_{\mathrm{spec}} = 7.884$) from the final photometric sample of MIRI/F770W detections described in Section~\ref{SectionThreeOne}. From left to right: stellar mass, mass-weighted stellar age, sSFR averaged over the last $10$ Myr, SFR averaged over the same timescale, rest-frame EW of $\mathrm{[OIII]}+\mathrm{H}\beta$, and diffuse dust attenuation as measured in the $V$-band. \textit{Upper right}: Example of the derived SFHs. The solid lines (shaded regions) represent the median ($68\%$ confidence interval) of the \texttt{Prospector} models. \textit{Lower left and upper right}: The fiducial \texttt{Prospector} model is described in Section~\ref{SectionFourTwo} and assumes the constant SFH model. The blue (pink) lines and shaded regions represent results from fitting to the full JADES filter set, including (excluding) MIRI/F770W. Similar to Figure~\ref{figset:ExampleSEDs_1}, it is evident that for this typical galaxy, MIRI/F770W has little to no effect on the inferred stellar population properties and SFH.}
\figsetgrpend

\figsetgrpstart
\figsetgrpnum{C1.14}
\figsetplot{Corner_177322_ConstantPrior_JADES_fixedRedshift.pdf}
\figsetgrpnote{\textit{Lower left}: Example of joint posterior distributions for the most relevant stellar populations properties for a typical galaxy (JADES$-$GS$-$ID$-$$177322$, or JADES$-$GS$+53.06036$$-$$27.86355$, at $z_{\mathrm{spec}} = 7.885$) from the final photometric sample of MIRI/F770W detections described in Section~\ref{SectionThreeOne}. From left to right: stellar mass, mass-weighted stellar age, sSFR averaged over the last $10$ Myr, SFR averaged over the same timescale, rest-frame EW of $\mathrm{[OIII]}+\mathrm{H}\beta$, and diffuse dust attenuation as measured in the $V$-band. \textit{Upper right}: Example of the derived SFHs. The solid lines (shaded regions) represent the median ($68\%$ confidence interval) of the \texttt{Prospector} models. \textit{Lower left and upper right}: The fiducial \texttt{Prospector} model is described in Section~\ref{SectionFourTwo} and assumes the constant SFH model. The blue (pink) lines and shaded regions represent results from fitting to the full JADES filter set, including (excluding) MIRI/F770W. Similar to Figure~\ref{figset:ExampleSEDs_1}, it is evident that for this typical galaxy, MIRI/F770W has little to no effect on the inferred stellar population properties and SFH.}
\figsetgrpend

\figsetgrpstart
\figsetgrpnum{C1.15}
\figsetplot{Corner_180446_ConstantPrior_JADES_fixedRedshift.pdf}
\figsetgrpnote{\textit{Lower left}: Example of joint posterior distributions for the most relevant stellar populations properties for a typical galaxy (JADES$-$GS$-$ID$-$$180446$, or JADES$-$GS$+53.08626$$-$$27.85932$, at $z_{\mathrm{spec}} = 7.956$) from the final photometric sample of MIRI/F770W detections described in Section~\ref{SectionThreeOne}. From left to right: stellar mass, mass-weighted stellar age, sSFR averaged over the last $10$ Myr, SFR averaged over the same timescale, rest-frame EW of $\mathrm{[OIII]}+\mathrm{H}\beta$, and diffuse dust attenuation as measured in the $V$-band. \textit{Upper right}: Example of the derived SFHs. The solid lines (shaded regions) represent the median ($68\%$ confidence interval) of the \texttt{Prospector} models. \textit{Lower left and upper right}: The fiducial \texttt{Prospector} model is described in Section~\ref{SectionFourTwo} and assumes the constant SFH model. The blue (pink) lines and shaded regions represent results from fitting to the full JADES filter set, including (excluding) MIRI/F770W. Similar to Figure~\ref{figset:ExampleSEDs_1}, it is evident that for this typical galaxy, MIRI/F770W has little to no effect on the inferred stellar population properties and SFH.}
\figsetgrpend

\figsetgrpstart
\figsetgrpnum{C1.16}
\figsetplot{Corner_300287_ConstantPrior_JADES_freeRedshift.pdf}
\figsetgrpnote{\textit{Lower left}: Example of joint posterior distributions for the most relevant stellar populations properties for a typical galaxy (JADES$-$GS$-$ID$-$$300287$, or JADES$-$GS$+53.08812$$-$$27.90817$, at $z_{\mathrm{phot}} = 8.08 \pm 0.10$) from the final photometric sample of MIRI/F770W detections described in Section~\ref{SectionThreeOne}. From left to right: stellar mass, mass-weighted stellar age, sSFR averaged over the last $10$ Myr, SFR averaged over the same timescale, rest-frame EW of $\mathrm{[OIII]}+\mathrm{H}\beta$, and diffuse dust attenuation as measured in the $V$-band. \textit{Upper right}: Example of the derived SFHs. The solid lines (shaded regions) represent the median ($68\%$ confidence interval) of the \texttt{Prospector} models. \textit{Lower left and upper right}: The fiducial \texttt{Prospector} model is described in Section~\ref{SectionFourTwo} and assumes the constant SFH model. The blue (pink) lines and shaded regions represent results from fitting to the full JADES filter set, including (excluding) MIRI/F770W. Similar to Figure~\ref{figset:ExampleSEDs_1}, it is evident that for this typical galaxy, MIRI/F770W has little to no effect on the inferred stellar population properties and SFH.}
\figsetgrpend

\figsetgrpstart
\figsetgrpnum{C1.17}
\figsetplot{Corner_300391_ConstantPrior_JADES_freeRedshift.pdf}
\figsetgrpnote{\textit{Lower left}: Example of joint posterior distributions for the most relevant stellar populations properties for a typical galaxy (JADES$-$GS$-$ID$-$$300391$, or JADES$-$GS$+53.08513$$-$$27.90636$, at $z_{\mathrm{phot}} = 8.29 \pm 0.08$) from the final photometric sample of MIRI/F770W detections described in Section~\ref{SectionThreeOne}. From left to right: stellar mass, mass-weighted stellar age, sSFR averaged over the last $10$ Myr, SFR averaged over the same timescale, rest-frame EW of $\mathrm{[OIII]}+\mathrm{H}\beta$, and diffuse dust attenuation as measured in the $V$-band. \textit{Upper right}: Example of the derived SFHs. The solid lines (shaded regions) represent the median ($68\%$ confidence interval) of the \texttt{Prospector} models. \textit{Lower left and upper right}: The fiducial \texttt{Prospector} model is described in Section~\ref{SectionFourTwo} and assumes the constant SFH model. The blue (pink) lines and shaded regions represent results from fitting to the full JADES filter set, including (excluding) MIRI/F770W. Similar to Figure~\ref{figset:ExampleSEDs_1}, it is evident that for this typical galaxy, MIRI/F770W has little to no effect on the inferred stellar population properties and SFH.}
\figsetgrpend

\figsetgrpstart
\figsetgrpnum{C1.18}
\figsetplot{Corner_380203_ConstantPrior_JADES_freeRedshift.pdf}
\figsetgrpnote{\textit{Lower left}: Example of joint posterior distributions for the most relevant stellar populations properties for a typical galaxy (JADES$-$GS$-$ID$-$$380203$, or JADES$-$GS$+53.08172$$-$$27.89881$, at $z_{\mathrm{phot}} = 8.03 \pm 0.26$) from the final photometric sample of MIRI/F770W detections described in Section~\ref{SectionThreeOne}. From left to right: stellar mass, mass-weighted stellar age, sSFR averaged over the last $10$ Myr, SFR averaged over the same timescale, rest-frame EW of $\mathrm{[OIII]}+\mathrm{H}\beta$, and diffuse dust attenuation as measured in the $V$-band. \textit{Upper right}: Example of the derived SFHs. The solid lines (shaded regions) represent the median ($68\%$ confidence interval) of the \texttt{Prospector} models. \textit{Lower left and upper right}: The fiducial \texttt{Prospector} model is described in Section~\ref{SectionFourTwo} and assumes the constant SFH model. The blue (pink) lines and shaded regions represent results from fitting to the full JADES filter set, including (excluding) MIRI/F770W. Similar to Figure~\ref{figset:ExampleSEDs_1}, it is evident that for this typical galaxy, MIRI/F770W has little to no effect on the inferred stellar population properties and SFH.}
\figsetgrpend

\figsetend

\figsetstart
\figsetnum{C}
\graphicspath{{./}{files/ExampleCorners/}}

\figsetgrpstart
\figsetgrpnum{C2.1}
\figsetplot{Corner_021468_ConstantPrior_JADES_fixedRedshift.pdf}
\figsetgrpnote{Similar to Figure~\ref{figset:ExampleCorners_1}, but for an atypical galaxy (JADES$-$GS$-$ID$-$$21468$, or JADES$-$GS$+53.10108$$-$$27.88310$, at $z_{\mathrm{spec}} = 8.808$) from the final photometric sample of MIRI/F770W detections described in Section~\ref{SectionThreeOne}. It is evident that for this atypical galaxy, MIRI/F770W has a significant effect on the inferred stellar population properties and SFH. The complete figure set of these atypical galaxies (four figures) is available in the online journal.}
\figsetgrpend

\figsetgrpstart
\figsetgrpnum{C2.2}
\figsetplot{Corner_066293_ConstantPrior_JADES_fixedRedshift.pdf}
\figsetgrpnote{Similar to Figure~\ref{figset:ExampleCorners_1}, but for an atypical galaxy (JADES$-$GS$-$ID$-$$66293$, or JADES$-$GS$+53.04601$$-$$27.85399$, at $z_{\mathrm{spec}} = 8.065$) from the final photometric sample of MIRI/F770W detections described in Section~\ref{SectionThreeOne}. It is evident that for this atypical galaxy, MIRI/F770W has a significant effect on the inferred stellar population properties and SFH. The complete figure set of these atypical galaxies (four figures) is available in the online journal.}
\figsetgrpend

\figsetgrpstart
\figsetgrpnum{C2.3}
\figsetplot{Corner_174693_ConstantPrior_JADES_fixedRedshift.pdf}
\figsetgrpnote{Similar to Figure~\ref{figset:ExampleCorners_1}, but for an atypical galaxy (JADES$-$GS$-$ID$-$$174693$, or JADES$-$GS$+53.06058$$-$$27.86795$, at $z_{\mathrm{spec}} = 7.882$) from the final photometric sample of MIRI/F770W detections described in Section~\ref{SectionThreeOne}. It is evident that for this atypical galaxy, MIRI/F770W has a significant effect on the inferred stellar population properties and SFH. The complete figure set of these atypical galaxies (four figures) is available in the online journal.}
\figsetgrpend

\figsetgrpstart
\figsetgrpnum{C2.4}
\figsetplot{Corner_179485_ConstantPrior_JADES_fixedRedshift.pdf}
\figsetgrpnote{Similar to Figure~\ref{figset:ExampleCorners_1}, but for an atypical galaxy (JADES$-$GS$-$ID$-$$179485$, or JADES$-$GS$+53.08738$$-$$27.86031$, at $z_{\mathrm{spec}} = 7.955$) from the final photometric sample of MIRI/F770W detections described in Section~\ref{SectionThreeOne}. It is evident that for this atypical galaxy, MIRI/F770W has a significant effect on the inferred stellar population properties and SFH. The complete figure set of these atypical galaxies (four figures) is available in the online journal.}
\figsetgrpend

\figsetend

\end{document}